\documentclass[12pt,twoside]{article}
\pdfoutput=1
\usepackage{latexsym}
\usepackage{graphicx}
\usepackage{epsfig}
\usepackage{url}
\usepackage{ifthen}
\usepackage{cite}
\begin{document}

\newcommand{\TschCurrentVersion}{Version 4.0 (January 6, 2009)}
\newcommand{\TschFirstVersion}{replaces Version 1.0 (July 7, 2007) and later}
\newcommand{\TschReferencesPageMode}{FromTo}
\newcommand{\TschStyle}{arXiv}
\newcommand{\TschStyles}{arXivOrTeX4ht}
\newcommand{\TschSpace}{\ }
\newcommand{\TschGerlich}{Gerhard Gerlich}
\newcommand{\TschTscheuschner}{Ralf D. Tscheuschner}
\newcommand{\TschAuthors}{\TschGerlich\ and \TschTscheuschner}
\newcommand{\TschTitleLineA}{Falsification Of}
\newcommand{\TschTitleLineB}{The Atmospheric CO$_2$ Greenhouse Effects}
\newcommand{\TschTitleLineBbf}{The Atmospheric
                               CO$_{\mbox{{\small\textbf{2}}}}$
                               Greenhouse Effects}
\newcommand{\TschTitleLineC}{Within The Frame Of Physics}
\newcommand{\TschTitleLineCdots}{$\dots$}
\newcommand{\TschTitle}{\TschTitleLineA\ \TschTitleLineB\ \TschTitleLineCdots}

%
%
\pagestyle{empty}
%
%
%
\begin{center}
\ \\[-1.0cm]
{\Large\bf \ }                                            \\[0.30cm]
{\Large\bf \TschTitleLineA}                               \\[0.30cm] 
{\Large\bf \TschTitleLineBbf}                             \\[0.30cm]
{\Large\bf \TschTitleLineC}                               \\[1.50cm]
%
%
{\large\bf \TschCurrentVersion}                           \\[0.00cm]
{\scriptsize\it \TschFirstVersion}                        \\[2.00cm]
%
%
{\Large\sc Gerhard Gerlich}                               \\[0.30cm]
{\large\rm Institut f\"ur Mathematische Physik}           \\[0.30cm]
{\large\rm Technische Universit\"at Carolo-Wilhelmina zu Braunschweig}    \\[0.30cm]
{\large\rm Mendelssohnstra\ss e 3}                        \\[0.30cm]
{\large\rm D-38106 Braunschweig}                          \\[0.30cm] 
{\large\rm Federal Republic of Germany}                   \\[0.30cm]
{\large\rm g.gerlich@tu-bs.de}                            \\[1.60cm]
%
%
{\Large\sc Ralf D. Tscheuschner}                          \\[0.30cm]
{\large\rm Postfach 60\,27\,62}                           \\[0.30cm]
{\large\rm D-22237 Hamburg}                               \\[0.30cm]
{\large\rm Federal Republic of Germany}                   \\[0.30cm]
{\large\rm ralfd@na-net.ornl.gov}                         \\[2.40cm]
\end{center}
%

%
\newpage
%
%
\pagestyle{myheadings}
\markboth{\TschAuthors}{\TschAuthors}
\markright{\TschTitle}
\addcontentsline{toc}{section}{Abstract}
\begin{abstract}
The atmospheric greenhouse effect, an idea that many authors trace back to the
traditional works of 
Fourier (1824), 
Tyndall (1861), 
and 
Arrhenius (1896), 
and
which is still supported in global climatology, essentially describes a fictitious
mechanism, in which a planetary atmosphere acts as a heat pump driven by an
environment that is
\textit{radiatively interacting with}
but
\textit{radiatively equilibrated to}
the atmospheric system. According to the second law of
thermodynamics such a planetary machine can never exist. Nevertheless, in
almost all texts of global climatology and in a widespread secondary literature
it is taken for granted that such mechanism is real and stands on a firm
scientific foundation. In this paper the popular conjecture is analyzed and the
underlying physical principles are clarified. By showing that (a) there are no
common physical laws between the warming phenomenon in glass houses and the
fictitious atmospheric greenhouse effects, (b) there are no calculations to
determine an average surface temperature of a planet, (c) the frequently
mentioned difference of 33~$^\circ{\rm C}$ is a meaningless number calculated
wrongly, (d) the formulas of cavity radiation are used inappropriately, (e) the
assumption of a radiative balance is unphysical, (f) thermal conductivity and
friction must not be set to zero, the atmospheric greenhouse conjecture is falsified.
\\
\\
\ifthenelse{\equal{arXivOrTeX4ht}{\TschStyles}}{%
Electronic version of an article published as
{\it International Journal of Modern Physics B, \mbox{Vol.\ 23},
\mbox{No.\ 3} (2009) 275--364\/},
DOI No: 10.1142/S021797920904984X,
\copyright\ World Scientific Publishing Company,
\url{http://www.worldscinet.com/ijmpb}.
}{}
\end{abstract}

\newpage
%
%
\tableofcontents
\newpage
%
%
\newpage%
\section{Introduction}
%
\subsection{Problem background}
\label{Sec:Background}
Recently, there have been lots of discussions 
regarding the economic and political implications
of climate variability, in particular global warming
as a measurable effect of an anthropogenic, 
i.e.\ human-made, climate change\TschSpace%
%
\ifthenelse{\equal{IJMPB}{\TschStyle}}
{%
\cite{%
Stockholm2006,%
IPCC2007summary,%
Svensmark1997,%
Heiss1999,%
Mann2003,%
Soon2003,%
Weart2004,%
Hardy2005,%
Avery2006,%
Khilyuk2006,%
Wegman2006,%
Jaworowski2007,%
Vatican2007%
}.
}{}%
\ifthenelse{\equal{arXiv}{\TschStyle}}
{%
\cite{%
Stockholm2006,%
IPCC2007summary,%
Svensmark1997,%
Heiss1999,%
Mann2003,%
Soon2003,%
Weart2004,%
Hardy2005,%
Avery2006,%
Khilyuk2006,%
Wegman2006,%
Jaworowski2007,%
Vatican2007%
}.
}{}%
\ifthenelse{\equal{TeX4ht}{\TschStyle}}
{%
\cite{%
Stockholm2006,%
IPCC2007summary,%
Svensmark1997,%
Heiss1999,%
Mann2003,%
Soon2003,%
Weart2004,%
Hardy2005,%
Avery2006,%
Khilyuk2006,%
Wegman2006,%
Jaworowski2007,%
Vatican2007%
}.
}{}%
%
Many authors assume that carbon dioxide emissions 
from fossil-fuel consumption represent a serious 
danger to the health of our planet, 
since they are supposed to influence the climates, 
in particular the average temperatures of the surface 
and lower atmosphere of the Earth.
However, carbon dioxide is a rare trace gas, 
a very small part of the atmosphere 
found in concentrations 
as low as 
$0,03\,{\rm Vol}\,\%$ 
(cf.\ Tables \ref{table:PB001} 
and  \ref{table:PB002},
see also
\ifthenelse{\equal{IJMPB}{\TschStyle}}
           {Ref.~\refcite{EngineeringToolbox2007}} 
           {}%
\ifthenelse{\equal{arXiv}{\TschStyle}}
           {Ref.~\cite{EngineeringToolbox2007}}
           {}%
\ifthenelse{\equal{TeX4ht}{\TschStyle}}
           {Ref.~\cite{EngineeringToolbox2007}}
           {}%
).%
\footnote{In a recent paper on 
\lq\lq 180 Years accurate CO2 Gas 
analysis of Air by Chemical Methods\rq\rq\ 
the German biologist Ernst-Georg Beck argues
that the IPCC reliance of ice core ${\rm CO}_2$ 
figures is wrong\TschSpace%
\cite{Beck2007,Beck2007Erratum}.
Though interesting on its own
that even the ${\rm CO}_2$ data themselves
are subject to a discussion it does not influence 
the rationale of this paper which is 
to show that the concentration of ${\rm CO}_2$
is \textit{completely} irrelevant.} 
\ifthenelse{\equal{IJMPB}{\TschStyle}}{%
\begin{table}[htbp] 
\tbl{Atmospheric concentration of carbon dioxide 
     in volume parts per million (1958 - 2007)}
{
\begin{tabular}{@{}ccc@{}} \Hline 
\\[-1.8ex] 
Date             & ${\rm CO}_2$ concentration & Source                     \\
                 & $[{\rm ppmv}]$        &                            \\[0.8ex] 
\hline                                        &                            \\[-1.8ex] 
March 1958       & $ 315.56 $                 & Ref.~\refcite{Lide2002}    \\
March 1967       & $ 322.88 $                 & Ref.~\refcite{Lide2002}    \\
March 1977       & $ 334.53 $                 & Ref.~\refcite{Lide2002}    \\
March 1987       & $ 349.24 $                 & Ref.~\refcite{Lide2002}    \\ 
March 1996       & $ 363.99 $                 & Ref.~\refcite{Lide2002}    \\
March 2007       & $ 377.3\phantom{0} $       & Ref.~\refcite{Blasing2007} \\[0.8ex]
\hline                                                                     \\[-1.8ex] 
\end{tabular}
}
\label{table:PB001}
\end{table}
}{}
\ifthenelse{\equal{arXivOrTeX4ht}{\TschStyles}}{%
\begin{table}[htbp] 
{
\begin{center}
\vspace*{0.5cm}
\begin{tabular}{|c|c|c|}
\hline 
Date             & ${\rm CO}_2$ concentration & Source                  \\
                 & $[{\rm ppmv}]$             &                         \\
\hline                                                                   
March 1958       & $ 315.56 $                 & Ref.~\cite{Lide2002}    \\
March 1967       & $ 322.88 $                 & Ref.~\cite{Lide2002}    \\
March 1977       & $ 334.53 $                 & Ref.~\cite{Lide2002}    \\
March 1987       & $ 349.24 $                 & Ref.~\cite{Lide2002}    \\ 
March 1996       & $ 363.99 $                 & Ref.~\cite{Lide2002}    \\
March 2007       & $ 377.3\phantom{0} $       & Ref.~\cite{Blasing2007} \\
\hline
\end{tabular}
\vspace*{0.0cm}
\end{center}
}
\caption{Atmospheric concentration of carbon dioxide 
         in volume parts per million (1958 - 2007).}
\label{table:PB001}
\vspace*{1.0cm}
\end{table}
}{}

\ifthenelse{\equal{IJMPB}{\TschStyle}}{%
\begin{table}[htbp] 
\tbl{Three versions of an idealized Earth's atmosphere
     and the associated gas volume concentrations, 
     including the working hypothesis chosen for this paper}
{
\begin{tabular}{@{}lcccc@{}} \Hline 
\\[-1.8ex] 
Gas            & Formula                     & U.S.\ Standard 1976             & Hardy \textit{et al.} 2005 & Working                 \\
               &                             & Ref.~\refcite{Lide2002}         & Ref.~\refcite{Hardy2005}  & hypothesis              \\
               &                             & [{\rm Vol}\,\%]                 & [{\rm Vol}\,\%]           & [{\rm Vol}\,\%]         \\[0.8ex] 
\hline                                                                                                                               \\[-1.8ex] 
Nitrogen       & ${\rm N}_2$                 & $\phantom{}78.084\phantom{0}$   & $\phantom{}78.09$         & $\phantom{}78.09$       \\
Oxygen         & ${\rm O}_2$                 & $\phantom{}20.9476\phantom{}$   & $\phantom{}20.95$         & $\phantom{}20.94$       \\
Argon          & ${\rm Ar}$                  & $\phantom{0}0.934\phantom{0}$   & $\phantom{0}0.93$         & $\phantom{0}0.93$       \\
Carbon dioxide & ${\rm CO}_2$                & $\phantom{0}0.0314\phantom{}$   & $\phantom{0}0.03$         & $\phantom{0}0.04$       \\[0.8ex]
\hline                                                                                                                               \\[-1.8ex] 
\end{tabular}
}
\label{table:PB002}
\end{table}
}{}
\ifthenelse{\equal{arXivOrTeX4ht}{\TschStyles}}{%
\begin{table}[htbp] 
{
\begin{center}
\vspace*{0.5cm}
\begin{tabular}{|l|c|c|c|c|}
\hline
Gas            & Formula                     & U.S.\ Standard 1976             & Hardy \textit{et al.} 2005 & Working                 \\
               &                             & Ref.~\cite{Lide2002}            & Ref.~\cite{Hardy2005}     & hypothesis              \\
               &                             & [{\rm Vol}\,\%]                 & [{\rm Vol}\,\%]           & [{\rm Vol}\,\%]         \\ 
\hline                                                                                                                                
Nitrogen       & ${\rm N}_2$                 & $\phantom{}78.084\phantom{0}$   & $\phantom{}78.09$         & $\phantom{}78.09$       \\
Oxygen         & ${\rm O}_2$                 & $\phantom{}20.9476\phantom{}$   & $\phantom{}20.95$         & $\phantom{}20.94$       \\
Argon          & ${\rm Ar}$                  & $\phantom{0}0.934\phantom{0}$   & $\phantom{0}0.93$         & $\phantom{0}0.93$       \\
Carbon dioxide & ${\rm CO}_2$                & $\phantom{0}0.0314\phantom{}$   & $\phantom{0}0.03$         & $\phantom{0}0.04$       \\
\hline                                                                                                                                
\end{tabular}
\vspace*{0.0cm}
\end{center}
}
\caption{Three versions of an idealized Earth's atmosphere
         and the associated gas volume concentrations, 
         including the working hypothesis chosen for this paper.}
\label{table:PB002}
\vspace*{0.5cm}
\end{table}
}{}

A physicist starts his analysis of the problem by pointing 
his attention to two fundamental thermodynamic properties, 
namely
\begin{itemize}
\item the \textit{thermal conductivity} $\lambda$, 
      a property that determines how much heat per time 
      unit and temperature difference flows in a medium;
\item the \textit{isochoric thermal diffusivity} 
      $a_{\rm v}$,
      a property that determines how rapidly
      a temperature change will spread,
      expressed in terms of  an area per time unit.
\end{itemize}      
Both quantities are related by 
\begin{equation}
a_{\rm v}
=
\frac{\lambda}{\varrho \, c_{\rm v}}
\end{equation}
the proportionality constant 
of the heat equation
\begin{equation}
\frac{\partial T}{\partial t}
=
a_{\rm v}
\cdot
\Delta T
\end{equation}
where
$T$ is the temperature,
$\varrho$ the mass density, 
and
$c_{\rm v}$ the isochoric specific heat.

To calculate the relevant data from the gaseous components of the air
one has to use their mass concentrations as weights to calculate the
properties of the mixture \lq\lq air\rq\rq\ 
according to Gibbs thermodynamics\TschSpace%
\cite{Callen1985,Huang1987}.%
%
\footnote{The thermal conductivity of a mixture of two gases
           does not, in general, vary linearly with the composition 
           of the mixture.
           However for comparable molecular weight and small 
           concentrations the non-linearity is negligible\TschSpace%
\cite{Evans1964}.%
}       
Data on volume concentrations
(Table \ref{table:PB002})
can be converted into mass concentrations 
with the aid of known mass densities 
(Table \ref{table:PB003}).
\ifthenelse{\equal{IJMPB}{\TschStyle}}{%
\begin{table}[htbp] 
\tbl{Mass densities of gases at normal atmospheric pressure 
     (101.325 ${\rm kPa}$) and standard temperature ($298\,{\rm K}$).}
{
\begin{tabular}{@{}lccc@{}} \Hline 
\\[-1.8ex] 
Gas              & Formula      & mass density\,$\varrho$  & Source                     \\
                 &              & $[{\rm kg}/{m}^3]$       &                            \\[0.8ex] 
\hline                                                                                  \\[-1.8ex] 
Nitrogen         & ${\rm N}_2$  & $ 1.1449 $               & Ref.~\refcite{Lide2002}    \\
Oxygen           & ${\rm O}_2$  & $ 1.3080 $               & Ref.~\refcite{Lide2002}    \\
Argon            & ${\rm Ar}$   & $ 1.6328 $               & Ref.~\refcite{Lide2002}    \\
Carbon Dioxide   & ${\rm CO}_2$ & $ 1.7989 $               & Ref.~\refcite{Lide2002}    \\[0.8ex] 
\hline                                                                                  \\[-1.8ex] 
\end{tabular}
}
\label{table:PB003}
\end{table}
}{}
\ifthenelse{\equal{arXivOrTeX4ht}{\TschStyles}}{%
\begin{table}[htbp] 
{
\begin{center}
\vspace*{0.5cm}
\begin{tabular}{|l|c|c|c|}
\hline 
Gas              & Formula      & mass density\,$\varrho$  & Source                     \\
                 &              & $[{\rm kg}/{m}^3]$       &                            \\ 
\hline                                                                                   
Nitrogen         & ${\rm N}_2$  & $ 1.1449 $               & Ref.~\cite{Lide2002}       \\
Oxygen           & ${\rm O}_2$  & $ 1.3080 $               & Ref.~\cite{Lide2002}       \\
Argon            & ${\rm Ar}$   & $ 1.6328 $               & Ref.~\cite{Lide2002}       \\
Carbon Dioxide   & ${\rm CO}_2$ & $ 1.7989 $               & Ref.~\cite{Lide2002}       \\ 
\hline                                                                                   
\end{tabular}
\end{center}
}
\caption{Mass densities of gases at normal atmospheric pressure 
         (101.325 ${\rm kPa}$) and standard temperature ($298\,{\rm K}$).}
\label{table:PB003}
\vspace*{0.5cm}
\end{table}
}{}

A comparison of volume percents and mass percents for ${\rm CO}_2$
shows that the current mass concentration, which is the physically 
relevant concentration, is approximately $0.06\,\%$ and not the
often quoted $0.03\,\%$
(Table \ref{table:PB004}).
%
\ifthenelse{\equal{IJMPB}{\TschStyle}}
           {\pagebreak}
           {}
\ifthenelse{\equal{arXiv}{\TschStyle}}
           {\pagebreak}
           {}
\ifthenelse{\equal{TeX4ht}{\TschStyle}}
           {\pagebreak}
           {}
%
%
\ifthenelse{\equal{IJMPB}{\TschStyle}}{%
\begin{table}[htbp] 
\tbl{Volume percent versus mass percent:
     The volume concentration $x_v$ and
     the mass concentration $x_m$ 
     of the gaseous components 
     of an idealized Earth's atmosphere.}
{
\begin{tabular}{@{}lcccc@{}} \Hline 
\\[-1.8ex] 
Gas            & Formula                     & $x_v$             & $\varrho\,(298\,{\rm K})$ & $x_m$              \\
               &                             & $[{\rm Vol}\,\%]$ & $[{\rm kg}/{\rm m}^3]$    & $[{\rm Mass}\,\%]$ \\[0.8ex] 
\hline                                                                                                            \\[-1.8ex] 
Nitrogen       & ${\rm N}_2$                 & $\phantom{}78.09$ & 1.1449                    & $\phantom{}75.52$  \\
Oxygen         & ${\rm O}_2$                 & $\phantom{}20.94$ & 1.3080                    & $\phantom{}23.14$  \\
Argon          & ${\rm Ar}$                  & $\phantom{0}0.93$ & 1.6328                    & $\phantom{0}1.28$  \\
Carbon dioxide & ${\rm CO}_2$                & $\phantom{0}0.04$ & 1.7989                    & $\phantom{0}0.06$  \\[0.8ex]
\hline                                                                                                            \\[-1.8ex] 
\end{tabular}
}
\label{table:PB004}
\end{table}
}{}
\ifthenelse{\equal{arXivOrTeX4ht}{\TschStyles}}{%
\begin{table}[htbp] 
{
\begin{center}
\vspace{0.5cm}
\begin{tabular}{|l|c|c|c|c|} 
\hline 
Gas            & Formula                     & $x_v$             & $\varrho\,(298\,{\rm K})$ & $x_m$              \\
               &                             & $[{\rm Vol}\,\%]$ & $[{\rm kg}/{\rm m}^3]$    & $[{\rm Mass}\,\%]$ \\ 
\hline                                                                                                             
Nitrogen       & ${\rm N}_2$                 & $\phantom{}78.09$ & 1.1449                    & $\phantom{}75.52$  \\
Oxygen         & ${\rm O}_2$                 & $\phantom{}20.94$ & 1.3080                    & $\phantom{}23.14$  \\
Argon          & ${\rm Ar}$                  & $\phantom{0}0.93$ & 1.6328                    & $\phantom{0}1.28$  \\
Carbon dioxide & ${\rm CO}_2$                & $\phantom{0}0.04$ & 1.7989                    & $\phantom{0}0.06$  \\
\hline                                                                                                             
\end{tabular}
\end{center}
}
\caption{Volume percent versus mass percent:
         The volume concentration $x_v$ and
         the mass concentration $x_m$ 
         of the gaseous components 
         of an idealized Earth's atmosphere.}
\label{table:PB004}
\vspace{0.5cm}
\end{table}
}{}

From known thermal conductivities
(Table \ref{table:PB005}), 
isochoric heat capacities,
and mass densities 
the isochoric thermal diffusivities 
of the
components of the air are determined  
(Table \ref{table:PB006}).
This allows to estimate the change of the effective
thermal conductivity of the air in dependence of a 
doubling of the ${\rm CO}_2$ concentration, expected
to happen within the next 300 years 
(Table \ref{table:PB007}).%
\ifthenelse{\equal{IJMPB}{\TschStyle}}{%
\begin{table}[htbp] 
\tbl{Thermal conductivities
     of the gaseous components 
     of the Earth's atmosphere
     at normal pressure ($101.325\,{\rm kPa}$). }
{
\begin{tabular}{@{}lccccc@{}} \Hline 
\\[-1.8ex] 
Gas            & Formula         & $\lambda(200\,{\rm K})$ & $\lambda(298\,{\rm K})$ & $\lambda(300\,{\rm K})$ & $\lambda(400\,{\rm K})$      \\
               &                 & $[{\rm W}/{\rm mK}]$    & $[{\rm W}/{\rm mK}]$    & $[{\rm W}/{\rm mK}]$    & $[{\rm W}/{\rm mK}]$         \\ 
               &                 & Ref.~\refcite{Lide2002} & (interpolated)          & Ref.~\refcite{Lide2002} & Ref.~\refcite{Lide2002}      \\[0.8ex]                
\hline \\[-1.8ex] 
Nitrogen       & ${\rm N}_2$     & 0.0187                  & 0.0259                  & 0.0260                  & 0.0323                       \\
Oxygen         & ${\rm O}_2$     & 0.0184                  & 0.0262                  & 0.0263                  & 0.0337                       \\
Argon          & ${\rm Ar}$      & 0.0124                  & 0.0178                  & 0.0179                  & 0.0226                       \\
Carbon dioxide & ${\rm CO}_2$    & 0.0096                  & 0.0167                  & 0.0168                  & 0.0251                       \\[0.8ex]
\hline                                                                                                                                        \\[-1.8ex] 
\end{tabular}
}
\label{table:PB005}
\end{table}
}{}
\ifthenelse{\equal{arXivOrTeX4ht}{\TschStyles}}{%
\begin{table}[htbp] 
{
\begin{center}
\vspace*{0.5cm}
\begin{tabular}{|l|c|c|c|c|c|} 
\hline 
Gas            & Formula         & $\lambda(200\,{\rm K})$ & $\lambda(298\,{\rm K})$ & $\lambda(300\,{\rm K})$ & $\lambda(400\,{\rm K})$      \\
               &                 & $[{\rm W}/{\rm mK}]$    & $[{\rm W}/{\rm mK}]$    & $[{\rm W}/{\rm mK}]$    & $[{\rm W}/{\rm mK}]$         \\ 
               &                 & Ref.~\cite{Lide2002}    & (interpolated)          & Ref.~\cite{Lide2002}    & Ref.~\cite{Lide2002}         \\
\hline 
Nitrogen       & ${\rm N}_2$     & 0.0187                  & 0.0259                  & 0.0260                  & 0.0323                       \\
Oxygen         & ${\rm O}_2$     & 0.0184                  & 0.0262                  & 0.0263                  & 0.0337                       \\
Argon          & ${\rm Ar}$      & 0.0124                  & 0.0178                  & 0.0179                  & 0.0226                       \\
Carbon dioxide & ${\rm CO}_2$    & 0.0096                  & 0.0167                  & 0.0168                  & 0.0251                       \\
\hline           
\end{tabular}
\end{center}
}
\caption{Thermal conductivities
         of the gaseous components 
         of the Earth's atmosphere
         at normal pressure ($101.325\,{\rm kPa}$). }
\label{table:PB005}
\vspace*{0.5cm}
\end{table}
}{}

\ifthenelse{\equal{IJMPB}{\TschStyle}}{%
\begin{table}[htbp] 
\tbl{Isobaric heat capacities $c_p$,
     relative molar masses $M_r$,  
     isochoric heat capacities $c_v \approx c_p - R/M_r$
     with universal gas constant $R = 8.314472\ {\rm J}/{\rm mol}\,{\rm K}$, 
     mass densities $\varrho$,
     thermal conductivities $\lambda$,
     and effective thermal conductivities 
     $\lambda_{\mbox{\scriptsize\rm eff}} = \lambda/ ( \varrho\cdot c_v )$  
     of the gaseous components 
     of the Earth's atmosphere
     at normal pressure ($101.325\,{\rm kPa}$). }
{
\begin{tabular}{@{}cccccccc@{}} \Hline 
\\[-1.8ex] 
Gas            & $c_p$                  
               & $M_r$                                                                  
               & $R/M_r$                   
               & $c_v$   
               & $\varrho$               
               & $\lambda$              
               & $\lambda_{\mbox{\scriptsize\rm eff}}$ 
\\ 
               & $[{\rm J}/{\rm kg\,K}]$ 
               & $[{\rm g}/{\rm mol}]$ 
               & $[{\rm J}/{\rm kg\,K}]$ 
               & $[{\rm J}/{\rm kg\,K}]$                      
               & $[{\rm kg}/{\rm m}^3]$ 
               & $[{\rm Js}/{\rm mK}]$     
               & $[{\rm m}^2/{\rm s}]$  
\\[0.8ex]
\hline 
\\[-1.8ex] 
${\rm N}_2$    & \phantom{}1039          
               & 28.01                  
               & 297                     
               & 742                       
               & 1.1449                 
               & 0.0259                
               & $3.038 \cdot 10^{-5}$                 
\\
${\rm O}_2$    & \phantom{0}919          
               & 32.00                  
               & 260                     
               & 659                       
               & 1.3080                 
               & 0.0262                
               & $3.040 \cdot 10^{-5}$                 
\\
${\rm Ar}$     & \phantom{0}521          
               & 39.95                  
               & 208                     
               & 304                       
               & 1.6328                 
               & 0.0178                
               & $3.586 \cdot 10^{-5}$                 
\\
${\rm CO}_2$   & \phantom{0}843          
               & 44.01                  
               & 189                     
               & 654                       
               & 1.7989                  
               & 0.0167                
               & $1.427 \cdot 10^{-5}$                 
\\[0.8ex]
\hline
\\[-1.8ex] 
\end{tabular}
}
\label{table:PB006}
\end{table}
}{}
\ifthenelse{\equal{arXivOrTeX4ht}{\TschStyles}}{%
\begin{table}[htbp] 
{
\begin{center}
\vspace*{0.5cm}
\begin{tabular}{|c|c|c|c|c|c|c|c|} 
\hline 
Gas            & $c_p$                  
               & $M_r$                                                                  
               & $R/M_r$                   
               & $c_{\rm v}$   
               & $\varrho$               
               & $\lambda$              
               & $a_{\rm v}$ 
\\ 
               & $[{\rm J}/{\rm kg\,K}]$ 
               & $[{\rm g}/{\rm mol}]$ 
               & $[{\rm J}/{\rm kg\,K}]$ 
               & $[{\rm J}/{\rm kg\,K}]$                      
               & $[{\rm kg}/{\rm m}^3]$ 
               & $[{\rm Js}/{\rm mK}]$     
               & $[{\rm m}^2/{\rm s}]$  
\\
\hline 
${\rm N}_2$    & \phantom{}1039          
               & 28.01                  
               & 297                     
               & 742                       
               & 1.1449                 
               & 0.0259                
               & $3.038 \cdot 10^{-5}$                 
\\
${\rm O}_2$    & \phantom{0}919          
               & 32.00                  
               & 260                     
               & 659                       
               & 1.3080                 
               & 0.0262                
               & $3.040 \cdot 10^{-5}$                 
\\
${\rm Ar}$     & \phantom{0}521          
               & 39.95                  
               & 208                     
               & 304                       
               & 1.6328                 
               & 0.0178                
               & $3.586 \cdot 10^{-5}$                 
\\
${\rm CO}_2$   & \phantom{0}843          
               & 44.01                  
               & 189                     
               & 654                       
               & 1.7989                  
               & 0.0167                
               & $1.427 \cdot 10^{-5}$                 
\\
\hline
\end{tabular}
\end{center}
}
\caption{Isobaric heat capacities $c_{\rm p}$,
         relative molar masses $M_r$,  
         isochoric heat capacities 
         $c_{\rm v} \approx c_{\rm p} - R/M_r$
         with universal gas constant $R = 8.314472\ {\rm J}/{\rm mol}\,{\rm K}$, 
         mass densities $\varrho$,
         thermal conductivities $\lambda$,
         and isochoric thermal diffusivities 
         $a_{\rm v}$ 
         of the gaseous components 
         of the Earth's atmosphere
         at normal pressure ($101.325\,{\rm kPa}$). }
\label{table:PB006}
\end{table}
}{}

\ifthenelse{\equal{IJMPB}{\TschStyle}}{%
\begin{table}[htbp] 
\tbl{The calculation of 
     the effective thermal conductivity 
     $\lambda_{\mbox{\scriptsize\rm eff}} = \lambda/ ( \varrho\cdot c_v )$
     of the air and its gaseous components
     for the current 
     ${\rm CO}_2$ concentration ($0.06\,{\rm Mass}\,\%$)
     and for a \textbf{fictitiously doubled} 
     ${\rm CO}_2$ concentration ($0.12\,{\rm Mass}\,\%$) 
     at normal pressure ($101.325\,{\rm kPa}$). }
{
\begin{tabular}{@{}cccccccc@{}} \Hline 
\\[-1.8ex] 
Gas            & $x_m$                  
               & $M_r$                                                                  
               & $c_p$                   
               & $c_v$   
               & $\varrho$               
               & $\lambda$              
               & $\lambda_{\mbox{\scriptsize\rm eff}}$ 
\\ 
               & $[{\rm Mass}\,\%]$ 
               & $[{\rm g}/{\rm mol}]$ 
               & $[{\rm J}/{\rm kg\,K}]$ 
               & $[{\rm J}/{\rm kg\,K}]$                      
               & $[{\rm kg}/{\rm m}^3]$ 
               & $[{\rm Js}/{\rm mK}]$     
               & $[{\rm m}^2/{\rm s}]$  
\\[0.8ex]
\hline 
\\[-1.8ex] 
${\rm N}_2$    & \phantom{0}75.52          
               & 28.01                  
               & \phantom{}1039                     
               & \phantom{0}742                       
               & 1.1449                 
               & 0.0259\phantom{0}                
               & $3.038\phantom{0} \cdot 10^{-5}$                 
\\
${\rm O}_2$    & \phantom{0}23.14          
               & 32.00                  
               & \phantom{0}929                     
               & \phantom{0}659                       
               & 1.3080                 
               & 0.0262\phantom{0}                
               & $3.040\phantom{0} \cdot 10^{-5}$                 
\\
${\rm Ar}$     & \phantom{00}1.28          
               & 39.95                  
               & \phantom{0}512                     
               & \phantom{0}304                       
               & 1.6328                 
               & 0.0178\phantom{0}                
               & $3.586\phantom{0} \cdot 10^{-5}$                 
\\
${\rm CO}_2$   & \phantom{00}0.06          
               & 44.01                  
               & \phantom{0}843                     
               & \phantom{0}654                       
               & 1.7989                 
               & 0.0167\phantom{0}                
               & $1.427\phantom{0} \cdot 10^{-5}$                 
\\[0.8ex]
\hline 
\\[-1.8ex] 
${\rm Air}$    & \phantom{}100.00          
               & 29.10                  
               & \phantom{}1005                     
               & \phantom{0}719                       
               & 1.1923                 
               & 0.02586\phantom{}                
               & $3.0166 \cdot 10^{-5}$                 
\\[0.8ex]
\hline
\\[0ex] 
\end{tabular}
}
{
\begin{tabular}{@{}cccccccc@{}} \Hline 
\\[-1.8ex] 
Gas            & $x_m$                  
               & $M_r$                                                                  
               & $c_p$                   
               & $c_v$   
               & $\varrho$               
               & $\lambda$              
               & $\lambda_{\mbox{\scriptsize\rm eff}}$ 
\\ 
               & $[{\rm Mass}\,\%]$ 
               & $[{\rm g}/{\rm mol}]$ 
               & $[{\rm J}/{\rm kg\,K}]$ 
               & $[{\rm J}/{\rm kg\,K}]$                      
               & $[{\rm kg}/{\rm m}^3]$ 
               & $[{\rm Js}/{\rm mK}]$     
               & $[{\rm m}^2/{\rm s}]$  
\\[0.8ex]
\hline 
\\[-1.8ex] 
${\rm N}_2$    & \phantom{0}75.52          
               & 28.01                  
               & \phantom{}1039                     
               & \phantom{0}742                       
               & 1.1449                 
               & 0.0259\phantom{0}                
               & $3.038\phantom{0} \cdot 10^{-5}$                 
\\
${\rm O}_2$    & \phantom{0}\textbf{23.08}          
               & 32.00                  
               & \phantom{0}929                     
               & \phantom{0}659                       
               & 1.3080                 
               & 0.0262\phantom{0}                
               & $3.040\phantom{0} \cdot 10^{-5}$                 
\\
${\rm Ar}$     & \phantom{00}1.28          
               & 39.95                  
               & \phantom{0}512                     
               & \phantom{0}304                       
               & 1.6328                 
               & 0.0178\phantom{0}                
               & $3.586\phantom{0} \cdot 10^{-5}$                 
\\
${\rm CO}_2$   & \phantom{00}\textbf{0.12}          
               & 44.01                  
               & \phantom{0}843                     
               & \phantom{0}654                       
               & 1.7989                 
               & 0.0167\phantom{0}                
               & $1.427\phantom{0} \cdot 10^{-5}$                 
\\[0.8ex]
\hline 
\\[-1.8ex] 
${\rm Air}$    & \phantom{}100.00          
               & \textbf{29.10}                  
               & \phantom{}\textbf{1005}                     
               & \phantom{0}\textbf{719}                       
               & \textbf{1.1926}                 
               & \textbf{0.02585\phantom{}}                
               & $\textbf{3.0146}\phantom{} \cdot \textbf{10}^\textbf{{-5}}$                 
\\[0.8ex]
\hline
\\[-1.8ex] 
\end{tabular}
}
\label{table:PB007}
\end{table}
}{}
\ifthenelse{\equal{arXivOrTeX4ht}{\TschStyles}}{%
\begin{table}[htbp] 
{
\begin{center}
\vspace*{0.5cm}
\begin{tabular}{|c|c|c|c|c|c|c|c|} 
\hline 
Gas            & $x_m$                  
               & $M_r$                                                                  
               & $c_{\rm p}$                   
               & $c_{\rm v}$   
               & $\varrho$               
               & $\lambda$              
               & $a_{\rm v}$ 
\\ 
               & $[{\rm Mass}\,\%]$ 
               & $[{\rm g}/{\rm mol}]$ 
               & $[{\rm J}/{\rm kg\,K}]$ 
               & $[{\rm J}/{\rm kg\,K}]$                      
               & $[{\rm kg}/{\rm m}^3]$ 
               & $[{\rm Js}/{\rm mK}]$     
               & $[{\rm m}^2/{\rm s}]$  
\\
\hline 
${\rm N}_2$    & \phantom{0}75.52          
               & 28.01                  
               & \phantom{}1039                     
               & \phantom{0}742                       
               & 1.1449                 
               & 0.0259\phantom{0}                
               & $3.038\phantom{0} \cdot 10^{-5}$                 
\\
${\rm O}_2$    & \phantom{0}23.14          
               & 32.00                  
               & \phantom{0}929                     
               & \phantom{0}659                       
               & 1.3080                 
               & 0.0262\phantom{0}                
               & $3.040\phantom{0} \cdot 10^{-5}$                 
\\
${\rm Ar}$     & \phantom{00}1.28          
               & 39.95                  
               & \phantom{0}512                     
               & \phantom{0}304                       
               & 1.6328                 
               & 0.0178\phantom{0}                
               & $3.586\phantom{0} \cdot 10^{-5}$                 
\\
${\rm CO}_2$   & \phantom{00}0.06          
               & 44.01                  
               & \phantom{0}843                     
               & \phantom{0}654                       
               & 1.7989                 
               & 0.0167\phantom{0}                
               & $1.427\phantom{0} \cdot 10^{-5}$                 
\\
\hline 
${\rm Air}$    & \phantom{}100.00          
               & 29.10                  
               & \phantom{}1005                     
               & \phantom{0}719                       
               & 1.1923                 
               & 0.02586\phantom{}                
               & $3.0166 \cdot 10^{-5}$                 
\\
\hline
\end{tabular}
\end{center}
}
{
\begin{center}
\vspace*{0.5cm}
\begin{tabular}{|c|c|c|c|c|c|c|c|} 
\hline 
Gas            & $x_m$                  
               & $M_r$                                                                  
               & $c_{\rm p}$                   
               & $c_{\rm v}$   
               & $\varrho$               
               & $\lambda$              
               & $a_{\rm v}$ 
\\ 
               & $[{\rm Mass}\,\%]$ 
               & $[{\rm g}/{\rm mol}]$ 
               & $[{\rm J}/{\rm kg\,K}]$ 
               & $[{\rm J}/{\rm kg\,K}]$                      
               & $[{\rm kg}/{\rm m}^3]$ 
               & $[{\rm Js}/{\rm mK}]$     
               & $[{\rm m}^2/{\rm s}]$  
\\
\hline 
${\rm N}_2$    & \phantom{0}75.52          
               & 28.01                  
               & \phantom{}1039                     
               & \phantom{0}742                       
               & 1.1449                 
               & 0.0259\phantom{0}                
               & $3.038\phantom{0} \cdot 10^{-5}$                 
\\
${\rm O}_2$    & \phantom{0}{\bf 23.08}          
               & 32.00                  
               & \phantom{0}929                     
               & \phantom{0}659                       
               & 1.3080                 
               & 0.0262\phantom{0}                
               & $3.040\phantom{0} \cdot 10^{-5}$                 
\\
${\rm Ar}$     & \phantom{00}1.28          
               & 39.95                  
               & \phantom{0}512                     
               & \phantom{0}304                       
               & 1.6328                 
               & 0.0178\phantom{0}                
               & $3.586\phantom{0} \cdot 10^{-5}$                 
\\
${\rm CO}_2$   & \phantom{00}{\bf 0.12}          
               & 44.01                  
               & \phantom{0}843                     
               & \phantom{0}654                       
               & 1.7989                 
               & 0.0167\phantom{0}                
               & $1.427\phantom{0} \cdot 10^{-5}$                 
\\
\hline 
${\rm Air}$    & \phantom{}100.00          
               & {\bf 29.10}                  
               & \phantom{}{\bf 1005}                     
               & \phantom{0}{\bf 719}                       
               & {\bf 1.1926}                 
               & {\bf 0.02585\phantom{}}                
               & ${\bf 3.0146}\phantom{} \cdot {\bf 10} ^{\bf -5}$                 
\\
\hline
\end{tabular}
\end{center}
}
\caption{The calculation of 
         the isochoric thermal diffusivity 
         $a_{\rm v} = \lambda/ ( \varrho \, c_{\rm v} )$
         of the air and its gaseous components
         for the current 
         ${\rm CO}_2$ concentration ($0.06\,{\rm Mass}\,\%$)
         and for a \textbf{fictitiously doubled} 
         ${\rm CO}_2$ concentration ($0.12\,{\rm Mass}\,\%$) 
         at normal pressure ($101.325\,{\rm kPa}$). }
\label{table:PB007}
\vspace*{0.5cm}
\end{table}
}{}

It is obvious that a doubling of the concentration 
of the trace gas ${\rm CO}_2$,
whose thermal conductivity is approximately 
one half than that of nitrogen and oxygen,
does change 
the thermal conductivity at the most by $0,03\,\%$ 
and 
the isochoric thermal diffusivity at the most by $0.07\,\%$. 
These numbers lie within the range of the measuring inaccuracy 
and other uncertainties such as rounding errors and therefore have 
no significance at all.
%
\ifthenelse{\equal{IJMPB}{\TschStyle}}{}{}
\ifthenelse{\equal{arXiv}{\TschStyle}}
           {\pagebreak}
           {}
\ifthenelse{\equal{TeX4ht}{\TschStyle}}
           {\pagebreak}
           {}
%

\subsection{The greenhouse effect hypothesis}
Among climatologists, 
in particular those who are 
affiliated with the Intergovernmental
Panel of Climate Change (IPCC)%
\footnote{The IPCC was created in 1988 
          by the World Meteorological Organization (WHO)
          and the United Nations Environmental Programme (UNEP).}%
, there is a \lq\lq scientific consensus\rq\rq\TschSpace%
%
\cite{AAAS2006}%
%
, that the relevant mechanism is the 
atmospheric greenhouse effect,
a mechanism heavily relying on
the assumption that  
radiative heat transfer 
clearly dominates over the other forms
of heat transfer such as thermal conductivity, 
convection, condensation \textit{et cetera}\TschSpace%
%
\cite{%
IPCC1990,%
IPCC1990SummaryForPolicymakers,%
IPCC1992,%
IPCC1994,%
IPCC1994RadiativeForcing,%
IPCC1996,%
IPCC2000,%
IPCC2001%
}.
%

In all past IPCC reports and 
other such scientific summaries 
the following point evocated 
in 
\ifthenelse{\equal{IJMPB}{\TschStyle}}
           {Ref.~\refcite{IPCC1990SummaryForPolicymakers}, p.\,5,}
           {}%
\ifthenelse{\equal{arXiv}{\TschStyle}}
           {Ref.~\cite{IPCC1990SummaryForPolicymakers}, p.\,5,}
           {}
\ifthenelse{\equal{TeX4ht}{\TschStyle}}
           {Ref.~\cite{IPCC1990SummaryForPolicymakers}, p.\,5,}
           {}
is central to the discussion:
\begin{quote}
\lq\lq 
One of the most important factors is the
\textbf{greenhouse effect}; 
a simplified explanation of which is as follows. 
Short-wave solar radiation can pass 
through the clear atmosphere relatively 
unimpeded.
But long-wave terrestrial radiation emitted 
by the warm surface of the Earth is partially 
absorbed and then re-emitted by a number of 
trace gases in the cooler atmosphere above. 
Since, on average, the outgoing long-wave 
radiation balances the incoming solar radiation, 
both the atmosphere and the surface 
will be warmer than they would be 
without the greenhouse gases
$\dots$
The greenhouse effect is real;
it is a well understood effect,
based on established scientific 
principles.\rq\rq
\end{quote}
To make things more precise, supposedly,     
the notion of 
\textit{radiative forcing} 
was introduced by the IPCC 
and related to the assumption of 
\textit{radiative equilibrium}.
In
\ifthenelse{\equal{IJMPB}{\TschStyle}}
           {Ref.~\refcite{IPCC1994RadiativeForcing}, pp.\,7-6,} 
           {}%
\ifthenelse{\equal{arXiv}{\TschStyle}}
           {Ref.~\cite{IPCC1994RadiativeForcing}, pp.\,7-6,}
           {}%
\ifthenelse{\equal{TeX4ht}{\TschStyle}}
           {Ref.~\cite{IPCC1994RadiativeForcing}, pp.\,7-6,}
           {}
one finds the statement:
\begin{quote}
\lq\lq
\textit{A change} 
in average net radiation
at the top of the troposphere 
(known as the tropopause), 
because of a change in either
solar or infrared radiation, 
is defined for the purpose of this report 
as a 
\textit{radiative forcing}. 
A radiative forcing perturbs the balance 
between incoming and outgoing radiation.
Over time climate responds to the perturbation
to re-establish the radiative balance.
A positive radiative forcing tends on average 
to warm the surface;
a negative radiative forcing on average
tends to cool the surface.
As defined here, the incoming solar
radiation is not considered a radiative forcing,
but a change in the amount of incoming solar radiation
would be a radiative forcing
$\dots$
For example, an increase in atmospheric 
${\rm CO}_2$ concentration leads to a
reduction in outgoing infrared radiation
and a positive radiative forcing.%
\rq\rq  
\end{quote}

\noindent%
However, in general \lq\lq scientific consensus\rq\rq\ 
is not related whatsoever to scientific truth 
as countless examples in history have shown. 
\lq\lq Consensus\rq\rq\ is a political term, 
not a scientific term.
In particular, from the viewpoint 
of theoretical physics the radiative approach, 
which uses physical laws
such as Planck's law and Stefan-Boltzmann's law
that only have a limited range of validity
that definitely does not cover the atmospheric 
problem, must be highly questioned\TschSpace%
\cite{Stefan1879,Boltzmann1884,Planck1900,Planck1901,Rybicki1979}.
For instance in many calculations 
climatologists perform calculations 
where idealized black surfaces 
e.g.\ representing 
a ${\rm CO}_2$ layer and
the ground, respectively, radiate
against each other. In reality, we must
consider a bulk problem, in which 
at concentrations of 300~ppmv
at normal state still 
\begin{eqnarray}
N 
&\approx&
3 \cdot 10^{-4} 
\,\cdot\,
V
\,\cdot\,
N_L
\nonumber \\
&\approx&
3 \cdot 10^{-4} 
\,\cdot\,
(10 \cdot 10^{-6}) ^3
\,\cdot\,
2.687 \cdot 10^{25}
\nonumber \\
&=&
3 \cdot 10^{-4} 
\,\cdot\,
10^{-15}
\,\cdot\,
2.687 \cdot 10^{25}
\nonumber \\
&\approx&
8 \cdot 10^{6}
\end{eqnarray}
${\rm CO}_2$ molecules are distributed within a cube $V$
with edge length $10\,\mu m$, 
a typical wavelength 
of the relevant infrared radiation.%
\footnote{$N_L$ is determined by the well-known Loschmidt number\TschSpace%
%
\cite{Virgo1933}%
%
.}
In this context an application of the formulas
of cavity radiation is sheer nonsense.

It cannot be overemphasized that 
a microscopic theory providing the base 
for a derivation of macroscopic quantities 
like thermal or electrical transport coefficients 
must be a highly involved many-body theory. 
Of course, heat transfer is due to interatomic 
electromagnetic interactions mediated 
by the electromagnetic field. 
But it is misleading to visualize a photon as 
a simple particle or wave packet travelling 
from one atom to another for example. 
Things are pretty much more complex and cannot be 
understood even in a (one-)particle-wave duality 
or Feynman graph picture.

On the other hand, the macroscopic thermodynamical 
quantities contain a lot of information and can be 
measured directly and accurately in the physics lab.
It is an interesting point that the thermal conductivity
of ${\rm CO}_2$ is only one half of that of nitrogen
or oxygen. In a 100 percent ${\rm CO}_2$ atmosphere
a conventional light bulb shines brighter
than in a nitrogen-oxygen atmosphere due to
the lowered thermal conductivity of its environment. 
But this has nothing to do 
with the supposed ${\rm CO}_2$ 
greenhouse effect which refers
to trace gas concentrations. Global climatologists
claim that the Earth's natural greenhouse effect 
keeps the Earth 33~$^\circ{\rm C}$ warmer than it 
would be without the trace gases in the atmosphere. 
About 80 percent of this warming is attributed to water vapor 
and 20 percent to the 0.03 volume percent ${\rm CO}_2$. 
If such an extreme effect existed, it would show up even in a
laboratory experiment involving concentrated ${\rm CO}_2$    
as a thermal conductivity anomaly. It would manifest itself 
as a new kind of \lq superinsulation\rq\
violating the conventional heat conduction equation.
However, for ${\rm CO}_2$  such anomalous heat transport
properties never have been observed.

Therefore, in this paper, the popular greenhouse 
ideas entertained by the global climatology 
community are reconsidered within the limits 
of theoretical and experimental physics. 
Authors trace back their origins  
to the works of Fourier\TschSpace%
\cite{Fourier1824a,Fourier1824b}
%
(1824), 
Tyndall\TschSpace%
\cite{Tyndall1861,Tyndall1863a,Tyndall1863b,Tyndall1873a,Tyndall1873b}
%
(1861)
and 
Arrhenius\TschSpace%
\cite{Arrhenius1896,Arrhenius1901,Arrhenius1906}
%
(1896). 
A careful analysis of the original papers shows
that Fourier's and Tyndall's works did not really 
include the concept of the atmospheric greenhouse 
effect, whereas Arrhenius's work fundamentally 
differs from the versions of today.
With exception of
%
\ifthenelse{\equal{IJMPB}{\TschStyle}}
           {Ref.~\refcite{Arrhenius1906},}
           {}%
\ifthenelse{\equal{arXiv}{\TschStyle}}
           {Ref.~\cite{Arrhenius1906},}
           {}
\ifthenelse{\equal{TeX4ht}{\TschStyle}}
           {Ref.~\cite{Arrhenius1906},}
           {}
%
the traditional works precede the 
seminal papers of modern physics,
such as Planck's work on the radiation
of a black body\TschSpace%
\cite{Planck1900,Planck1901}.
%
Although the arguments of Arrhenius were falsified 
by his contemporaries they were picked up by 
Callendar\TschSpace%
%
\cite{%
Callendar1938,%
Callendar1939,%
Callendar1940,%
Callendar1941,%
Callendar1949,%
Callendar1958,%
Callendar1961%
}
%
and Keeling\TschSpace%
%
\cite{%
Keeling1960,%
Keeling1973,%
Keeling1976,%
Keeling1978,%
Keeling1989,%
Keeling1996,%
Keeling1998%
}%
%
, the founders of the modern 
greenhouse hypothesis.%
\footnote{%
Recently, von\,Storch critized 
the anthropogenic global warming scepticism
by characterizing the discussion as 
\lq\lq a discussion of yesterday and the day before 
yesterday\rq\rq\TschSpace%
\cite{Stockholm2006}.
Ironically, it was Calendar and Keeling 
who once reactivated \lq\lq a discussion
of yesterday and the day before yesterday\rq\rq\
based on \textit{already falsified} arguments.
}
Interestingly, 
this hypothesis has been vague
ever since it has been used.
Even Keeling stated 1978\TschSpace%
%
\cite{Keeling1978}:
%
\begin{quote}
\lq\lq 
The idea that ${\rm CO}_2$ from fossil fuel burning 
might accumulate in air and cause warming of the 
lower atmosphere was speculated upon as early as the latter
the nineteenth century (Arrhenius, 1903). 
At that time the use of fossil fuel was
slight to expect a rise in atmospheric 
${\rm CO}_2$ to be detectable. The idea was
convincingly expressed by Callendar (1938, 1940) 
but still without solid evidence
rise in ${\rm CO}_2$.\rq\rq\
\end{quote} 
The influence of 
${\rm CO}_2$
on the climate was also discussed thoroughly 
in a number of publications that appeared
between 1909 and 1980, 
mainly in Germany\TschSpace%
\cite{%
Albrecht1933,%
Albrecht1935,%
Albrecht1951,%
Barker1933,%
Baur1934,%
Baur1935,%
Cess1990,%
Curtis1956,%
deBary1954,%
Gold1909,%
Gribbin1978,%
Hofmann1955,%
Manabe1964,%
Manabe1967,%
Manabe1969a,%
Manabe1969b,%
Manabe1980,%
Martin1932,%
Mecke1921,%
Moeller1933,%
Moeller1943a,%
Moeller1943b,%
Moeller1954,%
Moeller1959,%
Moeller1961,%
Muegge1932,%
Schaefer1926,%
Wimmer1926}.
The most influential authors were
M\"oller\TschSpace%
\cite{%
deBary1954,%
Moeller1933,%
Moeller1943a,%
Moeller1943b,%
Moeller1954,%
Moeller1959,%
Moeller1961,%
Muegge1932},
who also wrote a textbook on
meteorology\TschSpace%
\cite{%
Moeller1973a,%
Moeller1973b},
and Manabe\TschSpace%
\cite{%
Manabe1964,%
Manabe1967,%
Manabe1969a,%
Manabe1969b,%
Manabe1980,%
Moeller1961}. 
It seems, that the joint work of 
M\"oller and Manabe\TschSpace%
\cite{%
Moeller1961}
has had a significant influence 
on the formulation of the modern 
atmospheric ${\rm CO}_2$ 
greenhouse conjectures and hypotheses,
respectively.

In a very comprehensive report
of the US Department of Energy (DOE),
which appeared in 1985\TschSpace%
%
\cite{DOE1985},
%
the atmospheric greenhouse hypothesis
had been cast into its final form
and became the cornerstone in all
subsequent IPCC publications\TschSpace%
%
\cite{%
IPCC1990,%
IPCC1990SummaryForPolicymakers,%
IPCC1992,%
IPCC1994,%
IPCC1994RadiativeForcing,%
IPCC1996,%
IPCC2001,%
IPCC2000%
}.
%

Of course, it may be that even if the oversimplified picture
entertained in IPCC global climatology
is physically incorrect, a thorough discussion may
reveal a non-neglible influence of certain radiative
effects (apart from sunlight) on the weather, and hence 
on its local averages, the climates, which may be dubbed 
the ${\rm CO}_2$ greenhouse effect.
But then three key questions will remain, 
even if the effect is claimed to serve 
only as a genuine trigger of a network 
of complex reactions:

\begin{enumerate} 
\item Is there a fundamental ${\rm CO}_2$ 
      greenhouse effect in physics?
\item If so, what is the fundamental physical principle  
      behind this ${\rm CO}_2$ greenhouse effect?
\item Is it physically correct to consider 
      radiative heat transfer as the fundamental
      mechanism controlling the weather setting
      thermal conductivity and friction to zero?
\end{enumerate}        

The aim of this paper is to give an 
affirmative \textit{negative} answer 
to all of these questions rendering 
them rhetoric.  
%
%

\subsection{This paper}
In the language of physics 
\textit{an effect} is a not necessarily evident 
but a reproducible \textit{and} measurable phenomenon
\textit{together with} its theoretical explanation. 

\textit{Neither} the warming mechanism in a glass house
\textit{nor} the supposed anthropogenic warming
is due to an effect in the sense of this definition:
\begin{itemize}
\item In the first case (the glass house) one encounters 
      a straightforward phenomenon.
\item In the second case (the Earth's atmosphere) 
      one cannot measure something;
      rather, one only makes heuristic calculations.
\end{itemize}

The explanation of the warming mechanism
in a real greenhouse is a standard problem
in undergraduate courses, 
in which optics, nuclear physics 
and classical radiation theory are dealt with. 
On this level neither the mathematical 
formulation of the first and second law of thermodynamics 
nor the partial differential equations of hydrodynamics 
or irreversible thermodynamics are known; 
the phenomenon has thus to be analyzed
with comparatively elementary means.

However, looking up the search terms 
\lq\lq glass house effect\rq\rq, 
\lq\lq greenhouse effect\rq\rq, 
or the German word
\lq\lq Treibhauseffekt\rq\rq\ 
in classical textbooks 
on experimental physics or theoretical physics, 
one finds 
- possibly to one's surprise and disappointment - 
that this effect does not appear anywhere 
- with a few exceptions, where in updated editions
of some books publications in climatology are cited.
One prominent example is the textbook by Kittel
who added a \lq\lq supplement\rq\rq\ to the 1990 edition
of his Thermal Physics on page 115\TschSpace%
%
\cite{Kittel2000} :
%
\begin{quote} 
"The Greenhouse Effect describes the warming
of the surface of the Earth caused by the infrared
absorbent layer of water, as vapor and in clouds,
and of carbon dioxide on the atmosphere between
the Sun and the Earth. The water may contribute 
as much as 90 percent of the warming effect."   
\end{quote}
Kittel's \lq\lq supplement\rq\rq\ 
refers to the 1990 and 1992 books 
of J.T.\ Houghton \textit{et al.} on Climate Change,
which are nothing but the standard IPCC assessments\TschSpace%
%
\cite{IPCC1990,IPCC1992}.
%
In general, most climatologic texts do not refer
to any fundamental work of thermodynamics and 
radiation theory.
Sometimes the classical astrophysical work of 
Chandrasekhar\TschSpace%
%
\cite{Chandrasekhar1960}
%
is cited,
but it is not clear at all, which results are applied
where, and how the conclusions of Chandrasekhar fit 
into the framework of infrared radiation transfer 
in planetary atmospheres.

There seems to exist no source where an atmospheric greenhouse 
effect is introduced from fundamental university physics alone.
 
Evidently, the atmospheric greenhouse problem is not a fundamental 
problem of the philosophy of science, which is best described by 
the M\"unchhausen trilemma%
\footnote{%
The term was coined by the critical rationalist Hans Albert,
see e.g.\ 
%
\ifthenelse{\equal{IJMPB}{\TschStyle}}
           {Ref.~\refcite{HansAlbert1985}.}
           {}%
\ifthenelse{\equal{arXiv}{\TschStyle}}
           {Ref.~\cite{HansAlbert1985}.}
           {}
\ifthenelse{\equal{TeX4ht}{\TschStyle}}
           {Ref.~\cite{HansAlbert1985}.}
           {}
%
For the current discussion on global warming
Albert's work may be particularly interesting. 
According to Albert new insights are not easy 
to be spread, because there is often an ideological 
obstacle, for which Albert coined the 
notion of \textit{immunity against criticism}.}%
, stating that one is left 
with the ternary alternative%
\footnote{%
Originally, an \textit{alternative} is a choice
between two options, not one of the options itself.
A ternary alternative generalizes an ordinary alternative
to a threefold choice.}  
\begin{quote}
\textit{infinite regression} - 
\textit{dogma} - 
\textit{circular reasoning}
\end{quote}
Rather, the atmospheric greenhouse mechanism is a
conjecture, which may be proved or disproved
already in concrete engineering thermodynamics\TschSpace%
\cite{AlfredSchackBook,Kreith1999,Baukal2000}.
Exactly this was done well many years ago by an expert
in this field, namely Alfred Schack, who wrote
a classical textbook on this subject\TschSpace%
%
\cite{AlfredSchackBook}.
%
1972 he showed that the radiative component of heat transfer
of ${\rm CO}_2$, though relevant at the temperatures in 
combustion chambers, can be neglected at atmospheric
temperatures. The influence of carbonic acid on
the Earth's climates is definitively unmeasurable\TschSpace%
%
\cite{AlfredSchack1972}.

The remaining part of this paper is organized as follows:
\begin{itemize}
\item In Section 2 the warming effect in real greenhouses,
      which has to be distinguished strictly from the (in-)\,famous 
      conjecture of Arrhenius, is discussed.
\item Section 3 is devoted to the atmospheric greenhouse problem. 
      It is shown that this effect neither has experimental nor 
      theoretical foundations and must be considered as 
      \textit{fictitious.} 
      The claim that
      ${\rm CO}_2$ emissions give rise to anthropogenic climate changes
      has no physical basis. 
\item In Section 4 theoretical physics and climatology 
      are discussed in context of the philosophy of science.
      The question is raised, 
      how far global climatology 
      fits into the framework of exact 
      sciences such as physics.
\item The final Section 5 is a physicist's summary.
\end{itemize}

\newpage%
\section{The warming mechanism in real greenhouses}
%
%
\subsection{Radiation Basics}
%
%
\subsubsection{Introduction}
For years, the warming mechanism in real greenhouses,
paraphrased as \lq\lq the greenhouse effect\rq\rq,
has been commonly misused to explain the conjectured 
atmospheric greenhouse effect. 
In school books, in popular scientific
articles, and even in high-level scientific debates,
it has been stated that the mechanism observed
within a glass house bears some similarity to
the anthropogenic global warming.
Meanwhile, even mainstream climatologists admit
that the warming mechanism in real glass houses
has to be distinguished strictly from the 
claimed ${\rm CO}_2$ greenhouse effect.

Nevertheless, one should have a look 
at the classical glass house problem
to recapitulate some fundamental principles
of thermodynamics and radiation theory.
Later on, the relevant radiation dynamics 
of the atmospheric system will be elaborated on
and distinguished from the glass house set-up. 

\textit{Heat} is the kinetic energy of molecules and atoms
and will be transferred by contact or radiation.
Microscopically both interactions are mediated by photons.
In the former case, which is governed by the
Coulomb respective van\,der\,Waals interaction 
these are the virtual or off-shell photons,
in the latter case these are the real or on-shell
photons. The interaction between photons and
electrons (and other particles that are electrically
charged or have a non-vanishing magnetic momentum)
is microscopically described by the laws of quantum 
theory. Hence, in principle, thermal conductivity
and radiative transfer may be described in a unified framework.
However, the non-equilibrium many body problem is 
a highly non-trivial one and subject to the discipline
of physical kinetics unifying quantum theory
and non-equilibrium statistical mechanics.

Fortunately, an analysis of the problem by applying the
methods and results of classical radiation theory already leads 
to interesting insights.  
\subsubsection{The infinitesimal specific intensity}
\label{Sec:ISI}
\begin{figure}[htbp]
\ifthenelse{\equal{IJMPB}{\TschStyle}}{%
\centerline{\psfig{file=PictureLarge_specific_intensity_.eps,width=3.00in}}
}{}%
\ifthenelse{\equal{arXiv}{\TschStyle}}{%
\centerline{\includegraphics[scale=2.00]{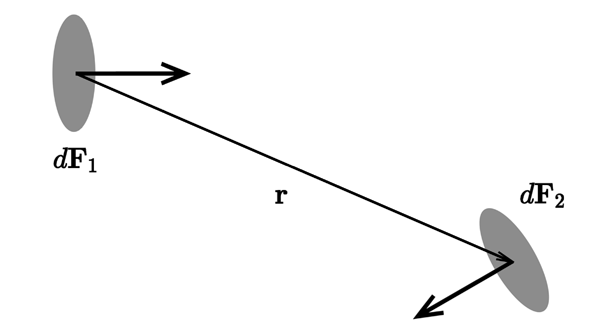}}
}{}%
\ifthenelse{\equal{TeX4ht}{\TschStyle}}{%
\centerline{\includegraphics[scale=0.66]{PictureLarge_specific_intensity_.png}}
}{}
\vspace*{8pt}
\caption{The geometry of classical radiation:
         A radiating infinitesimal area 
         $d\textbf{F}_1$
         and 
         an illuminated infinitesimal area 
         $d\textbf{F}_2$
         at distance 
         $\textbf{r}$.}
\label{Fig:dF1dF2}
\end{figure}

In classical radiation theory\TschSpace%
%
\cite{Chandrasekhar1960}
%
the main quantity is the 
\textit{specific intensity} 
$I_\nu$.
It is defined in terms of the 
\textit{amount of radiant energy}
$dE_\nu$ 
in a specified frequency interval 
$[\nu,\nu+d\nu]$
that is transported 
across an area element 
$d\textbf{F}_1$
in direction of another area element 
$d\textbf{F}_2 $ 
during a time
$dt$:

\begin{equation}
dE_\nu
=
I_\nu\,
d\nu\,
dt\,
\frac { (\textbf{r}\,d\textbf{F}_1)\,(\textbf{r}\,d\textbf{F}_2) } 
      { |\textbf{r}|^4 }
\label{Eq:dEv1}      
\end{equation}
where $\textbf{r}$ is the distance vector 
pointing from $d\textbf{F}_1$ to $d\textbf{F}_2$
(Figure \ref{Fig:dF1dF2}).

For a general radiation field one may write
\begin{equation}
I_\nu = I_\nu(x,y,z;l,m,n;t)
\end{equation}
where 
$(x,y,z)$ denote the coordinates,
$(l,m,n)$ the direction cosines,
$t$ the time, respectively, 
to which $I_\nu$ refers.

With the aid of the definition
of the scalar product
Equation (\ref{Eq:dEv1}) 
may be cast into the form 
\begin{equation}
dE_\nu
=
I_\nu\,
d\nu\,
dt\,
\cdot
\frac { (\cos\vartheta_1\,d{\rm F}_1) \cdot 
        (\cos\vartheta_2\,d{\rm F}_2) } 
      { {\rm r}^2 }
\label{Eq:dEv2}      
\end{equation}
A special case is given by
\begin{equation}
\cos\vartheta_2 = 1
\end{equation}
With 
\begin{eqnarray}
\vartheta &=& \vartheta_1 \nonumber\\
d\sigma   &=& dF_1        \nonumber\\
d\omega   &=& dF_2/r^2    \nonumber\\
\end{eqnarray}
Equation~(\ref{Eq:dEv2}) becomes
\begin{equation}
dE_\nu 
= 
I_\nu\,
d\nu\,
dt\,
\cos\vartheta\,
d\sigma\,
d\omega
\label{Eq:dEv3} 
\end{equation}
defining the \textit{pencil of radiation}%
%
\cite{Chandrasekhar1960}.
%

Equation (\ref{Eq:dEv2}), which will be used below,  
is slightly more general than Equation (\ref{Eq:dEv3}), 
which is more common in the literature.
Both ones can be simplified 
by introducing an \textit{integrated intensity}
\begin{equation}
I_0 = \int_0^\infty I_\nu\,d\nu
\end{equation}
and a \textit{radiant power} $dP$.
For example, Equation (\ref{Eq:dEv2})
may be cast into the form 
\begin{equation}
dP
=
I_0
\cdot
\frac { (\cos\vartheta_1\,d{\rm F}_1) \cdot 
        (\cos\vartheta_2\,d{\rm F}_2) } 
      { {\rm r}^2 }
\label{Eq:I}      
\end{equation}
%
%
\subsubsection{Integration}
When performing integration 
one has to bookkeep the dimensions 
of the physical quantities involved. 
Usually, the area $dF_1$ is integrated
and the equation is rearranged in such 
a way, that there is an intensity $I$
(resp.\ an intensity times 
an area element $IdF$) 
on both sides of the equation.
Three cases are particularly interesting:
\begin{itemize}
\item[(a)] 
\textit{Two parallel areas with distance $a$.}
\begin{figure}[htbp]
\begin{center}
\ifthenelse{\equal{IJMPB}{\TschStyle}}{%
\centerline{\psfig{file=PictureLarge_specific_intensity_parallel_areas_.eps,width=2.50in}}
}{}%
\ifthenelse{\equal{arXiv}{\TschStyle}}{%
\centerline{\includegraphics[scale=1.00]{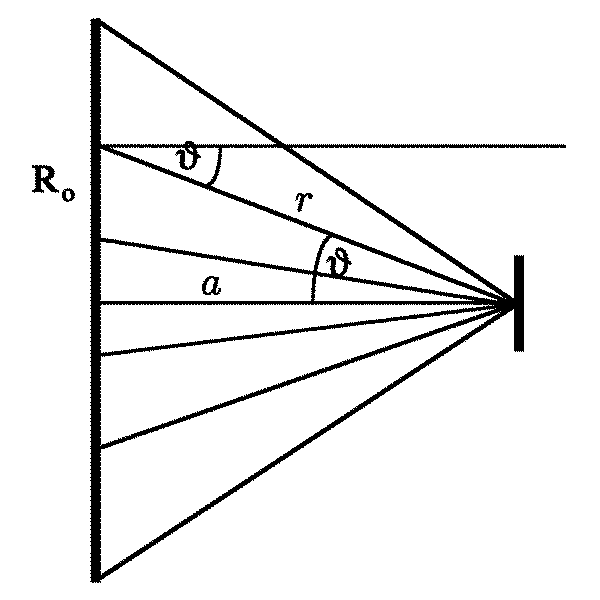}}
}{}%
\ifthenelse{\equal{TeX4ht}{\TschStyle}}{%
\centerline{\includegraphics[scale=0.33]{PictureLarge_specific_intensity_parallel_areas_.png}}
}{}
\vspace*{8pt}
\caption{Two parallel areas with distance $a$.}
\label{Fig:sipa}
\end{center}
\end{figure}
According to Figure~\ref{Fig:sipa} one may write
\begin{equation}
\vartheta_1 = \vartheta_2 =: \vartheta
\end{equation}
By setting
\begin{eqnarray}
r^2             &\,=& r_0^2+a^2 \\
2rdr            &\,=& 2r_0dr_0 \\
\cos\,\vartheta &\,=& \frac{a}{r}
\end{eqnarray}  
one obtains
\begin{eqnarray}
I_{\mbox{\scriptsize\rm parallel areas}}
&=&
\int_0^{2\pi} 
\int_0^{R_0}
I_0 \, \frac{(\cos\,\vartheta)^2}{r^2} \, r_0 dr_0 d\varphi 
\nonumber\\
&=&
\int_0^{2\pi} 
\int_0^{R_0}
I_0 \, \frac{a^2}{r^4} \, r_0 dr_0 d\varphi 
\nonumber\\
&=&
\int_0^{2\pi} 
\int_a^{\sqrt{R_0^2+a^2}}
I_0 \, \frac{a^2}{r^4} \, r dr d\varphi 
\nonumber\\
&=&
2\pi \cdot I_0 \cdot a^2 \cdot 
\int_a^{\sqrt{R_0^2+a^2}}
\frac{1}{r^3} \, dr 
\nonumber\\
&=&
2\pi \cdot I_0 \cdot a^2 \cdot 
\left(
\left. 
- \frac{1}{2r^2} 
\right|_a^{\sqrt{R_0^2+a^2}}
\right)
\nonumber\\
&=&
\pi \cdot I_0 \cdot a^2 \cdot 
\left( \frac{1}{a^2} - \frac{1}{R_0^2+a^2} \right)
\nonumber\\
&=&
\pi \cdot I_0 \cdot 
\frac{R_0^2}{R_0^2+a^2}
\end{eqnarray}
\item[(b)] 
\textit{Two parallel areas with distance $a \rightarrow 0$}
\\ 
If the distance $a$ is becoming very small whereas $R_0$ is kept finite
one will have
\begin{equation}
I_{{\mbox{\scriptsize\rm parallel areas\ }}{(a\rightarrow 0)}} 
=
\lim_{a\rightarrow 0}
\left(
\pi\cdot I_0 \cdot
\frac{R_0^2}{R_0^2+a^2}
\right)
=
\pi I_0
\end{equation}
This relation corresponds to 
the total half-space intensity 
for a radiation from a unit surface. 
\item[(c)] 
\textit{The Earth illuminated by the Sun}
\\
With $I_0^{\mbox{\scriptsize\rm Sun}}$ 
being the factor $I_0$ for the Sun
the solar total half-space intensity 
is given by
\begin{equation}
I_{\mbox{\scriptsize\rm Sun's surface}} = \pi \cdot I_0^{\mbox{\scriptsize\rm Sun}} 
\end{equation} 
Setting 
\begin{eqnarray}
a 
&=&
{{\rm R}_{\mbox{\scriptsize\rm Earth's orbit}}}
\\
R_0
&=&
{{\rm R}_{\mbox{\scriptsize\rm Sun}}}
\end{eqnarray}
one gets for the solar intensity at the Earth's orbit  
\begin{eqnarray}
I_{\mbox{\scriptsize\rm Earth's orbit}}
&=&
\pi \cdot
I_0^{\mbox{\scriptsize\rm Sun}}
\cdot
\frac{{\rm R}^2_{\mbox{\scriptsize\rm Sun}}}
     {{\rm R}^2_{\mbox{\scriptsize\rm Sun}} +
      {\rm R}^2_{\mbox{\scriptsize\rm Earth's orbit}}}
\nonumber \\
&=&
I_{\mbox{\scriptsize\rm Sun's surface}}
\cdot
\frac{{\rm R}^2_{\mbox{\scriptsize\rm Sun}}}
     {{\rm R}^2_{\mbox{\scriptsize\rm Sun}} +
      {\rm R}^2_{\mbox{\scriptsize\rm Earth's orbit}}}
\nonumber \\
&\approx&
I_{\mbox{\scriptsize\rm Sun's surface}}
\cdot
\frac{{\rm R}^2_{\mbox{\scriptsize\rm Sun}}}
     {{\rm R}^2_{\mbox{\scriptsize\rm Earth's orbit}}}
\nonumber \\
&\approx&
I_{\mbox{\scriptsize\rm Sun's surface}}
\cdot
\frac{1}{(215)^2}
\end{eqnarray}
\end{itemize}
%
%
\subsubsection{The Stefan-Boltzmann law}
\label{SBlaw}
For a perfect black body
and a unit area positioned
in its proximity we can compute
the intensity $I$ with the aid of
the the Kirchhoff-Planck-function,
which comes in two versions
\begin{eqnarray}
{\rm B}_\nu (T) 
&=&
\frac {2{\rm h}\nu^3} {c^2}
\left[ \exp 
\left( \frac { {\rm h}\nu } { {\rm k} T } \right) - 1 
\right]^{-1}
\\
{\rm B}_\lambda (T) 
&=&
\frac {2{\rm h}{\rm c}^2} {\lambda^5}
\left[ \exp 
\left( \frac { {\rm h}{\rm c} } { \lambda {\rm k} T } \right) - 1 
\right]^{-1}
\end{eqnarray}
that are related to each other by 
\begin{eqnarray}
{\rm B}_\nu (T) \, d\nu 
=
{\rm B}_\nu (T) \, \frac{d\nu}{d\lambda} \, d\lambda
=
- {\rm B}_\nu (T) \, \frac{{\rm c}}{\lambda^2} \, d\lambda
=:
- {\rm B}_\lambda (T) \,d\lambda   
\end{eqnarray}
with
\begin{equation}
\nu = c / \lambda
\end{equation}
where $c$ is the speed of light, $h$ the Planck constant,
$k$ the Boltzmann constant, $\lambda$ the wavelength,
$\nu$ the frequency, and $T$ the absolute temperature, 
respectively. 
Integrating over all frequencies or wavelengths
we obtain the Stefan-Boltzmann $T^4$ law
\begin{equation}
I
=
\pi \cdot
\int_0^\infty {\rm B}_\nu(T) \, d\nu 
=
\pi \cdot
\int_0^\infty {\rm B}_\lambda (T) \, d\lambda
=
\sigma\,T^4
\end{equation}
with 
\begin{equation}
\sigma
=
\pi 
\cdot
\frac{2\pi^4k^4}{15 c^2 h^3}
=
5.670400 
\cdot
10^{-8}
\,
\frac{{\rm W}}{{\rm m}^2{\rm K}^4}
\end{equation}
One conveniently writes
\begin{equation}
S(T)
=
5.67
\cdot
\left(
\frac{T}{100}
\right)^4
\,
\frac{{\rm W}}{{\rm m}^2}
\end{equation}
This is the net radiation energy 
per unit time per unit area 
placed in the neighborhood 
of a radiating plane surface 
of a black body.
%
%
\subsubsection{Conclusion}
\label{Sec:Conclusion}
\begin{figure}[htbp]
\ifthenelse{\equal{IJMPB}{\TschStyle}}{%
\centerline{\psfig{file=PictureLarge_nb001_.eps,width=5.00in}}
}{}%
\ifthenelse{\equal{arXiv}{\TschStyle}}{%
\centerline{\includegraphics[scale=1.00]{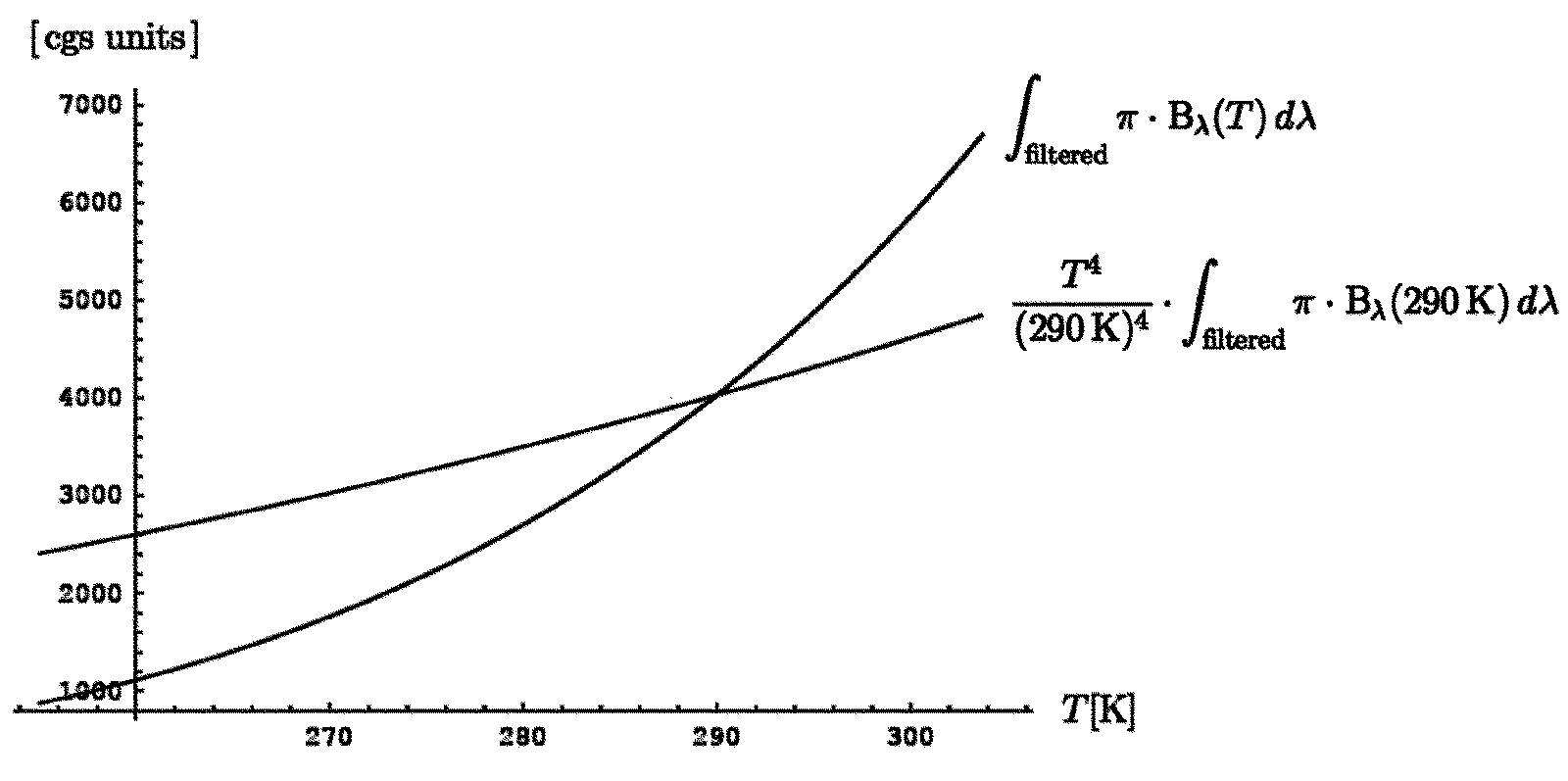}}
}{}%
\ifthenelse{\equal{TeX4ht}{\TschStyle}}{%
\centerline{\includegraphics[scale=1.00]{PictureLarge_nb001_.png}}
}{}
\vspace*{8pt}
\caption{Black body radiation compared to the 
         radiation of a sample coloured body.
         The non-universal constant $\sigma$ 
         is normalized in such a way that
         both curves coincide at $T=290\,{\rm K}$. 
         The Stefan-Boltzmann $T^4$ law does 
         no longer hold in the latter case,
         where only two bands are integrated
         over, namely that of visible light 
         and of infrared radiation from 
         $3\,\mu{\rm m}$ to $5\,\mu{\rm m}$,
         giving rise to a steeper curve.}
\label{Fig:nb001}
\end{figure}

Three facts should be emphasized here:
\begin{itemize}
\item In \textit{classical radiation theory}
      radiation is \textit{not} described by a vector field
      assigning to every space point a corresponding vector.
      Rather, with each point of space many rays
      are associated (Figure~\ref{Fig:radiation}). 
      This is in sharp contrast to the modern 
      description of the radiation field as an
      electromagnetic field with the Poynting 
      vector field as the relevant quantity\TschSpace%
      \cite{Jackson1962}.
\begin{figure}[htbp]
\ifthenelse{\equal{IJMPB}{\TschStyle}}{%
\centerline{\psfig{file=PictureLarge_specific_intensity_integrated_.eps,width=3.00in}}
}{}%
\ifthenelse{\equal{arXiv}{\TschStyle}}{%
\centerline{\includegraphics[scale=1.00]{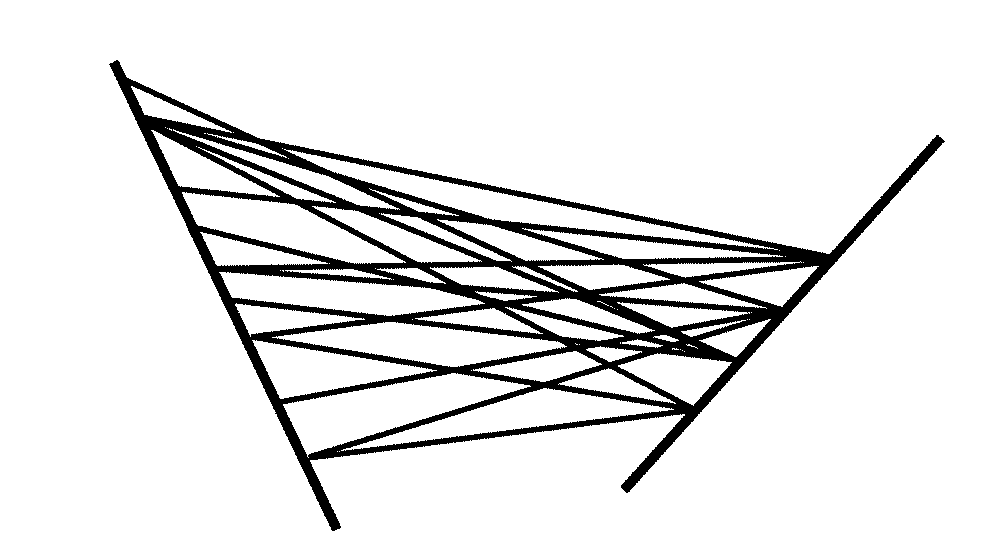}}
}{}%
\ifthenelse{\equal{TeX4ht}{\TschStyle}}{%
\centerline{\includegraphics[scale=1.00]{PictureLarge_specific_intensity_integrated_.png}}
}{}
\vspace*{8pt}
\caption{The geometry of classical radiation: 
         Two surfaces radiating against each other.}
\label{Fig:radiation}
\end{figure}
\item The constant $\sigma$ appearing in the $T^4$ law
      is \textit{not} a universal constant of physics.
      It strongly depends on the particular geometry 
      of the problem considered.%
\footnote{For instance, to compute the radiative transfer 
          in a multi-layer setup, the correct point of departure
          is the infinitesimal expression for the radiation intensity,
          not an integrated Stefan-Boltzmann expression already computed
          for an entirely different situation.}
\item The $T^4$-law will no longer hold
      if one integrates only over a filtered spectrum, 
      appropriate to real world situations. 
      This is illustrated in Figure~\ref{Fig:nb001} .
\end{itemize}
Many pseudo-explanations 
in the context of global climatology
are already falsified 
by these three fundamental observations 
of mathematical physics.  
%

\subsection{The Sun as a black body radiator}
\label{Sec:SunBlackBody}
\begin{figure}[htbp]
\ifthenelse{\equal{IJMPB}{\TschStyle}}{%
\centerline{\psfig{file=PictureLarge_sunlight_.eps,width=4.50in}}
}{}%
\ifthenelse{\equal{arXiv}{\TschStyle}}{%
\centerline{\includegraphics[scale=2.00]{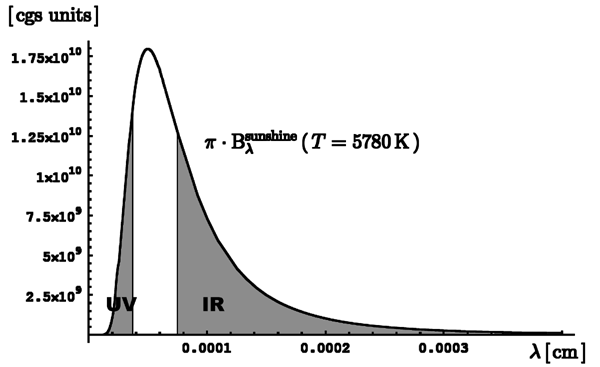}}
}{}%
\ifthenelse{\equal{TeX4ht}{\TschStyle}}{%
\centerline{\includegraphics[scale=0.66]{PictureLarge_sunlight_.png}}
}{}
\vspace*{8pt}
\caption{The spectrum of the sunlight assuming the Sun
         is a black body at $T=5780\,{\rm K}$.}
\label{Fig:Visible}
\end{figure}

The Kirchhoff-Planck function describes
an ideal black body radiator.
For matter of convenience one may define 
\begin{equation}
{\rm B}_\lambda^{\mbox{\scriptsize\rm sunshine}}
=
{\rm B}_\lambda^{\mbox{\scriptsize\rm Sun}}
\cdot
\frac{{\rm R}^2_{\mbox{\scriptsize\rm Sun}}}
     {{\rm R}^2_{\mbox{\scriptsize\rm Earth's orbit}}}
=
{\rm B}_\lambda^{\mbox{\scriptsize\rm Sun}}
\cdot
\frac{1}{(215)^2}
\end{equation}
Figure~\ref{Fig:Visible} shows the spectrum 
of the sunlight, assuming the Sun is a black
body of temperature $T=5780\,{\rm K}$. 

To compute the part of radiation
for a certain wave length interval 
$[\lambda_1,\lambda_2]$
one has to evaluate the expression 
\begin{equation}
\frac{ \int_{\lambda_1}^{\lambda_2} 
       {\rm B}_\lambda^{\mbox{\scriptsize\rm sunshine}} (5780)\,d\lambda }
     { \int_0^\infty 
       {\rm B}_\lambda^{\mbox{\scriptsize\rm sunshine}} (5780)\,d\lambda }
\end{equation}
Table~\ref{table:sunlight} shows the  
proportional portions of the 
ultraviolet, visible, and infrared sunlight,
respectively.
\ifthenelse{\equal{IJMPB}{\TschStyle}}{%
\begin{table}[htbp] 
\tbl{The proportional portion of the 
     ultraviolet, visible, and infrared sunlight,
     respectively.}
{
\begin{tabular}{@{}lcc@{}} \Hline 
\\[-1.8ex] 
Band             & Range                   & Portion  \\
                 & $[{\rm nm}]$            & $[\%]$   \\[0.8ex] 
\hline                                     &          \\[-1.8ex] 
ultraviolet      & $ \phantom{00}0 - 380 $ & 10.0     \\
visible          & $ \phantom{}380 - 760 $ & 44,8     \\
infrared         & $ 760\, - \,\infty\,    $ & 45,2   \\[0.8ex]
\hline                                                \\[-1.8ex] 
\end{tabular}
}
\label{table:sunlight}
\end{table}
}{}
\ifthenelse{\equal{arXivOrTeX4ht}{\TschStyles}}{%
\begin{table}[htbp] 
{
\begin{center}
\vspace*{0.5cm}
\begin{tabular}{|l|c|c|} 
\hline 
Band             & Range                   & Portion  \\
                 & $[{\rm nm}]$            & $[\%]$   \\ 
\hline                        
ultraviolet      & $ \phantom{00}0 - 380 $ & 10.0     \\
visible          & $ \phantom{}380 - 760 $ & 44,8     \\
infrared         & $ 760\, - \,\infty\,  $ & 45,2     \\
\hline                                                 
\end{tabular}
\end{center}
}
\caption{The proportional portion of the 
         ultraviolet, visible, and infrared sunlight,
         respectively.}
\label{table:sunlight}
\vspace*{0.5cm}
\end{table}
}{}

Here the visible range of the light is assumed
to lie between 380\,{\rm nm} and 760\,{\rm nm}.
It should be mentioned that the visible range 
depends on the individuum. 

In any case,
a larger portion of the incoming sunlight 
lies in the infrared range than in the 
visible range. 
In most papers discussing 
the supposed greenhouse effect 
this important fact is completely 
ignored.   

%
\subsection{The radiation on a very nice day}
\subsubsection{The phenomenon}
Especially after a year's hot summer every car driver 
knows a sort of a glass house or greenhouse effect: 
If he parks his normally tempered car in the morning 
and the Sun shines into the interior of the car
until he gets back into it at noon, 
he will almost burn his fingers at the steering wheel, if the 
dashboard area had been subject to direct Sun radiation. 
Furthermore, the air inside the car is unbearably hot, 
even if it is quite nice outside. One opens the window 
and the slide roof, but unpleasant hot air may still 
hit one from the dashboard while driving. One can notice 
a similar effect in the winter, only then one will probably 
welcome the fact that it is warmer inside the car than outside. 
In greenhouses or glass houses this effect is put to use: 
the ecologically friendly solar energy, for which 
no energy taxes are probably going to be levied 
even in the distant future, is used for heating. 
Nevertheless, glass houses have not replaced conventional 
buildings in our temperate climate zone not only because 
most people prefer 
to pay energy taxes, 
to heat in the winter,
and 
to live in a cooler apartment on summer days, 
but because glass houses have other disadvantages as well.
%
%
\subsubsection{The sunshine}
\begin{figure}[htbp]
\ifthenelse{\equal{IJMPB}{\TschStyle}}{%
\centerline{\psfig{file=PictureLarge_nb002_.eps,width=2.50in}\psfig{file=PictureLarge_nb003_.eps,width=2.50in}}
}{}%
\ifthenelse{\equal{arXiv}{\TschStyle}}{%
\centerline{\includegraphics[scale=0.60]{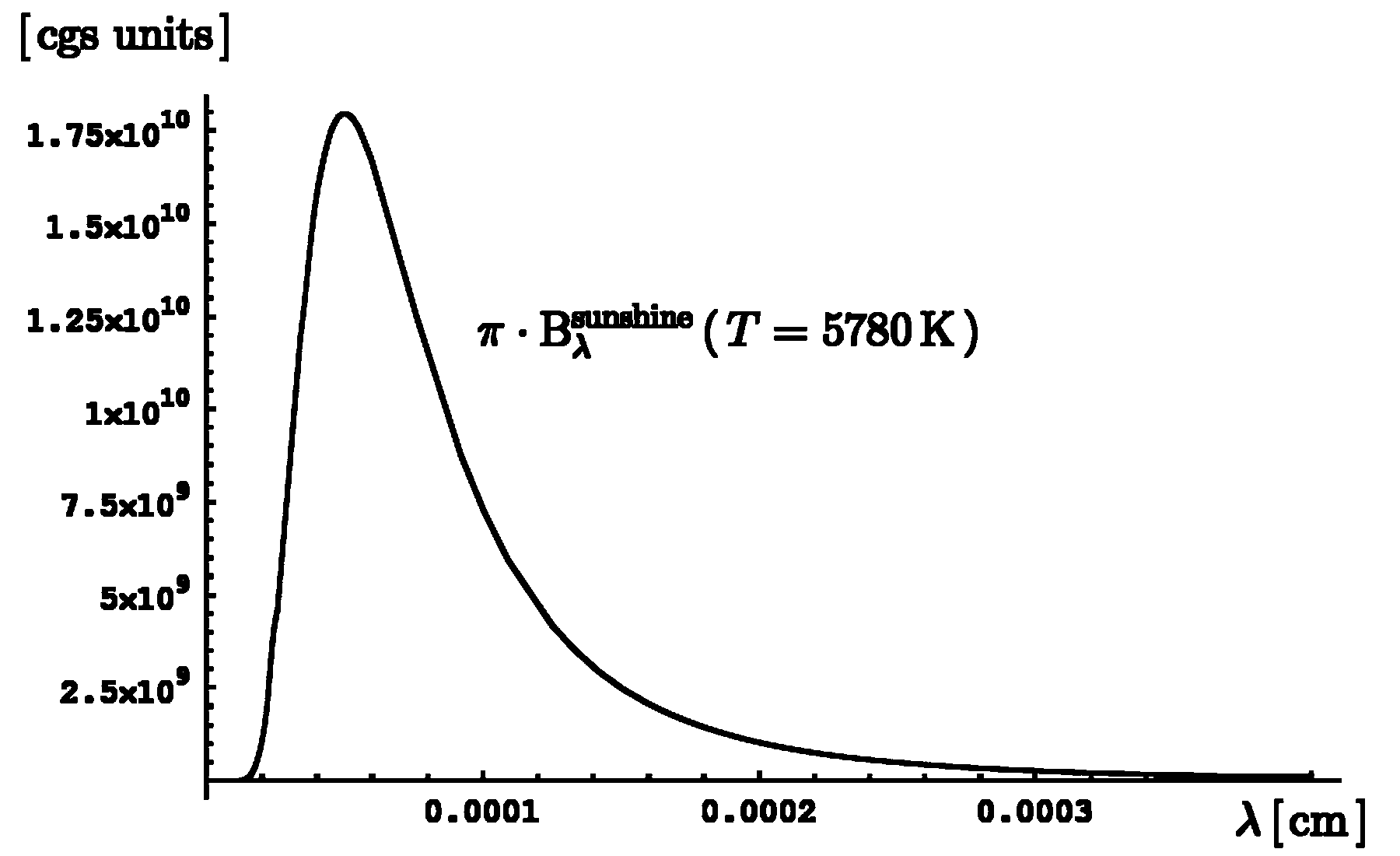}\includegraphics[scale=0.60]{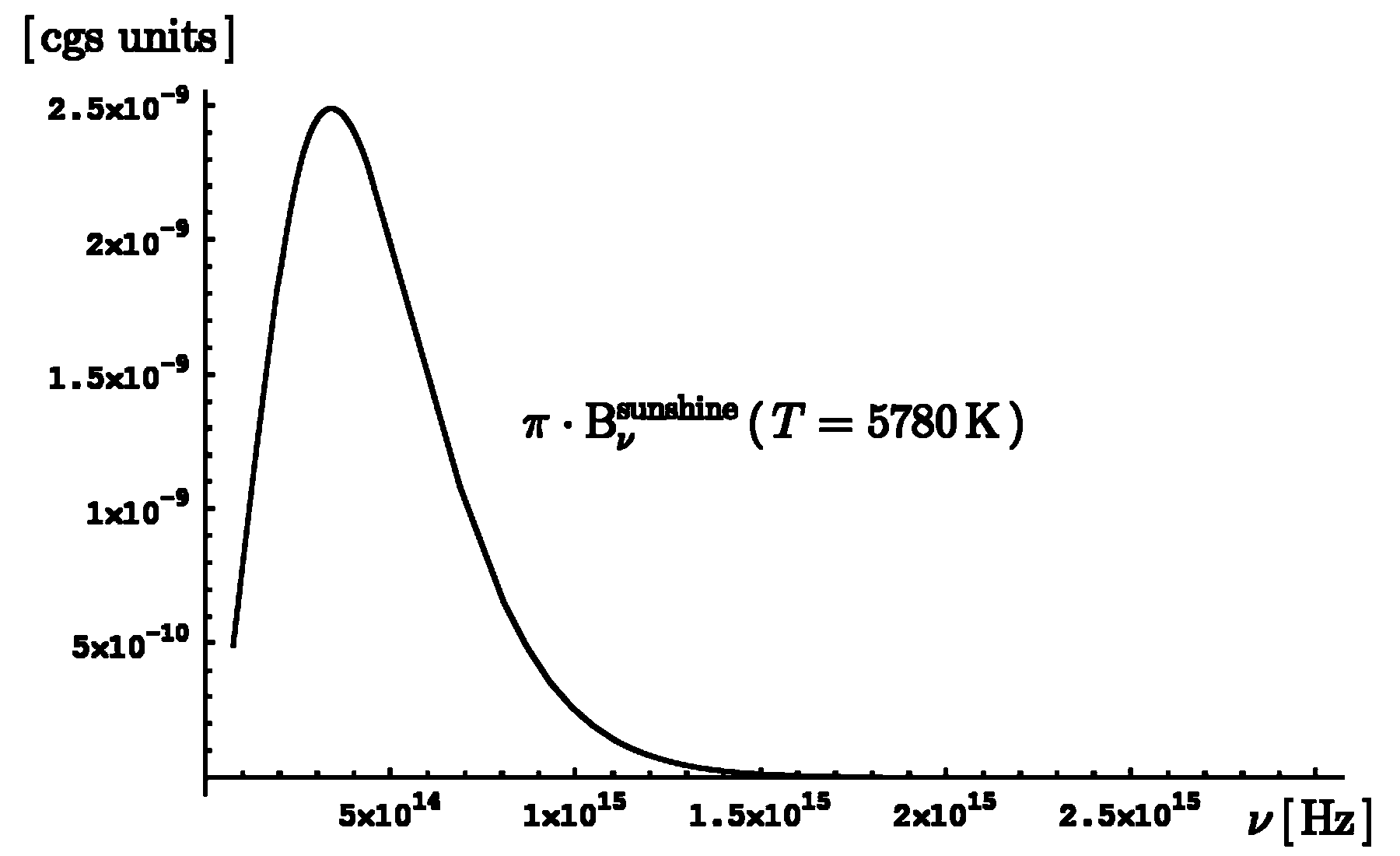}}
}{}%
\ifthenelse{\equal{TeX4ht}{\TschStyle}}{%
\centerline{\includegraphics[scale=0.60]{PictureLarge_nb002_.png}\includegraphics[scale=0.60]{PictureLarge_nb003_.png}}
}{}
\vspace*{8pt}
\caption{The unfiltered spectral distribution 
         of the sunshine on Earth
         under the assumption that the Sun is a black body
         with temperature $T = {\rm 5780}\,{\rm K}$
         (left: in wave length space,
          right: in frequency space).} 
\label{Fig:nb002nb003}
\end{figure}

\begin{figure}[htbp]
\ifthenelse{\equal{IJMPB}{\TschStyle}}{%
\centerline{\psfig{file=PictureLarge_nb012_.eps,width=2.50in}\psfig{file=PictureLarge_nb013_.eps,width=2.50in}}
}{}%
\ifthenelse{\equal{arXiv}{\TschStyle}}{%
\centerline{\includegraphics[scale=0.60]{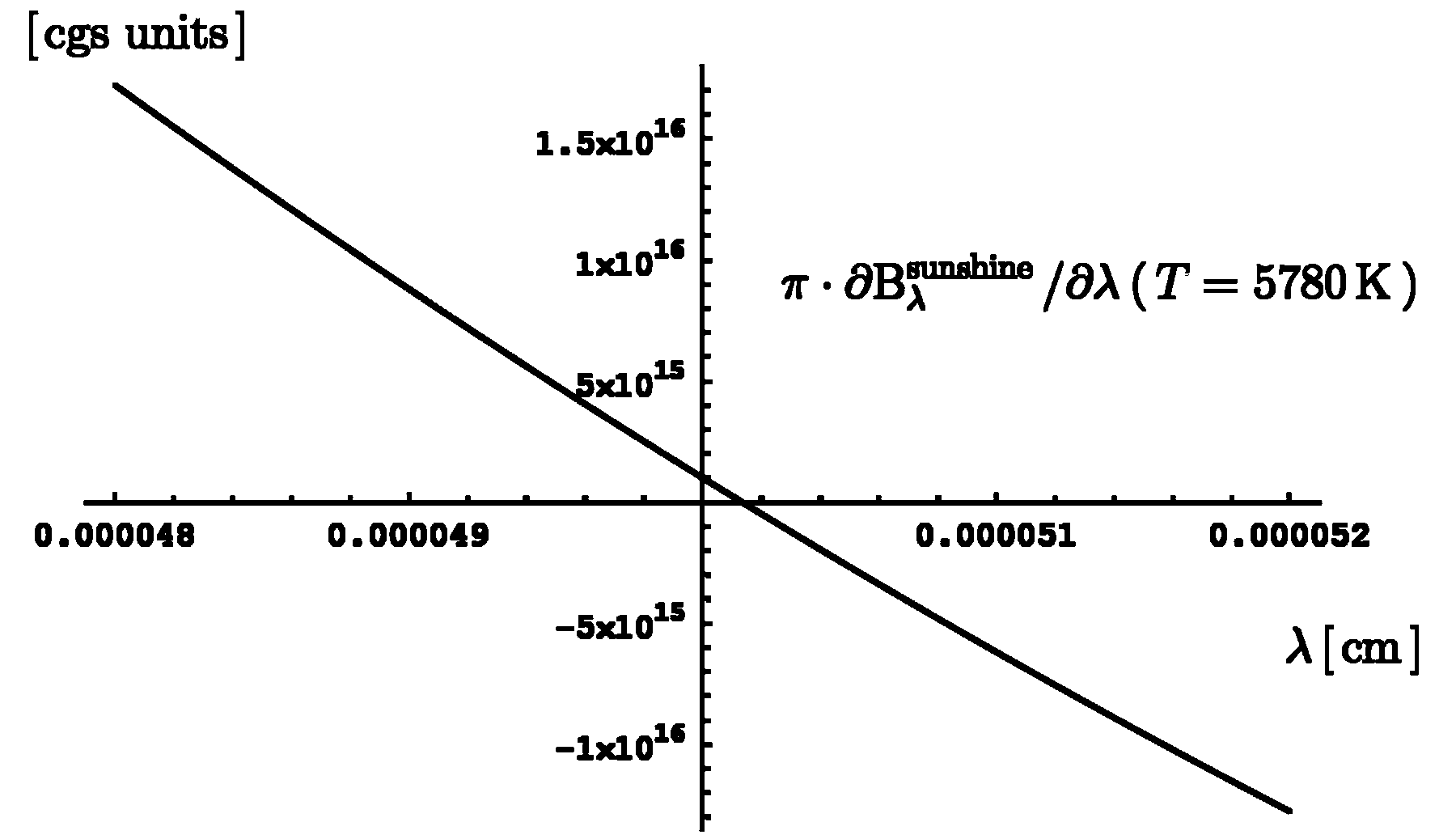}\includegraphics[scale=0.60]{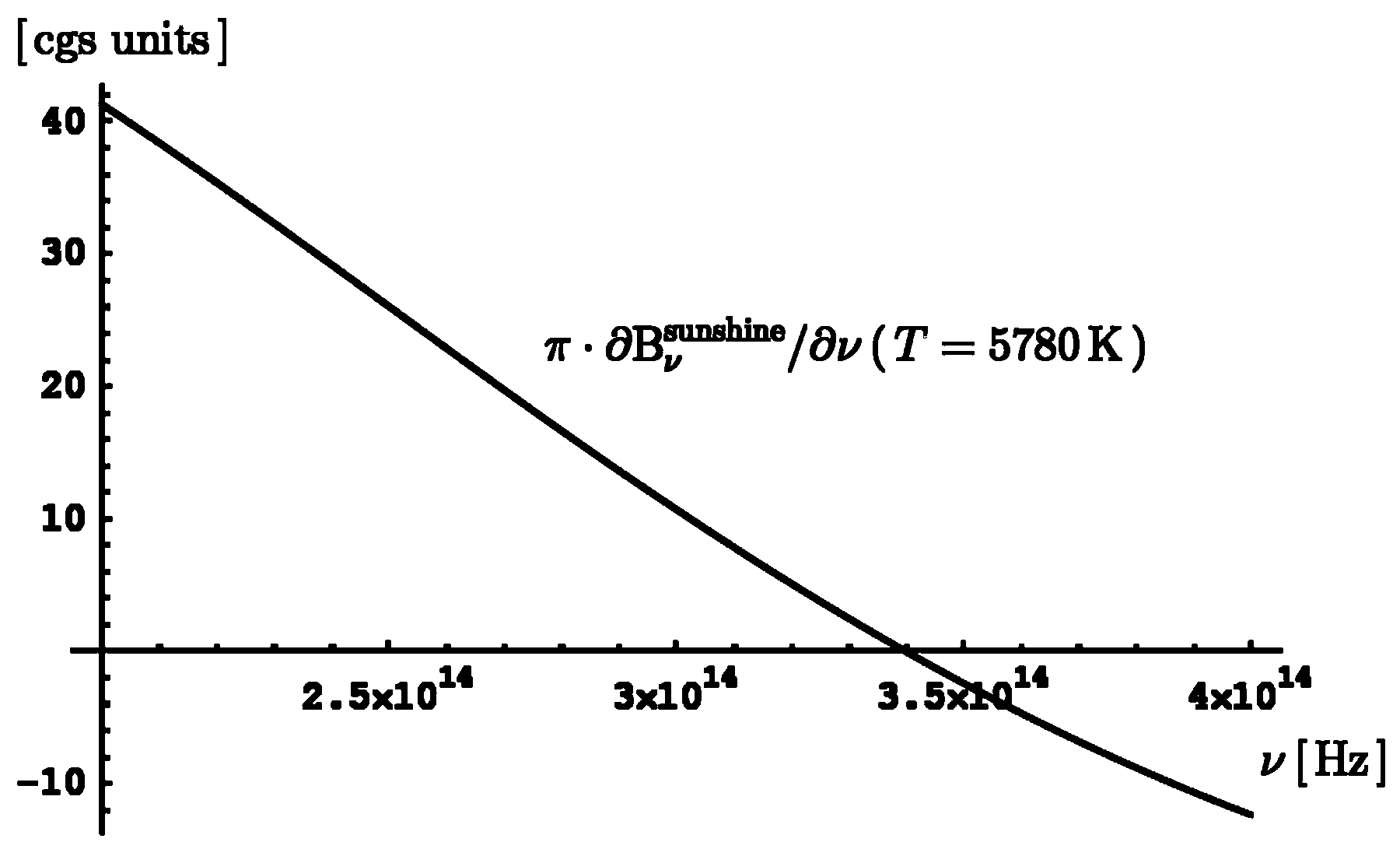}}
}{}%
\ifthenelse{\equal{TeX4ht}{\TschStyle}}{%
\centerline{\includegraphics[scale=0.60]{PictureLarge_nb012_.png}\includegraphics[scale=0.60]{PictureLarge_nb013_.png}}
}{}
\vspace*{8pt}
\caption{The exact location of the zero 
         of the partial derivatives 
         of the radiation intensities
         of the sunshine on Earth
         (left: in wave length space,
          right: in frequency space).}
\label{Fig:nb012nb013}
\end{figure}

One does not need to be an expert in physics 
to explain immediately why the car is so hot inside: 
It is the Sun, which has heated the car inside like this. 
However, it is a bit harder to answer the question why 
it is not as hot outside the car, although there 
the Sun shines onto the ground without obstacles. 
Undergraduate students with their standard physical 
recipes at hand can easily \lq\lq explain\rq\rq\ 
this kind of a greenhouse effect: 
The main part of the Sun's radiation 
(Figure~\ref{Fig:nb002nb003})
%
%
passes 
through the glass, as the maximum 
(Figure~\ref{Fig:nb012nb013})
%
%
of the solar radiation is of bluegreen wavelength
\begin{equation}
\lambda_{\mbox{\scriptsize\rm bluegreen}} = 0.5\ \mu{\rm m}
\end{equation}
which the glass lets through. This part can be 
calculated with the Kirchhoff-Planck-function.

Evidently, the result depends on the type of glass. 
For instance, if it is transparent to electromagnetic
radiation in the 
$300\,{\rm nm}$\,-\,$1000\,{\rm nm}$ 
range 
one will have
\begin{equation}
\frac{ \int_{0.3\,\mu{\rm m}}^{1\,\mu{\rm m}} {\rm B}_\lambda^{\mbox{\scriptsize\rm sunshine}} (5780)\,d\lambda }
     { \int_0^\infty {\rm B}_\lambda^{\mbox{\scriptsize\rm sunshine}} (5780)\,d\lambda }
= 77,2\,\%
\end{equation}
In the case of a glass, 
which is assumed to be transparent
only to visible light
($380\,{\rm nm}$\,-\,$760\,{\rm nm}$) 
one gets
\begin{equation}
\frac{ \int_{0.380\,\mu{\rm m}}^{0.760\,\mu{\rm m}} {\rm B}_\lambda^{\mbox{\scriptsize\rm sunshine}} (5780)\,d\lambda }
     { \int_0^\infty {\rm B}_\lambda^{\mbox{\scriptsize\rm sunshine}} (5780)\,d\lambda }
= 44,8\,\%
\end{equation}
Because of the Fresnel reflection\TschSpace%
%
\cite{Jackson1962}
%
at both pane boundaries 
one has to subtract 
$8$\,-\,$10$ percent
and only 
$60$\,-\,$70$ percent 
(resp.\ $40$ percent)  
of the solar radiation 
reach the interior of the vehicle.
High performance tinted glass which is also referred to 
as \textit{spectrally selective tinted glass} reduces 
solar heat gain typically by a factor of $0.50$ 
(only by a factor of $0.69$ in the visible range) 
compared to standard glass\TschSpace%
%
\cite{EfficientWindowsOrg}.
%
%
\subsubsection{The radiation of the ground}
\begin{figure}[htbp]
\ifthenelse{\equal{IJMPB}{\TschStyle}}{%
\centerline{\psfig{file=PictureLarge_nb004_.eps,width=2.50in}\psfig{file=PictureLarge_nb005_.eps,width=2.50in}}
}{}%
\ifthenelse{\equal{arXiv}{\TschStyle}}{%
\centerline{\includegraphics[scale=0.60]{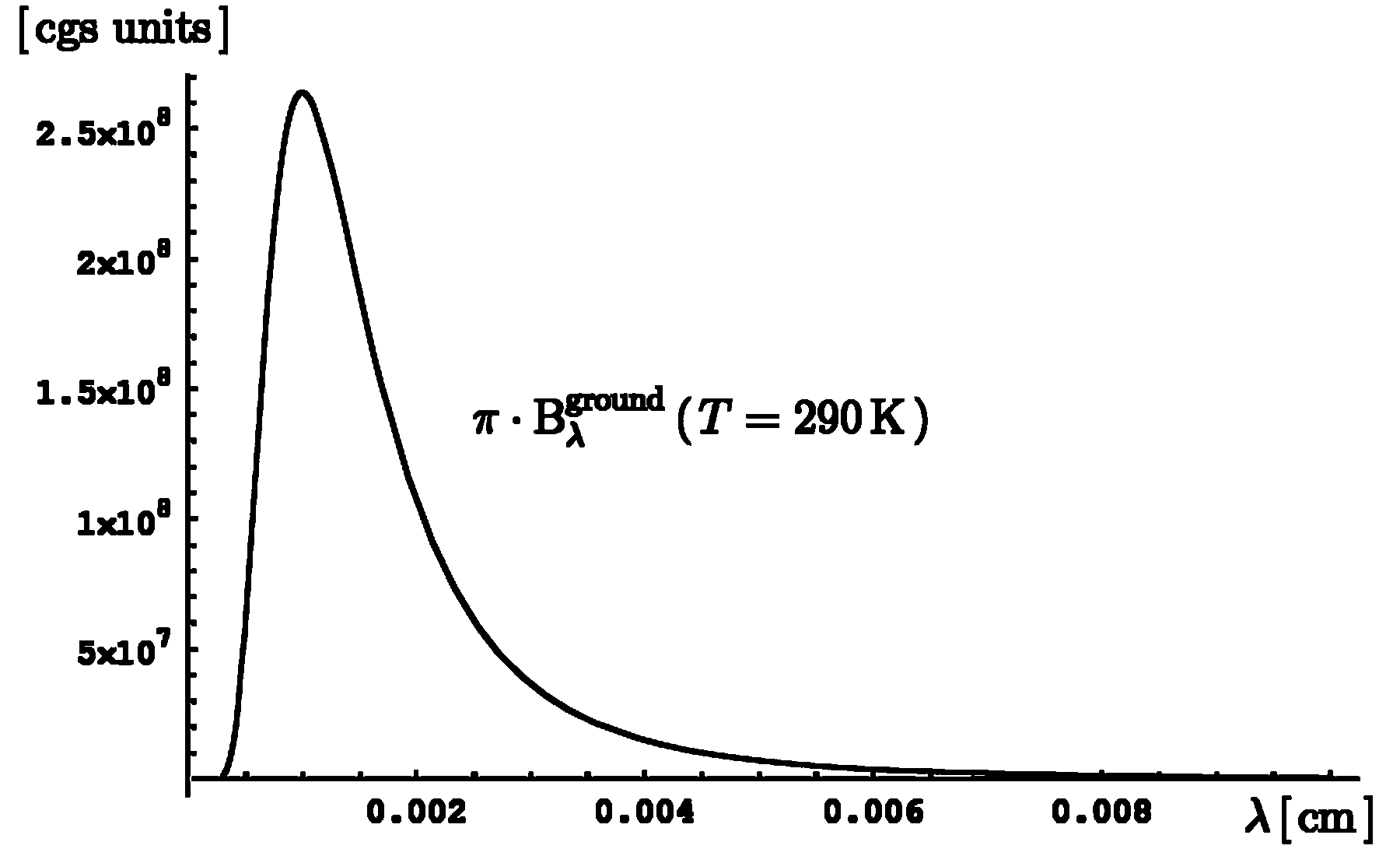}\includegraphics[scale=0.60]{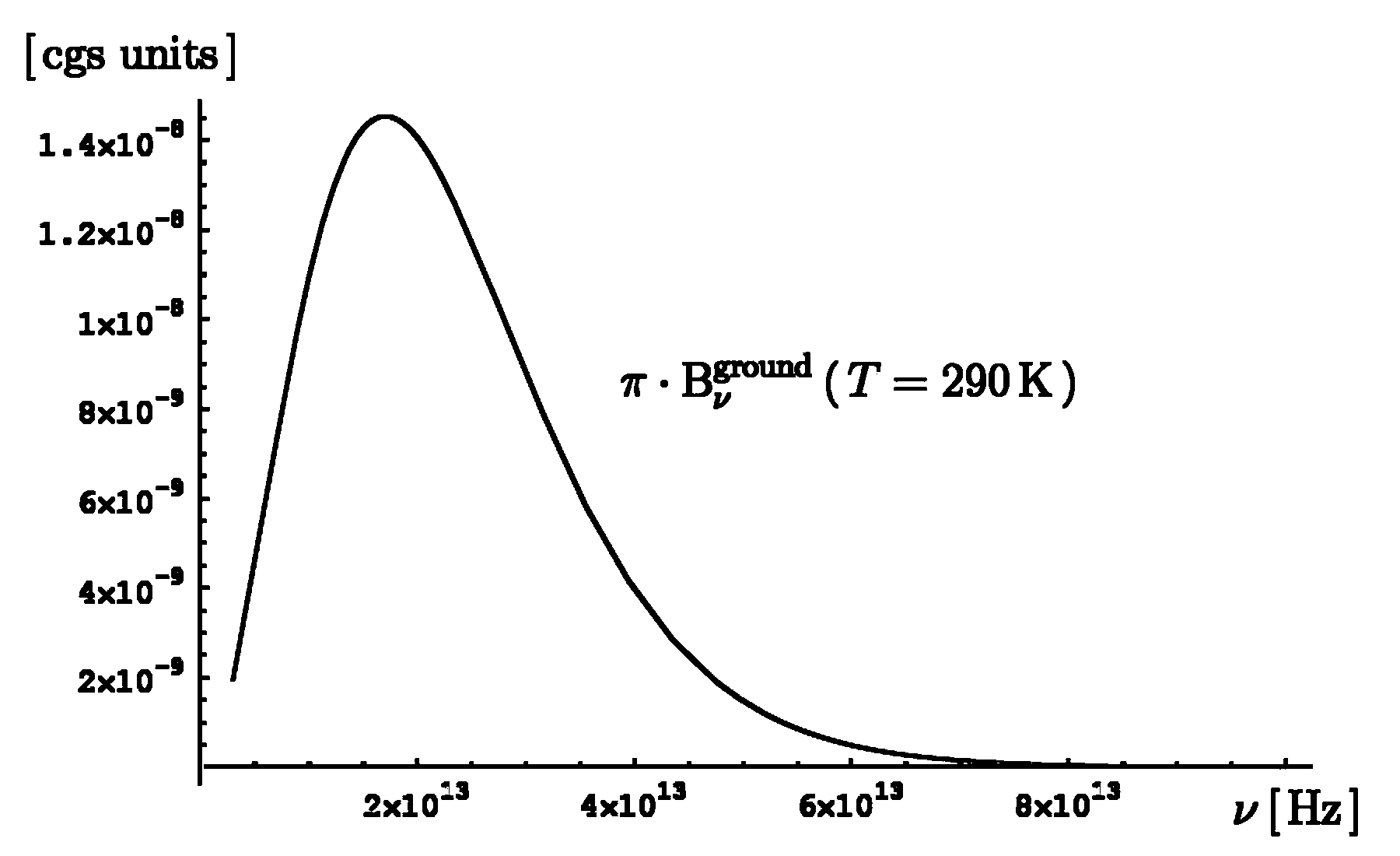}}
}{}%
\ifthenelse{\equal{TeX4ht}{\TschStyle}}{%
\centerline{\includegraphics[scale=0.60]{PictureLarge_nb004_.png}\includegraphics[scale=0.60]{PictureLarge_nb005_.png}}
}{}
\vspace*{8pt}
\caption{The unfiltered spectral distribution 
         of the radiation of the ground
         under the assumption that the Earth is a black body
         with temperature $T = {\rm 290}\,{\rm K}$
         (left: in wave length space,
          right: in frequency space).}   
\label{Fig:nb004nb005}
\end{figure}

\begin{figure}[htbp]
\ifthenelse{\equal{IJMPB}{\TschStyle}}{%
\centerline{\psfig{file=PictureLarge_nb014and015and016and017_.eps,width=5.00in}}
}{}%
\ifthenelse{\equal{arXiv}{\TschStyle}}{%
\centerline{\includegraphics[scale=0.60]{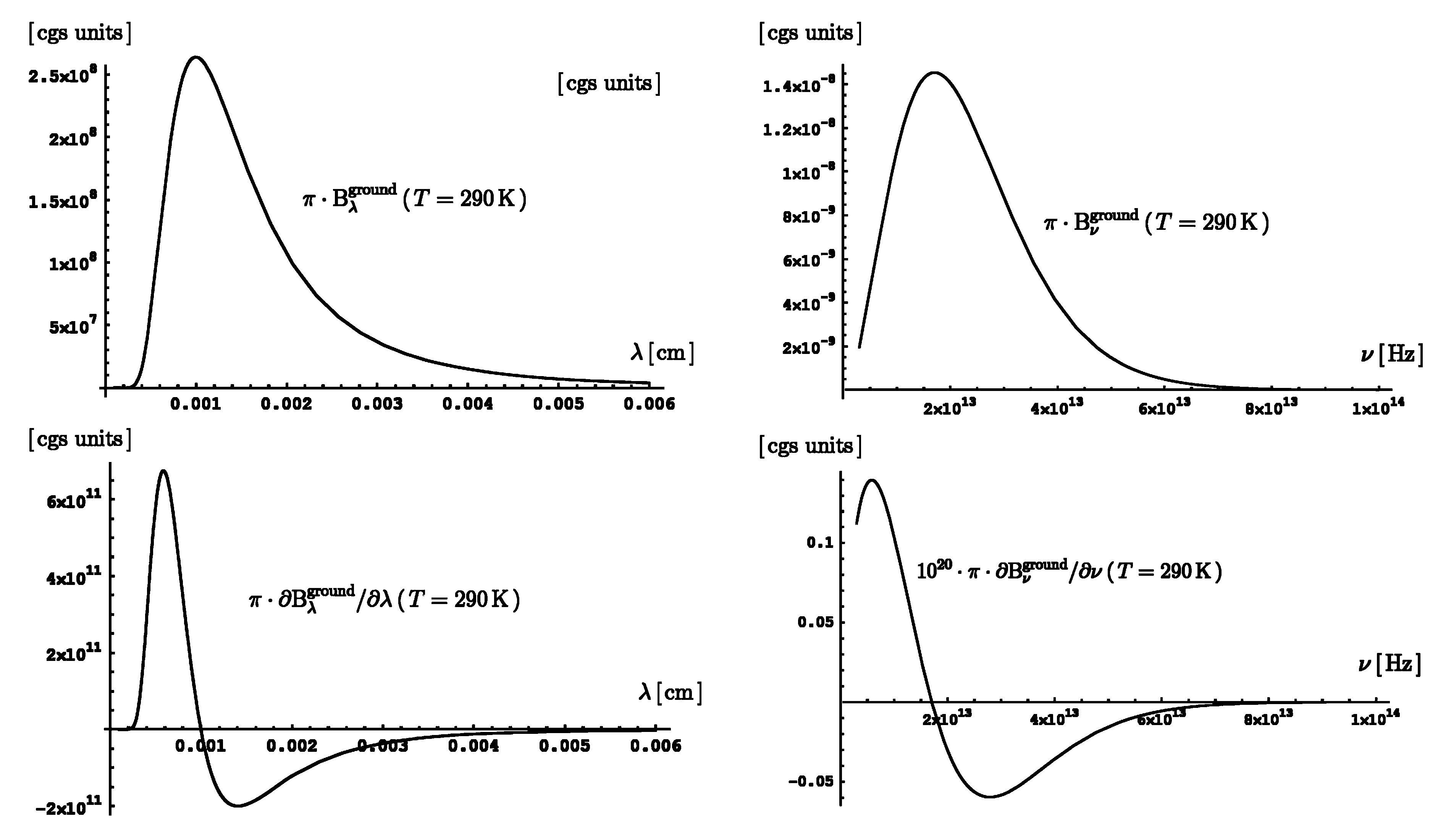}}
}{}%
\ifthenelse{\equal{TeX4ht}{\TschStyle}}{%
\centerline{\includegraphics[scale=0.60]{PictureLarge_nb014and015and016and017_.png}}
}{}
\vspace*{8pt}
\caption{The radiation intensity of the ground
         \textit{and} 
         its partial derivative as a function
         of the wave length $\lambda$ (left column)
         and of the frequency $\nu$ (right column).}
\label{Fig:nb016nb017nb018nb019}
\end{figure}

\begin{figure}[htbp]
\ifthenelse{\equal{IJMPB}{\TschStyle}}{%
\centerline{\psfig{file=PictureLarge_nb021_.eps,width=5.00in}}
}{}%
\ifthenelse{\equal{arXiv}{\TschStyle}}{%
\centerline{\includegraphics[scale=1.00]{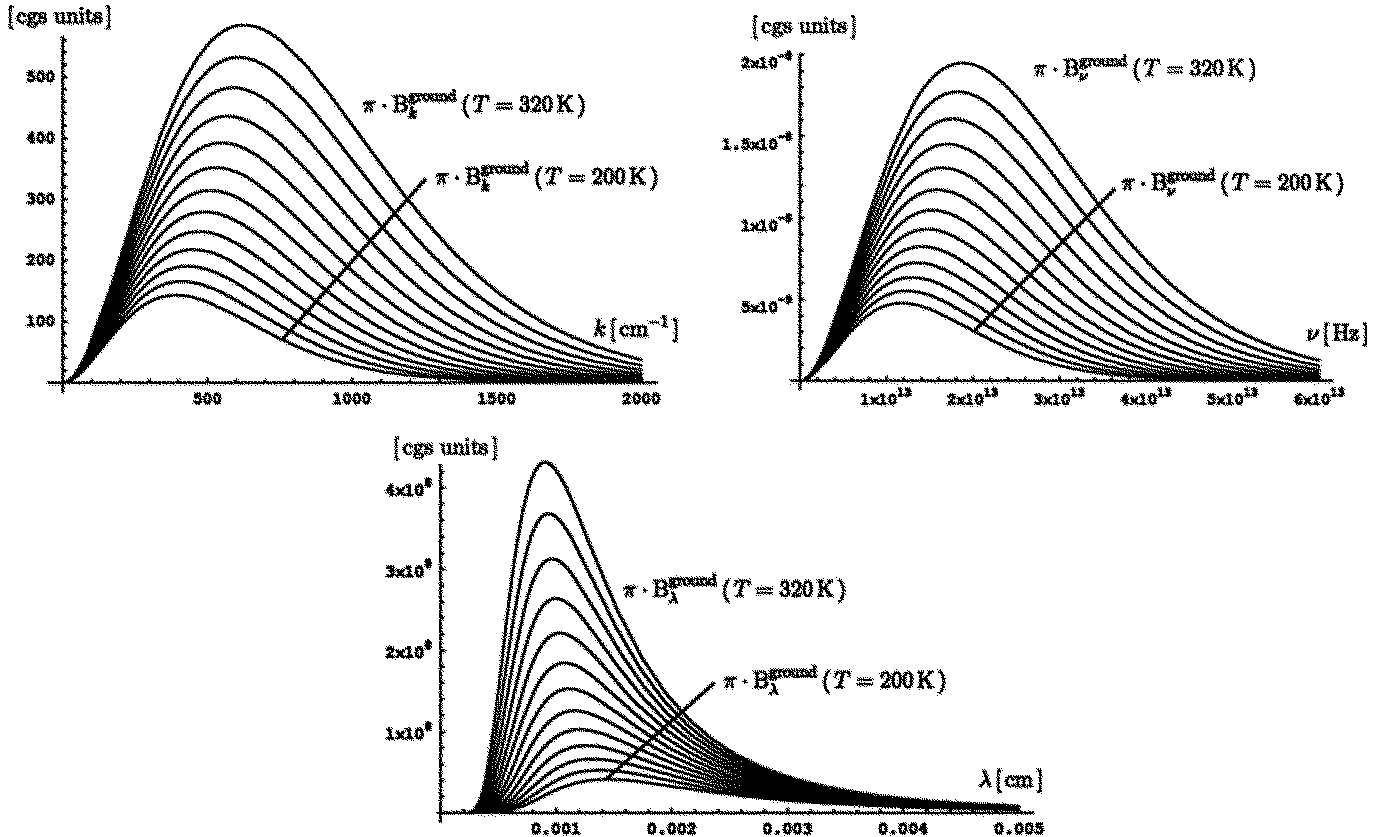}}
}{}%
\ifthenelse{\equal{TeX4ht}{\TschStyle}}{%
\centerline{\includegraphics[scale=0.33]{PictureLarge_nb021_.png}}
}{}
\vspace*{8pt}
\caption{Three versions of radiation curve families 
         of the radiation of the ground 
         (as a function
         of the wave number $k$,
         of the frequency $\nu$,
         of the wave length $\lambda$,
         respectively), assuming
         that the Earth is a black radiator.}
\label{Fig:nb021}
\end{figure}

The bottom of a glass house has 
a temperature of approximately $290\,{\rm K}$
(Figure~\ref{Fig:nb004nb005}). 
The maximum of a black body's radiation
can be calculated with the help of Wien's 
displacement law
(cf.\ Figure~\ref{Fig:nb016nb017nb018nb019} 
      and Figure~\ref{Fig:nb021})  
%
%
%
%
%
%
\begin{equation}
\lambda_{\mbox{\scriptsize\rm max}}(T) \cdot T = {\rm const.}
\end{equation}
giving 
\begin{equation}
\lambda_{\mbox{\scriptsize\rm max}}(300\,{\rm K})
=
\frac{6000\,{\rm K}}{300\,{\rm K}}
\cdot
\lambda_{\mbox{\scriptsize\rm max}}(6000\,{\rm K})
=
10\,\mu{\rm m}
\end{equation}
This is far within the infrared wave range, 
where glass reflects practically all light, 
according to Beer's formula\TschSpace%
%
\cite{Weizel1963}.
%
Practically 100 percent of a black body's radiation 
at ground temperatures lie above the wavelengths 
of 3.5 $\mu{\rm m}$. 
The thermal radiation of the ground is thus 
\lq \lq trapped\rq\rq\ by the panes. 
 
According to Wien's power law 
describing the intensity of the maximum wave-length
\begin{equation}
{\rm B}_{\lambda_{\mbox{\scriptsize\rm max}}} (T) \propto T^5
\end{equation}
the intensity of the radiation 
on the ground at the maximum is
\begin{equation}
\frac{T^5_{\rm Sun}}{T^5_{\mbox{\scriptsize\rm Earth's ground}}} 
\approx
\frac{6000^5}{300^5} 
=
20^5
= 
3.2 \cdot 10^6
\end{equation}
times smaller than on the Sun and
\begin{equation}
\frac{T^5_{\rm Sun}}{T^5_{\mbox{\scriptsize\rm Earth's ground}}} 
\cdot
\frac{{\rm R}^2_{\mbox{\scriptsize\rm Sun}}}
     {{\rm R}^2_{\mbox{\scriptsize\rm Earth's orbit}}}
\approx
20^5
\cdot
\frac{1}{215^2}
\approx
70
\end{equation}
times smaller than the solar radiation on Earth.

The \textit{total radiation} can be calculated 
from the Stefan-Boltzmann law
\begin{equation}
{\rm B}_{\mbox{\scriptsize\rm total}} (T) = \sigma \cdot T^4
\end{equation}
Hence, the ratio of the intensities
of the sunshine and the ground radiation
is given by
\begin{equation}
\frac{T^4_{\rm Sun}}{T^4_{\mbox{\scriptsize\rm Earth's ground}}} 
\cdot
\frac{{\rm R}^2_{\mbox{\scriptsize\rm Sun}}}
     {{\rm R}^2_{\mbox{\scriptsize\rm Earth's orbit}}}
\approx
20^4
\cdot
\frac{1}{215^2}
\approx 3.46
\end{equation}
Loosely speaking, the radiation
of the ground is about four times weaker 
than the incoming solar radiation. 
%
%
\subsubsection{Sunshine versus ground radiation}
\begin{figure}[htbp]
\ifthenelse{\equal{IJMPB}{\TschStyle}}{%
\centerline{\psfig{file=PictureLarge_nb006_.eps,width=2.50in}\psfig{file=PictureLarge_nb007_.eps,width=2.50in}}
}{}%
\ifthenelse{\equal{arXiv}{\TschStyle}}{%
\centerline{\includegraphics[scale=0.60]{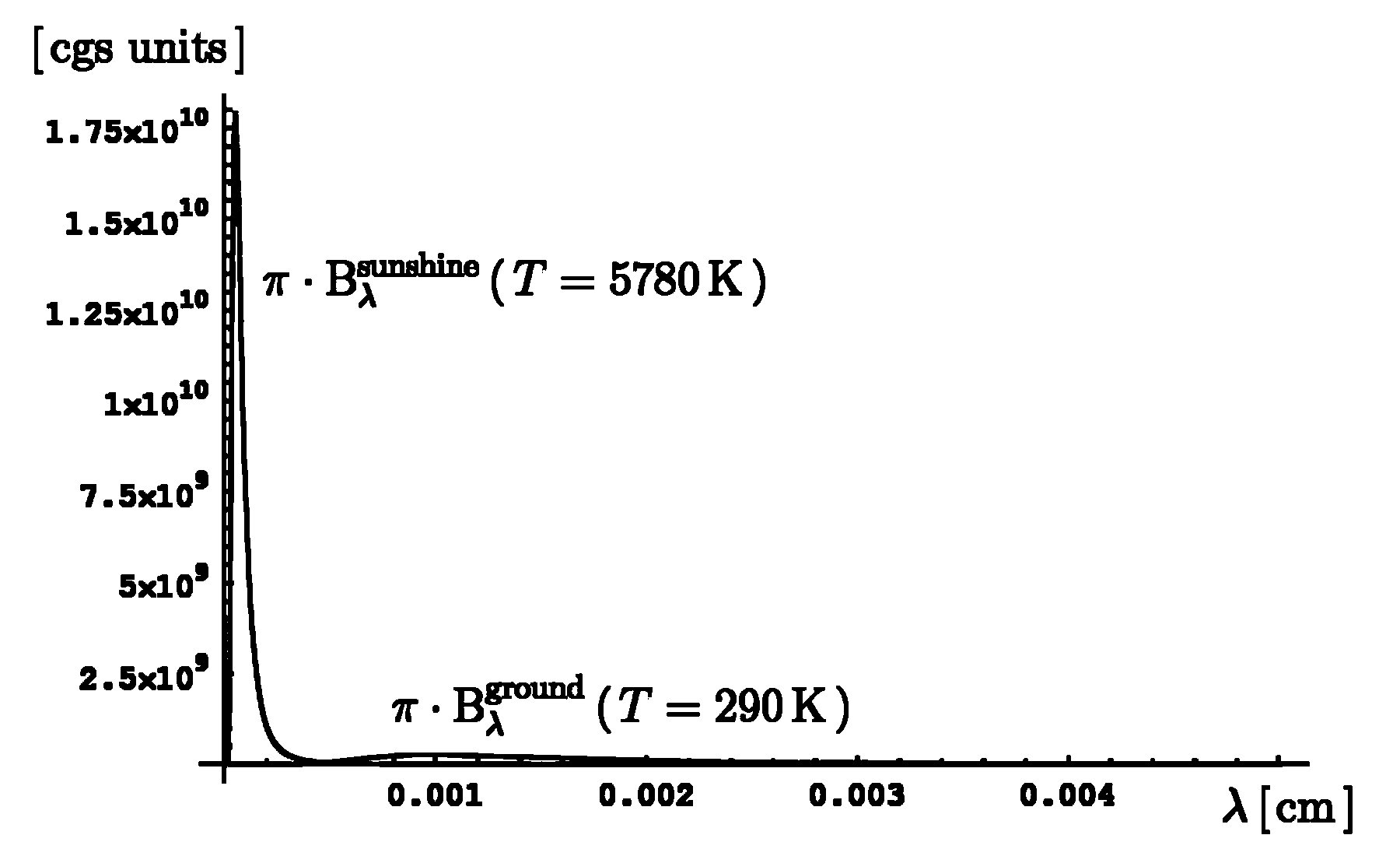}\includegraphics[scale=0.60]{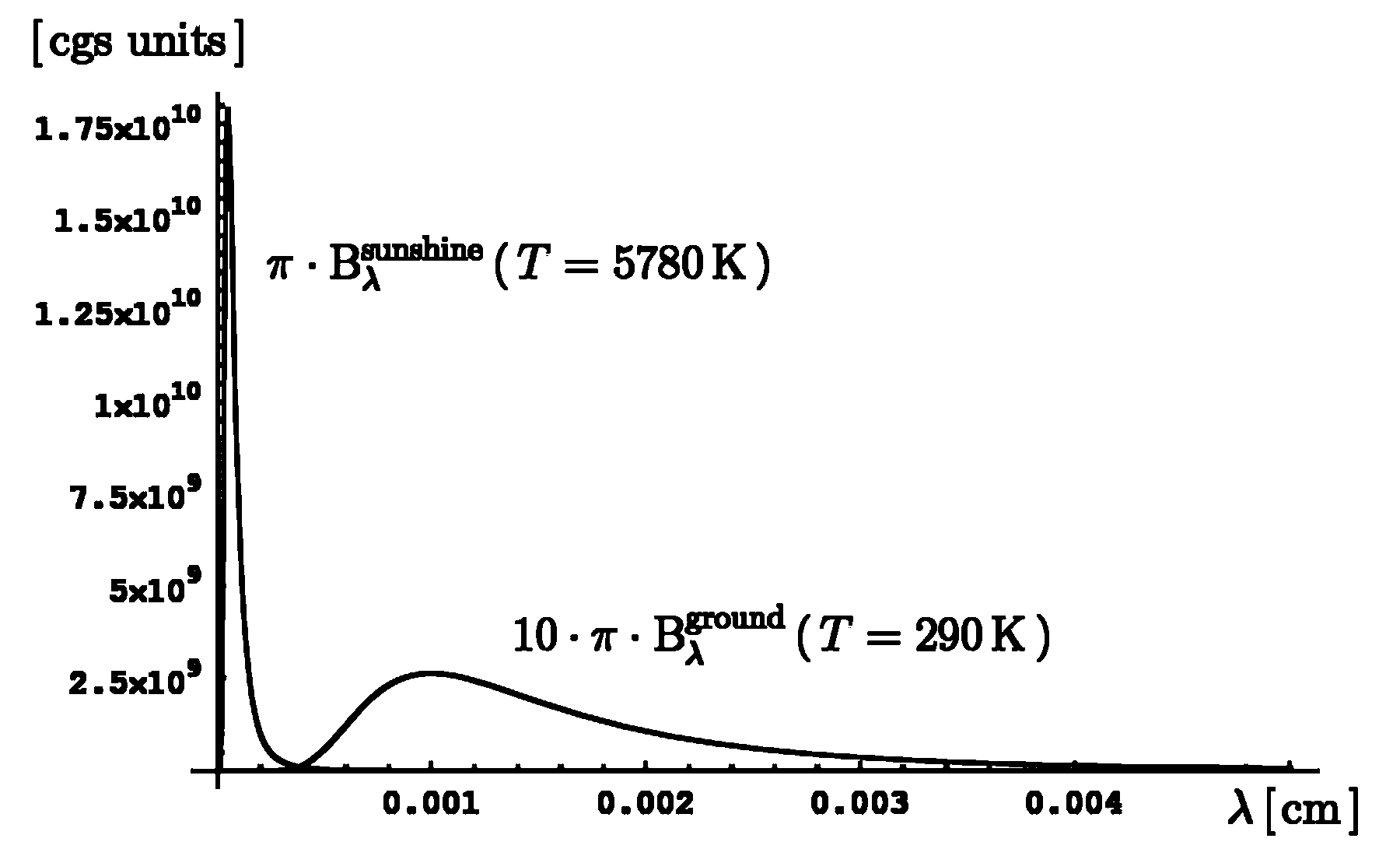}}
}{}%
\ifthenelse{\equal{TeX4ht}{\TschStyle}}{%
\centerline{\includegraphics[scale=0.60]{PictureLarge_nb006_.png}\includegraphics[scale=0.60]{PictureLarge_nb007_.png}}
}{}
\vspace*{8pt}
\caption{The unfiltered spectral distribution 
         of the sunshine on Earth
         under the assumption that the Sun is a black body
         with temperature $T = {\rm 5780}\,{\rm K}$
         \textit{and} the unfiltered spectral distribution 
         of the radiation of the ground
         under the assumption that the Earth is a black body
         with temperature $T = {\rm 290}\,{\rm K}$, 
         \textit{both} in one diagram 
         (left: normal, 
          right: super elevated
          by a factor of 10 for the 
          radiation of the ground).}
\label{Fig:nb006nb007}
\end{figure}

\begin{figure}[htbp]
\ifthenelse{\equal{IJMPB}{\TschStyle}}{%
\centerline{\psfig{file=PictureLarge_nb008_.eps,width=2.50in}\psfig{file=PictureLarge_nb009_.eps,width=2.50in}}
}{}%
\ifthenelse{\equal{arXiv}{\TschStyle}}{%
\centerline{\includegraphics[scale=0.60]{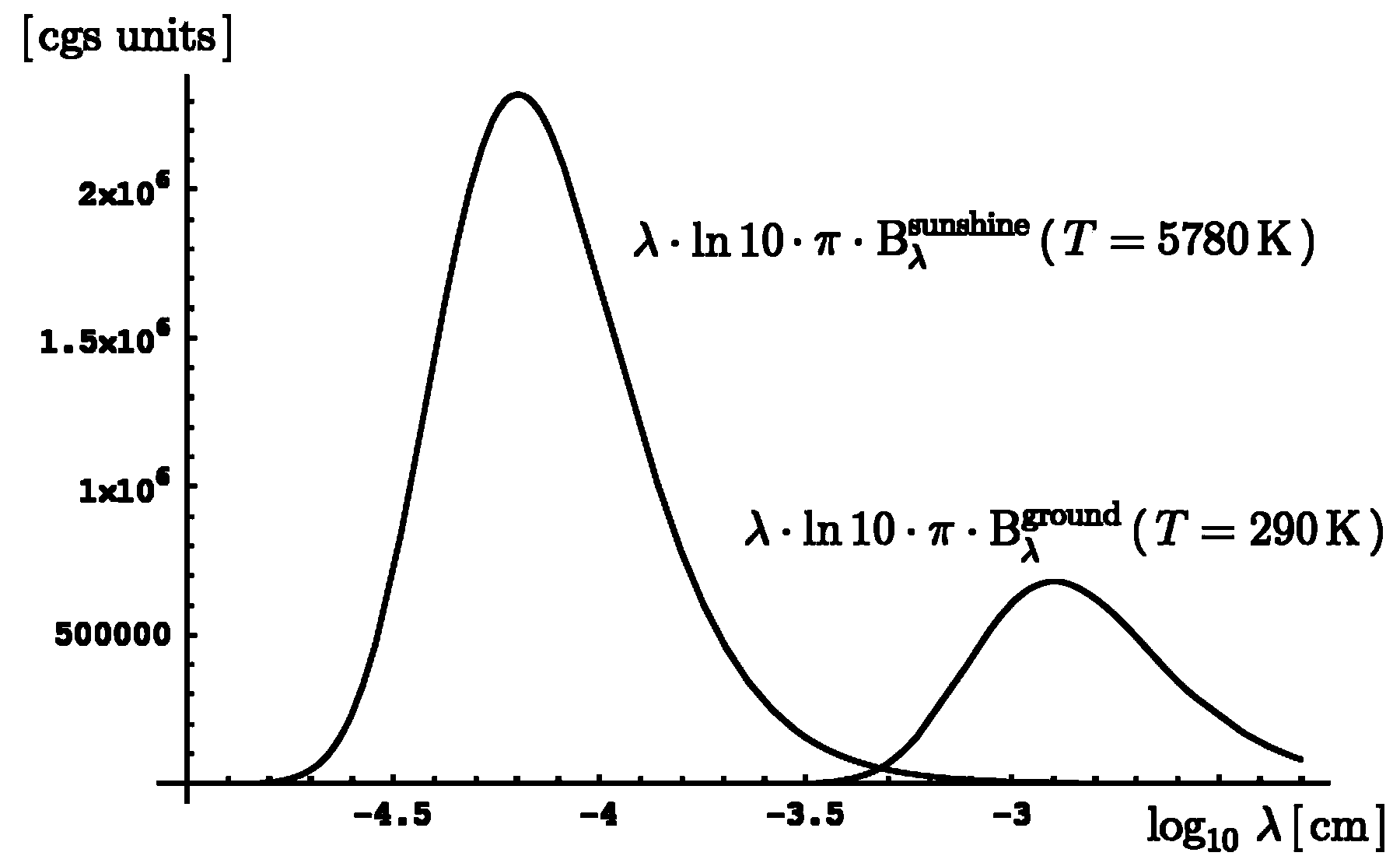}\includegraphics[scale=0.60]{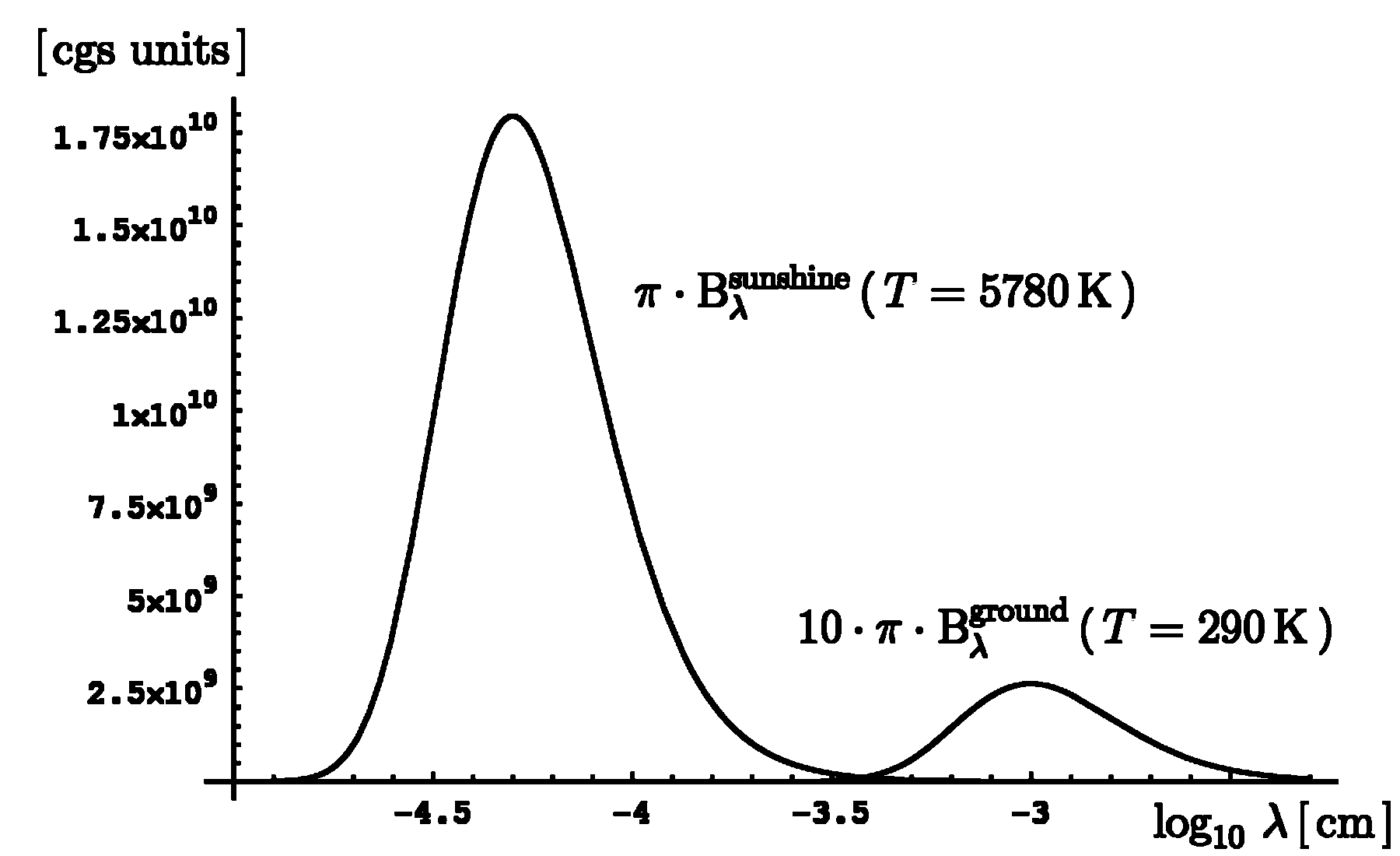}}
}{}%
\ifthenelse{\equal{TeX4ht}{\TschStyle}}{%
\centerline{\includegraphics[scale=0.60]{PictureLarge_nb008_.png}\includegraphics[scale=0.60]{PictureLarge_nb009_.png}}
}{}
\vspace*{8pt}
\caption{The unfiltered spectral distribution 
         of the sunshine on Earth
         under the assumption that the Sun is a black body
         with temperature $T = {\rm 5780}\,{\rm K}$
         \textit{and} the unfiltered spectral distribution 
         of the radiation of the ground
         under the assumption that the Earth is a black body
         with temperature $T = {\rm 290}\,{\rm K}$, 
         \textit{both} in one semi-logarithmic diagram 
         (left: normalized in such a way that
          equal areas correspond to equal intensities, 
          right: super elevated 
          by a factor of 10 for the 
          radiation of the ground).} 
\label{Fig:nb008nb009}
\end{figure}

\begin{figure}[htbp]
\ifthenelse{\equal{IJMPB}{\TschStyle}}{%
\centerline{\psfig{file=PictureLarge_nb010_.eps,width=2.50in}\psfig{file=PictureLarge_nb011_.eps,width=2.50in}}
}{}%
\ifthenelse{\equal{arXiv}{\TschStyle}}{%
\centerline{\includegraphics[scale=0.60]{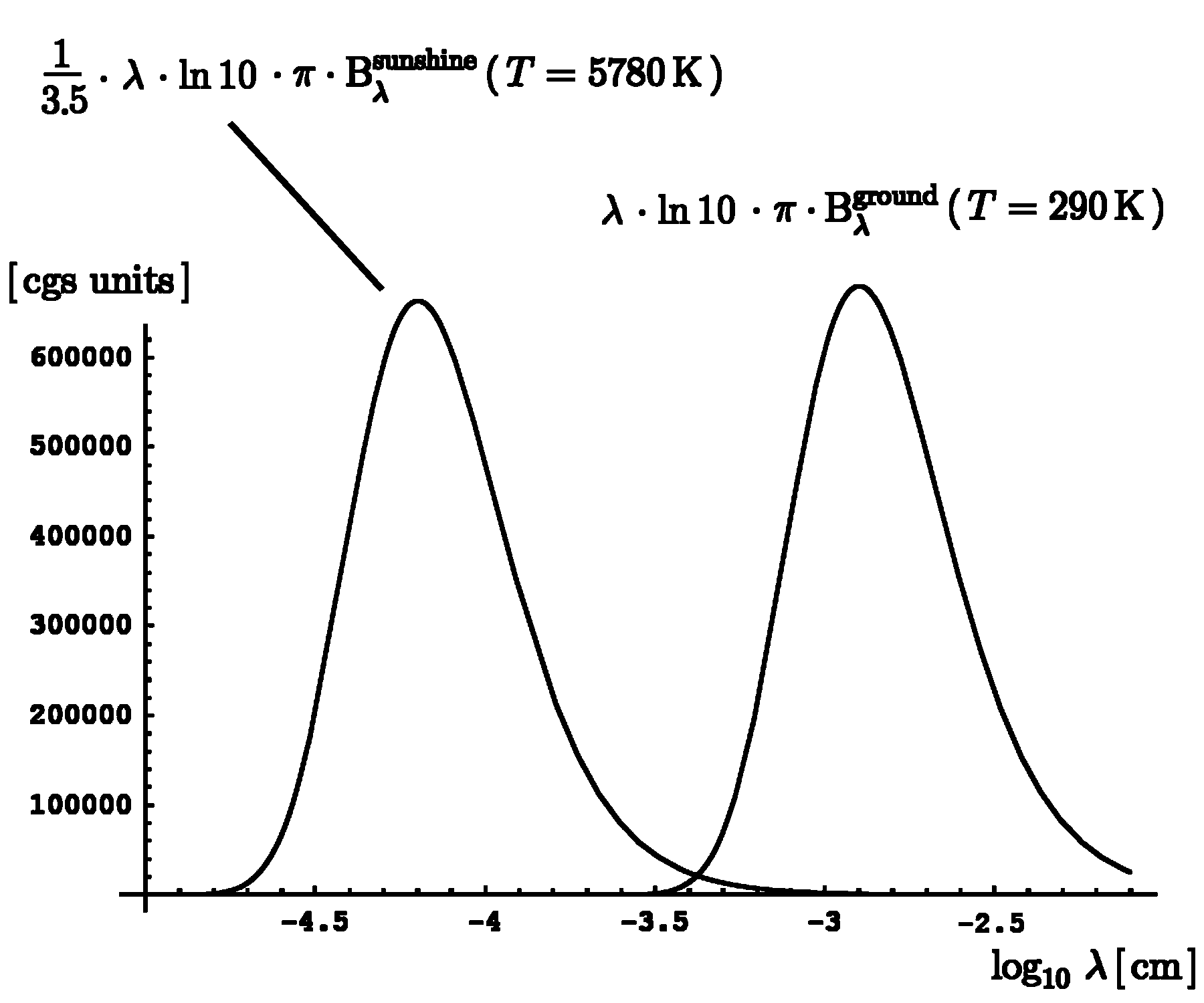}\includegraphics[scale=0.60]{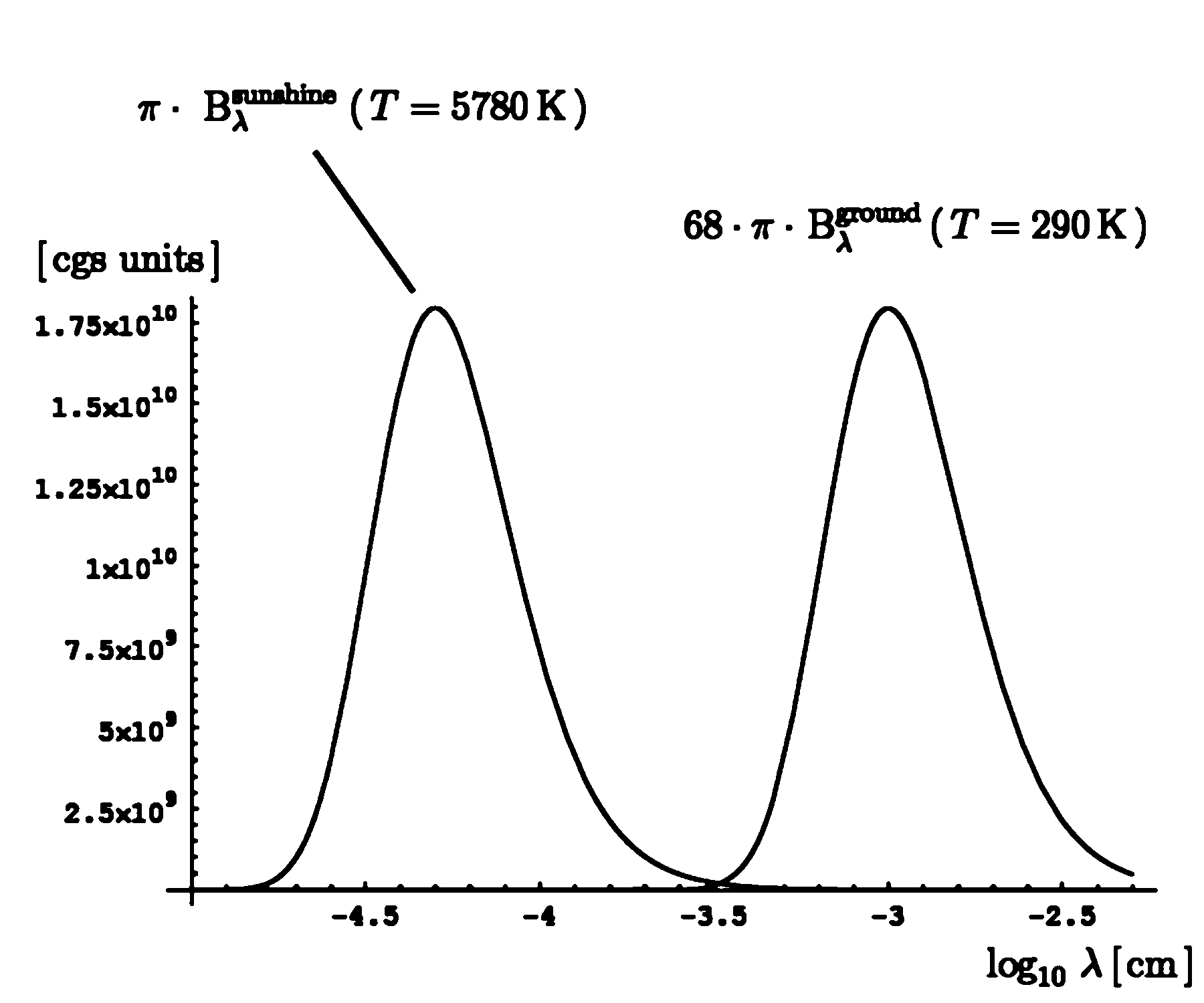}}
}{}%
\ifthenelse{\equal{TeX4ht}{\TschStyle}}{%
\centerline{\includegraphics[scale=0.60]{PictureLarge_nb010_.png}\includegraphics[scale=0.60]{PictureLarge_nb011_.png}}
}{}
\vspace*{8pt}
\caption{The unfiltered spectral distribution
         of the sunshine on Earth
         under the assumption that the Sun is a black body
         with temperature $T = {\rm 5780}\,{\rm K}$
         \textit{and} the unfiltered spectral distribution 
         of the radiation of the ground
         under the assumption that the Earth is a black body
         with temperature $T = {\rm 290}\,{\rm K}$, 
         \textit{both} in one semi-logarithmic diagram 
         (left: normalized in such a way that
          equal areas correspond to equal intensities
          with an additional re-scaling of the sunshine curve 
          by a factor of $1/3.5$,
          right: super elevated 
          by a factor of 68 for the 
          radiation of the ground).} 
\label{Fig:nb010nb011}
\end{figure}

To make these differences even clearer, 
it is convenient to graphically represent 
the spectral distribution of intensity 
at the Earth's orbit and of a black radiator 
of $290\,{\rm K}$, respectively, 
in relation to the wavelength
(Figures \ref{Fig:nb006nb007}, 
\ref{Fig:nb008nb009}, 
and \ref{Fig:nb010nb011}).
To fit both curves into one drawing,
one makes use of the technique of super-elevation
and/or applies an appropriate re-scaling.
%
%
%
%
%
%
\ifthenelse{\equal{arXivOrTeX4ht}{\TschStyles}}{\newpage}{}

\noindent%
It becomes clearly visible, 
\begin{itemize}
\item that the maxima are at 
      0.5 $\mu{\rm m}$ 
      or 
      10 $\mu{\rm m}$, 
      respectively; 
\item that
      the intensities of the maxima 
      differ by more than an order 
      of 10; 
\item that above 0.8 $\mu{\rm m}$   
      (infrared) the solar luminosity 
      has a notable intensity.
\end{itemize}
Figure~\ref{Fig:nb010nb011} 
is an obscene picture,
since it is physically misleading.
The obscenity will not remain in the eye of the beholder,
if the latter takes a look at the obscure scaling factors
already applied by Bakan and Raschke in an undocumented 
way in their paper on the so-called natural greenhouse effect\TschSpace%
%
%
\cite{Bakan2002}.  
%
%
This is scientific misconduct as is the missing citation. 
Bakan and Raschke borrowed this figure from  
%
\ifthenelse{\equal{IJMPB}{\TschStyle}}
           {Ref.~\refcite{Luther1985}}
           {}%
\ifthenelse{\equal{arXiv}{\TschStyle}}
           {Ref.~\cite{Luther1985}}
           {}%
\ifthenelse{\equal{TeX4ht}{\TschStyle}}
           {Ref.~\cite{Luther1985}}
           {} 
%
where the scaling factors,
which are of utmost importance 
for the whole discussion, 
are left unspecified. 
This is scientific misconduct as well.  
\subsubsection{Conclusion}
Though in most cases the preceding 
\lq\lq explanation\rq\rq\ 
suffices to provide an accepted solution 
to the standard problem,
presented in the undergraduate course,
the analysis leaves the main question untouched, 
namely, why the air inside the car is warmer 
than outside and why the dashboard is hotter 
than the ground outside the car. 
Therefore, in the following, the situation 
inside the car is approached experimentally.
%

\subsection{High School Experiments}  
\label{Sec:Experiments}
On a hot summer afternoon, 
temperature measurements
were performed 
with a standard digital thermometer 
by the first author\TschSpace%
\cite{%
Gerlich1995,%
Gerlich2004,%
Gerlich2005a,%
Gerlich2005b,%
Gerlich2007}
and were recently reproduced  
by the other author. 

In the summertime, such measurements 
can be reproduced by everyone very 
easily. 
The results are
listed in Table~\ref{table:GerlichMeasurements}.
\ifthenelse{\equal{IJMPB}{\TschStyle}}{%
\begin{table}[htbp] 
\tbl{Measured temperatures inside and outside a car 
     on a hot summer day.}
{
\begin{tabular}{@{}lc@{}} \Hline 
\\[-1.8ex] 
Thermometer located $\dots$                      & Temperature        \\[0.8ex] 
\hline                                                                        \\[-1.8ex] 
inside the car, in direct Sun                    & $71\, ^\circ{\rm C}$ \\
inside the car, in the shade                     & $39\, ^\circ{\rm C}$ \\
next to the car, in direct Sun, above the ground & $31\, ^\circ{\rm C}$ \\
next to the car, in the shade, above the ground  & $29\, ^\circ{\rm C}$ \\
in the living room                               & $25\, ^\circ{\rm C}$ \\[0.8ex] 
\hline                                                   \\[-1.8ex] 
\end{tabular}
}
\label{table:GerlichMeasurements}
\end{table}
}{}
\ifthenelse{\equal{arXivOrTeX4ht}{\TschStyles}}{%
\begin{table}[htbp] 
{
\begin{center}
\vspace*{0.5cm}
\begin{tabular}{|l|c|} 
\hline 
Thermometer located $\dots$                      & Temperature        \\ 
\hline                                                                   
inside the car, in direct Sun                    & $71\, ^\circ{\rm C}$ \\
inside the car, in the shade                     & $39\, ^\circ{\rm C}$ \\
next to the car, in direct Sun, above the ground & $31\, ^\circ{\rm C}$ \\
next to the car, in the shade, above the ground  & $29\, ^\circ{\rm C}$ \\
in the living room                               & $25\, ^\circ{\rm C}$ \\ 
\hline                                                
\end{tabular}
\end{center}
}
\caption{Measured temperatures inside and outside a car 
         on a hot summer day.}
\label{table:GerlichMeasurements}
\vspace*{0.5cm}
\end{table}
}{}  %

Against these measurements one may object 
that one had to take the dampness of the ground 
into account: at some time during the year the 
stones certainly got wet in the rain.
The above mentioned measurements 
were made at a time, when it had not 
rained for weeks. They are real measured values, 
not average values over all breadths and lengths 
of the Earth, day and night and all seasons and 
changes of weather.
These measurements are recommended to 
every climatologist, who believes in the 
${\rm CO}_2$-greenhouse effect, because 
he feels already while measuring, that 
the just described effect 
\textbf{has nothing to do with} 
trapped thermal radiation. One can touch 
the car's windows and notice that the panes, 
which absorb the infrared light, 
are rather cool and do not heat the inside 
of the car in any way. If one holds his hand 
in the shade next to a very hot part 
of the dashboard that lies in the Sun, 
one will practically feel no thermal radiation 
despite the high temperature of 
70~$^\circ{\rm C}$, 
whereas one clearly 
feels the hot air. 
Above the ground one sees why it is cooler 
there than inside the car: the air inside 
the car \lq\lq stands still\rq\rq, above 
the ground one always feels a slight movement 
of the air. The ground is never completely plain, 
so there is always light and shadow, which keep the 
circulation going. This effect was formerly used for many 
old buildings in the city of Braunschweig, Germany. 
The south side of the houses had convexities.
Hence, for most of the time during the day, 
parts of the walls are in the shade and, 
because of the thus additionally stimulated circulation,
the walls are heated less. 

\begin{figure}[hbtp]
\ifthenelse{\equal{IJMPB}{\TschStyle}}{%
\centerline{\psfig{file=PictureLarge_parallelepiped_.eps,width=2.00in}}
}{}%
\ifthenelse{\equal{arXiv}{\TschStyle}}{%
\centerline{\includegraphics[scale=1.00]{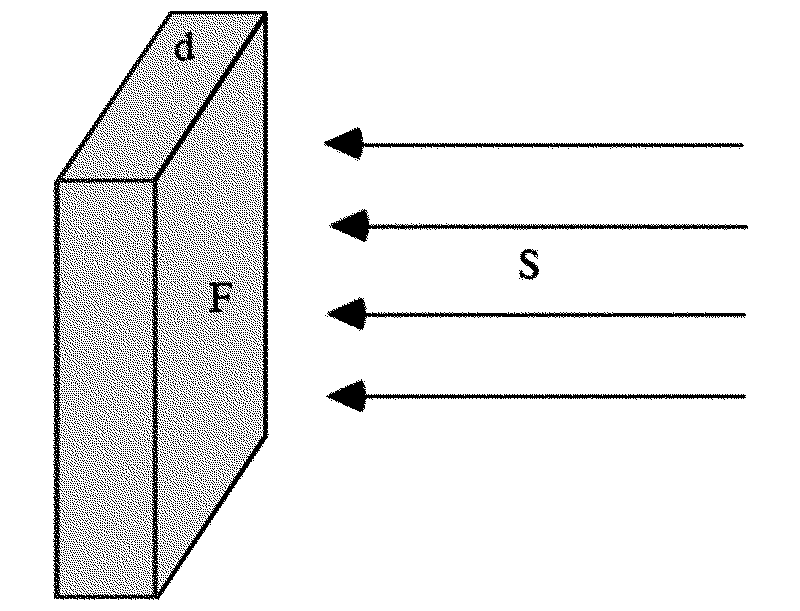}}
}{}%
\ifthenelse{\equal{TeX4ht}{\TschStyle}}{%
\centerline{\includegraphics[scale=1.00]{PictureLarge_parallelepiped_.png}}
}{}
\vspace*{8pt}
\caption{A solid parallelepiped of thickness ${\rm d}$
         and cross section ${\rm F}$ subject to 
         solar radiation.}
\label{fig:Parallelepiped}
\end{figure}

In order to study the warming effect
one can look at a body of 
specific heat 
$c_{\rm v}$ 
and width 
${\rm d}$, 
whose cross section 
${\rm F}$ 
is subject to the radiation 
intensity 
${\rm S}$
(see Figure \ref{fig:Parallelepiped}).
\noindent%
One has
\begin{equation}
\varrho\,
{\rm F}\,
{\rm d}\,
c_{\rm v}
\frac{dT}{dt}
=
{\rm F}{\rm S}
\end{equation}
or, respectively,
\begin{equation}
\frac{dT}{dt}
=
\frac{\rm S}{{\varrho}\,c_{\rm v}\,{\rm d}}\,
\end{equation}
which may be integrated yielding
\begin{equation}
T = T_0 +
\frac{\rm S}{{\varrho}\,c_{\rm v}\,{\rm d}}\,
(t-t_0)
\end{equation}
In this approximation, there is 
a linear rise of the temperature 
in time because of the 
irradiated intensity. One sees 
that the temperature rises particularly 
fast in absorbing bodies of small diameter: 
Thin layers are heated especially fast 
to high temperatures by solar radiation. 
The same applies to the heat capacity   
per unit volume: 
\begin{itemize}
\item 
If the heat capacity is large
the change of temperature will 
be slow.
\item 
If the heat capacity is small
the change in temperature will 
be fast.
\end{itemize}
Thus the irradiated intensity is 
responsible for the quick 
change of temperature, 
\textit{not} for its value. 
This rise in temperature 
is stopped by the heat transfer
of the body to its environment.

Especially in engineering thermodynamics
the different kinds of heat transfer and
their interplay are discussed thoroughly\TschSpace%
%
\cite{AlfredSchackBook,Kreith1999,Baukal2000}.
%
A comprehensive source is the classical 
textbook by Schack\TschSpace%
\cite{AlfredSchackBook}.
The results have been tested e.g.\ in combustion 
chambers and thus have a strong experimental
background. 

One has to distinguish between
\begin{itemize}
\item Conduction
\item Convection
\item Radiation
\item Transfer of latent heat in phase transitions 
      such as condensation and sublimation%
\footnote{%
Among those phenomena governed by the exchange of latent
heat there is \textit{radiation frost}, an striking
example for a cooling of the Earth's surface through 
emission of infrared radiation.} 
\end{itemize}
Conduction, condensation and radiation, 
which slow down the rise in temperature, 
work practically the same inside and outside 
the car. 
Therefore, the only possible reason for 
\textit{a difference} in final temperatures 
must be convection: 
A volume element of air above the ground, 
which has been heated by radiation, 
is heated up (by heat transfer
through conduction), rises and 
is replaced by cooler air. 
This way, there is, in the average, 
a higher difference of temperatures 
between the ground and the air and a higher 
heat transmission compared to a situation, 
where the air would not be replaced. 
This happens inside the car as well, 
but there the air stays locked in and the air 
which replaces the rising air is getting 
warmer and warmer, which causes lower heat 
transmission. Outside the car, there is 
of course a lot more cooler air than inside. 
On the whole, there is a higher temperature 
for the sunlight absorbing surfaces as well 
as for the air.

Of course, the exposed body loses energy 
by thermal radiation as well.
The warmer body inside the car would lose 
more heat in unit of time than the colder 
ground outside, which would lead to a 
higher temperature outside, 
if this temperature rise were not absorbed 
by another mechanism! If one considers, 
that only a small part of the formerly 
reckoned 
$60$\,-\,$70$ 
percent of solar radiation 
intensity reaches the inside of the car 
through its metal parts, this effect would 
contribute far stronger to the temperature 
outside! The \lq\lq explanation\rq\rq\ 
of the physical greenhouse effect 
only with attention to the radiation 
balance would therefore 
lead to the reverse effect!
The formerly discussed effect of the 
\lq\lq trapped\rq\rq\ heat radiation 
by reflecting glass panes remains, 
which one can read as hindered heat 
transmission in this context. 
So this means a deceleration of the cooling 
process. 
However, as this heat transmission 
is less important compared to the convection, 
nothing remains of the absorption and reflection 
properties of glass for infrared radiation 
to explain the physical greenhouse effect.
Neither the absorption nor the reflection 
coefficient of glass for the infrared light 
is relevant for this explanation of the 
physical greenhouse effect, but only the 
movement of air, hindered by the panes 
of glass.
 
Although meteorologists have known this 
for a long time\TschSpace%
%
\cite{Lee1973,Berry1974},
%
some of them still use the physical greenhouse 
effect to explain temperature effects of planetary 
atmospheres.
For instance in their book 
on the atmospheric greenhouse effect,
Sch\"onwiese and Diekmann 
build their arguments 
upon the glass house effect\TschSpace%
%
\cite{Schoenwiese1987}.
%
Their list of references
contains a seminal publication
that clearly shows that this is
inadmissable\TschSpace%
%
\cite{DOE1985}.
%

\subsection{Experiment by Wood}
Although the warming phenomenon in a glass house
is due to the suppression of convection,
say air cooling%
\footnote{A problem familiar to those who are 
          involved in PC hardware problems.},
it remains true that most glasses 
absorb infrared light at wavelength 
$1\,\mu{\rm m}$ and higher almost completely.

An \textit{experimentum crucis} therefore
is to build a glass house with panes consisting
of ${\rm NaCl}$ or ${\rm KCl}$, which are 
transparent to visible light as well as
infrared light. For rock salt (${\rm NaCl}$) 
such an experiment was realized as early 
as 1909 by Wood\TschSpace%
\cite{Wood1909,Jones1990,Schloerer2007,Connolley2007}:

\begin{quote}
\lq\lq
There appears to be a widespread belief 
that the comparatively high temperature 
produced within a closed space covered 
with glass, and exposed to solar radiation, 
results from a transformation of wave-length, 
that is, that the heat waves from the Sun, 
which are able to penetrate the glass, 
fall upon the walls of the enclosure 
and raise its temperature: the heat energy 
is re-emitted by the walls in the form 
of much longer waves, which are unable 
to penetrate the glass, the greenhouse 
acting as a radiation trap. 
\\[2mm]
I have always felt some doubt as to 
whether this action played any very 
large part in the elevation of temperature. 
It appeared much more probable that the 
part played by the glass was the prevention 
of the escape of the warm air heated by the 
ground within the enclosure. If we open 
the doors of a greenhouse on a cold and 
windy day, the trapping of radiation 
appears to lose much of its efficacy. 
As a matter of fact I am of the opinion 
that a greenhouse made of a glass transparent 
to waves of every possible length would show 
a temperature nearly, if not quite, as high 
as that observed in a glass house. The 
transparent screen allows the solar 
radiation to warm the ground, and the 
ground in turn warms the air, but only 
the limited amount within the enclosure. 
In the \lq\lq open\rq\rq, the ground is 
continually brought into contact with 
cold air by convection currents. 
\\[2mm]
To test the matter I constructed two enclosures 
of dead black cardboard, one covered with a 
glass plate, the other with a plate of rock-salt 
of equal thickness. The bulb of a thermometer 
was inserted in each enclosure and the whole 
packed in cotton, with the exception of the 
transparent plates which were exposed. When 
exposed to sunlight the temperature rose 
gradually to 65~$^\circ{\rm C}$, the enclosure 
covered with the salt plate keeping a little 
ahead of the other, owing to the fact that 
it transmitted the longer waves from the Sun, 
which were stopped by the glass. In order 
to eliminate this action the sunlight 
was first passed through a glass plate. 
\\[2mm]
There was now scarcely a difference of one 
degree between the temperatures of the two 
enclosures. The maximum temperature reached 
was about 55~$^\circ{\rm C}$. From what we know 
about the distribution of energy in the spectrum 
of the radiation emitted by a body at 55~$^\circ{\rm C}$, 
it is clear that the rock-salt plate is capable 
of transmitting practically all of it, while the 
glass plate stops it entirely. This shows us that 
the loss of temperature of the ground by radiation 
is very small in comparison to the loss by convection, 
in other words that we gain very little from the 
circumstance that the radiation is trapped. 
\\[2mm]
Is it therefore necessary to pay attention 
to trapped radiation in deducing the temperature 
of a planet as affected by its atmosphere? 
The solar rays penetrate the atmosphere, 
warm the ground which in turn warms the 
atmosphere by contact and by convection 
currents. The heat received is thus stored 
up in the atmosphere, remaining there on 
account of the very low radiating power 
of a gas. It seems to me very doubtful 
if the atmosphere is warmed to any great 
extent by absorbing the radiation from 
the ground, even under the most favourable 
conditions. 
\\[2mm]
I do not pretend to have gone very deeply 
into the matter, and publish this note merely 
to draw attention to the fact that trapped 
radiation appears to play but a very small 
part in the actual cases with which we are familiar.\rq\rq 
\end{quote}

This text is a recommended reading 
for all global climatologists referring 
to the greenhouse effect.  

\subsection{Glass house summary}
It is not the \lq\lq trapped\rq\rq\ infrared 
radiation, which explains the warming phenomenon 
in a real greenhouse, but it is the suppression 
of air cooling.%
\footnote{As almost everybody knows, this is
           also a standard problem in PCs.}

\newpage%
\section{The fictitious atmospheric greenhouse effects} 
\subsection{Definition of the problem} 
After it has been thoroughly discussed, 
that the physical greenhouse effect 
is essentially the explanation, 
why air temperatures in a closed glass house 
or in a closed car are higher than outside, 
one should have a closer look at the 
fictitious atmospheric greenhouse effects.
 
Meanwhile there are many different phenomena 
and different explanations for these effects,
so it is justified to pluralize here.

Depending on the particular school
and the degree of popularization,
the assumption that the atmosphere is transparent
for visible light but opaque for infrared
radiation is supposed to lead to
\begin{itemize}
\item a warming of the Earth's surface
      \textit{and/or}
\item a warming of the lower atmosphere
      \textit{and/or}
\item a warming of a certain layer of the atmosphere
      \textit{and/or}
\item a slow-down of the natural cooling
      of the Earth's surface
\end{itemize}
and so forth.

Unfortunately, there is no source
in the literature, where the greenhouse effect
is introduced in harmony with the scientific 
standards of theoretical physics.
As already emphasized,
the \lq\lq supplement\rq\rq\ to Kittel's
book on thermal physics\TschSpace%
\cite{Kittel2000} 
only refers to the IPCC assessments\TschSpace%
\cite{IPCC1990,IPCC1992}. 
Prominent global climatologists
(as well as \lq\lq climate sceptics\rq\rq) often
present their ideas in handbooks,
encyclopedias, and in secondary 
and tertiary literature.
%

\subsection{Scientific error versus scientific fraud} 
Recently, the German climatologist Gra\ss l emphasized
that errors in science are unavoidable, even in 
climate research\TschSpace%
%
\cite{Grassl2007}.
%
And the IPCC weights most of its official statements 
with a kind of a \lq\lq probability measure\rq\rq\TschSpace%
%
\cite{IPCC2007summary}.
%
So it seems that,
even in the mainstream discussion on the supposed
anthropogenic global warming, there is room left 
for scientific errors and their corrections.   
  
However, some authors and filmmakers have argued 
that the greenhouse effect hypothesis is not 
based on an error, but clearly is a kind of a scientific fraud. 

Five examples:

\begin{itemize}
\item As early as 1990 the Australian movie
      entitled \lq\lq The Greenhouse Conspiracy\rq\rq\
      showed that the case for the greenhouse effect 
      rests on four pillars\TschSpace%
%
\cite{AustralianMovie1990}:
%
      %
      \begin{enumerate}
      \item the \textit{factual evidence}, 
            i.e.\ the climate records,
            that supposedly suggest that a global warming
            has been observed and is exceptional; 
      \item the \textit{assumption} that carbon dioxide is the 
            cause of these changes;
      \item the \textit{predictions of climate models} 
            that claim that a doubling of ${\rm CO}_2$
            leads to a predictable global warming;
      \item the \textit{underlined physics}.
      \end{enumerate}
      In the movie these four pillars were dismantled
      bringing the building down. The speaker states:
      \begin{quote}  
      \lq\lq In a recent paper on the effects of carbon dioxide, 
             Professor Ellsaesser of the Lawrence Livermore Laboratories, 
             a major US research establishment in California, 
             concluded that a doubling of carbon dioxide 
             would have little or no effect on the 
             temperature at the surface and, 
             if anything, might cause the surface to cool.\rq\rq
      \end{quote}    
The reader is referred to Ellsaesser's original work\TschSpace%
%
\cite{Elsaesser1984}.
%
\item Two books by the popular German meteorologist and
      sociologist Wolfgang Th\"une, entitled
      \textit{The Greenhouse Swindle} 
      (In German, 1998)\TschSpace%
%
\cite{Thuene1998}
%
      and
      \textit{Aquittal for $\textit{CO}_2$} 
      (In German, 2002)\TschSpace%
%
\cite{Thuene2002}
%
      tried to demonstrate that the 
      ${\rm CO}_2$ greenhouse effect hypothesis 
      is pure nonsense.
\item A book written by Heinz Hug entitled
      \textit{Those who play the trumpet of fear}
      (In German, 2002)
      elucidated the history and the background of the 
      current greenhouse business\TschSpace%
%
\cite{Hug2006}
%
\item Another movie was shown recently on Channel 4 (UK)
      entitled
      \lq\lq The great global warming swindle\rq\rq\
      supporting the thesis that the supposed
      ${\rm CO}_2$ induced anthropogenic global warming
      has no scientific basis\TschSpace%
%
\cite{GGWS2007}.
%
\item In his paper 
      \lq\lq ${\rm CO}_2$: The Greatest Scientific 
      Scandal of Our Time\rq\rq\ 
      the eminent atmospheric scientist Jaworowski
      made a well-founded statement\TschSpace%
%
\cite{Jaworowski2007}.
%
\end{itemize} 
On the other hand,
Sir David King, the science advisor 
of the British government, stated that
\lq\lq global warming is a greater threat 
to humanity than terrorism\rq\rq\ (Singer)%
\footnote{cf.\ Singer's summary at the Stockholm 2006 conference\TschSpace%
%
\cite{Stockholm2006}.%
%
},  
other individuals put anthropogenic 
global warming deniers in the same category 
as holocaust deniers, and so on.
In an uncountable number of contributions 
to newspapers and TV shows in Germany the 
popular climatologist Latif%
\footnote{%
Some time ago one of the authors (R.D.T.), in his role as a physics löab research assistant, instructed his student Mojib Latif in fundamental university physics.}
continues to warn the public 
about the consequences of rising 
greenhouse gas (GHG) emissions\TschSpace%
%
\cite{Mopo2007}.
%
But until today it is \textit{impossible} to find 
a book on non-equilibrium thermodynamics or 
radiation transfer where this effect is derived 
from first principles.  

The main objective of this paper is not to draw 
the line between error and fraud, but to find 
out where the greenhouse effect appears or
\textit{disappears} within the frame of physics. 
Therefore, 
in 
Section~\ref{Sec:Versions}
several different 
variations of the atmospheric greenhouse hypotheses
will be analyzed and disproved.
The authors restrict themselves
on statements that appeared 
\textit{after} a publication 
by Lee in the well-known
\textit{Journal of Applied Meteorology} 
1973, see
%
\ifthenelse{\equal{IJMPB}{\TschStyle}}
           {Ref.~\refcite{Lee1973}}
           {}%
\ifthenelse{\equal{arXiv}{\TschStyle}}
           {Ref.~\cite{Lee1973}}
           {}%
\ifthenelse{\equal{TeX4ht}{\TschStyle}}
           {Ref.~\cite{Lee1973}}
           {}
%
and references therein.

Lee's 1973 paper is a milestone.
In the beginning Lee writes:
\begin{quote}
\lq\lq
The so-called radiation \lq greenhouse\rq\
effect is a misnomer. Ironically, while the 
concept is useful in describing what occurs
in the Earth's atmosphere, it is invalid for 
cryptoclimates created when space is enclosed
with glass, e.g.\ in greenhouses and solar 
energy collectors. Specifically, elevated 
temperatures observed under glass cannot be 
traced to the spectral absorbtivity of glass.
\\[2.00mm]
The misconception  was demonstrated experimentally
by R.\ W. Wood more than 60 years ago (Wood, 1909)\TschSpace%
%
\cite{Wood1909}
%
and recently in an analytical manner by Businger (1963)\TschSpace%
%
\cite{Businger1963}.
%
Fleagle and Businger (1963)\TschSpace%
%
\cite{Fleagle1963}
%
devoted a section of their text to the point, 
and suggested that radiation trapping 
by the Earth's atmosphere 
should be called \lq atmosphere effect\rq\ to 
discourage use of the misnomer. Munn (1966)\TschSpace%
%
\cite{Munn1966}
%
reiterated that the analogy between \lq atmosphere\rq\
and \lq greenhouse\rq\ effect \lq is not correct
because a major factor in greenhouse climate is the 
protection the glass gives against turbulent heat 
losses\rq.
In one instance, Lee (1966)\TschSpace%
%
\cite{Lee1966},
%
observed that the net
flux of radiant energy actually was diminished
be more than 10\,\%\ in a 6-mil polyvinyl enclosure.
\\[2.00mm]
In spite of the evidence, modern textbooks on
meteorology and climatology not only repeat the 
misnomer, but frequently support the false notion 
that      
\lq heat-retaining behavior of the atmosphere
is analogous to what happens in a greenhouse\rq\
(Miller, 1966)\TschSpace%
%
\cite{Miller1966},
%
or that 
\lq the function of the [greenhouse] glass is 
to form a radiation trap\rq\ (Peterssen, 1958)\TschSpace%
%
\cite{Pettersen1958}.
%
(see also Sellers, 1965, Chang, 1968, 
and Cole, 1970)\TschSpace%
%
\cite{Sellers1965,Chang1968,Cole1970}.
%
The mistake obviously is subjective, based on
similarities of the atmosphere and glass,
and on the \lq neatness\rq\ of the example 
in teaching. The problem can be rectified
through straightforward analysis, suitable
for classroom instruction.\rq\rq
\end{quote}
Lee continues his analysis with a calculation based on 
radiative balance equations, which are physically
questionable. 
The same holds for a comment by Berry\TschSpace%
%
\cite{Berry1974}
%
on Lee's work.
Nevertheless, Lee's paper
is a milestone marking \textit{the day after} 
which every serious scientist or science educator 
is no longer allowed to compare the
greenhouse with the atmosphere, 
even in the classroom, which Lee
explicitly refers to.

\subsection{Different versions of the atmospheric greenhouse conjecture}
\label{Sec:Versions}
%
%
\subsubsection{Atmospheric greenhouse effect after M\"oller (1973)}
In his popular textbook on meteorology\TschSpace%
\cite{Moeller1973a,Moeller1973b}
M\"oller claims: 
\begin{quote}
\lq\lq
In a real glass house (with no additional heating, 
i.e.\ no greenhouse) the window panes are transparent 
to sunshine, but opaque to terrestrial radiation. 
The heat exchange must take place through 
heat conduction within the glass, 
which requires a certain temperature gradient. 
Then the colder boundary surface of the window 
pane can emit heat. In case of the atmosphere 
water vapor and clouds play the role of the glass.\rq\rq \end{quote}
\textbf{Disproof:}
The existence of the greenhouse effect is considered 
as a necessary condition for thermal conductivity. 
This is a physical nonsense.
Furthermore it is implied that the spectral transmissivity
of a medium determines its thermal conductivity
straightforwardly. 
This is a physical nonsense as well.
%
%
\subsubsection{Atmospheric greenhouse effect after Meyer's encyclopedia (1974)}
In the 1974 edition of Meyer's Enzyklop\"adischem Lexikon 
one finds under \lq\lq glass house effect\rq\rq\TschSpace%
%
\cite{Meyers1974}:
%
\begin{quote}
\lq\lq
Name for the influence of the Earth's atmosphere 
on the radiation and heat budget of the Earth,
which compares to the effect of a glass house:
Water vapor and carbon dioxide in the atmosphere
let short wave solar radiation go through
down to the Earth's surface with a relative weak 
attenuation and, however, reflect the portion of
long wave (heat) radiation  which is emitted from the 
Earth's surface (atmospheric backradiation).\rq\rq
\end{quote}
\textbf{Disproof:}
Firstly, 
the main part of the solar radiation 
lies outside the visible light.
Secondly, 
reflection is confused with emission. 
Thirdly, 
the concept of atmospheric backradiation relies
on an inappropriate application of the formulas 
of cavity radiation. This will be discussed in
Section~\ref{Sec:Radiation}
%
%
\subsubsection{Atmospheric greenhouse effect after Sch\"onwiese (1987)}
The prominent climatologist Sch\"onwiese
states\TschSpace%
\cite{Schoenwiese1987}:
%
\begin{quote}
\lq\lq
$\dots$ we use the picture of a glass window
that is placed between the Sun and the Earth's surface.
The window pane lets pass the solar radiation unhindered 
but absorbs a portion of the heat radiation of the Earth.
The glass pane emits, corresponding to its own temperature,
heat in both directions: To the Earth's surface and to the
interplanetary space. Thus the radiative balance of the
Earth's surface is raised. 
The additional energy coming from the glass pane 
is absorbed almost completely by the Earth's surface
immediately warming up until a new radiative equilibrium
is reached.\rq\rq
\end{quote}
\textbf{Disproof:}
That the window pane lets pass the solar 
radiation unhindered is simply wrong.
Of course, some radiation goes sidewards.
As shown experimentally in 
Section~\ref{Sec:Experiments},
the panes of the car window are relatively cold.
This is only one out of many reasons, 
why the glass analogy is unusable.
Hence the statement is vacuous.  
%
%
\subsubsection{Atmospheric greenhouse effect after Stichel (1995)}
\label{Sec:Stichel1995}
Stichel
(the former deputy head of the German Physical Society)
stated once\TschSpace%
%
\cite{Stichel1995}:
%
\begin{quote}
\lq\lq
Now it is generally accepted textbook knowledge 
that the long-wave infrared radiation, emitted 
by the warmed up surface of the Earth, is 
partially absorbed and re-emitted by ${\rm CO}_2$ 
and other trace gases in the atmosphere. 
This effect leads to a warming of the lower 
atmosphere and, for reasons of the total 
radiation budget, to a cooling of the 
stratosphere at the same time.\rq\rq
\end{quote}
\textbf{Disproof:}
This would be a 
\textit{Perpetuum Mobile of the Second Kind}.
A detailed discussion is given in 
Section~\ref{Sec:Thermodynamics}.
Furthermore, there is no total radiation budget,
since there are no individual conservation laws 
for the different forms of energy participating 
in the game. The radiation energies in question 
are marginal compared to the relevant geophysical 
and astrophysical energies.
Finally, the radiation depends on the temperature
and not \textit{vice versa}.
%
%
\subsubsection{Atmospheric greenhouse effect after Anonymous 1 (1995)}
\label{Sec:WaterPot}
\begin{quote}
\lq\lq
The carbon dioxide in the atmosphere lets the radiation
of the Sun, whose maximum lies in the visible light, 
go through completely, while on the other hand 
it absorbs a part of the heat radiation emitted by the Earth 
into space because of its larger wavelength. 
This leads to higher near-surface air temperatures.\rq\rq
\end{quote}
\textbf{Disproof:}
The first statement is incorrect 
since the obviously non-neglible infrared part of the 
incoming solar radiation is being absorbed
(cf.\ Section~\ref{Sec:SunBlackBody}).
The second statement is falsified
by referring to a counterexample
known to every housewife:
The water pot on the stove.
Without water filled in, the bottom of the pot
will soon become glowing red. Water is an excellent
absorber of infrared radiation. However, with water
filled in, the bottom of the pot will be substantially 
colder. Another example would be the replacement of the 
vacuum or gas by glass in the space between two panes.
Conventional glass absorbs infrared radiation pretty well,
but its thermal conductivity shortcuts any thermal isolation. 
%
%
\subsubsection{Atmospheric greenhouse effect after Anonymous 2 (1995)}
\begin{quote}
\lq\lq
If one raises the concentration of carbon
dioxide, which absorbs the infrared light 
and lets visible light go through, in the 
Earth's atmosphere, the ground heated by the 
solar radiation and/or near-surface air 
will become warmer, 
because the cooling of the ground 
is slowed down.\rq\rq 
\end{quote}
\textbf{Disproof:}
It has already been shown in 
Section~\ref{Sec:Background}
that the thermal conductivity is changed
only marginally even by doubling the 
${\rm CO}_2$ concentration in the Earth's
atmosphere. 
%
%
\subsubsection{Atmospheric greenhouse effect after Anonymous 3 (1995)}
\begin{quote}
\lq\lq
If one adds to the Earth's atmosphere a gas,
which absorbs parts of the radiation of the 
ground into the atmosphere, the surface temperatures 
and near-surface air temperatures will become
larger.\rq\rq
\end{quote}
\textbf{Disproof:}
Again, the counterexample is the water pot on the stove;
see 
Section~\ref{Sec:WaterPot}.
%
%
\subsubsection{Atmospheric greenhouse effect 
               after German Meteorological Society (1995)}
In its 1995 statement, the German Meteorological Society says\TschSpace%
%
\cite{DMG1995}:
%
\begin{quote}
\lq\lq
As a point of a departure the radiation budget 
of the Earth is described. 
In this case the incident unweakened solar radiation 
at the Earth's surface is partly absorbed and partly 
reflected.
The absorbed portion is converted into heat and 
must be re-radiated in the infrared spectrum. 
Under such circumstances simple model calculations 
yield an average temperature of about $-18^\circ{\rm C}$
at the Earth's surface
$\dots$
Adding an atmosphere, the incident radiation 
at the Earth's surface is weakened only 
a little, because the atmosphere is essentially 
transparent in the visible range of the spectrum. 
Contrary to this, in the infrared range of the 
spectrum the radiation emitted form the ground 
is absorbed to a large extent by the atmosphere 
$\dots$ 
and, depending on the temperature, re-radiated 
in all directions.
Only in the so-called window ranges 
(in particular in the large atmospheric window 
$8$\,-\,$13$\,$\mu{\rm m}$) 
the infrared radiation can escape into space. 
The infrared radiation that is emitted downwards 
from the atmosphere (the so-called back-radiation) 
raises the energy supply of the Earth's surface.   
A state of equilibrium can adjust itself 
if the temperature of the ground rises and, 
therefore, a raised radiation according to 
Planck's law is possible. 
This undisputed natural greenhouse effect 
gives rise to an increase temperature of the 
Earth's surface.\rq\rq
\end{quote}
\textbf{Disproof:}
The concept of an radiation budget 
is physically wrong.
The average of the temperature 
is calculated incorrectly. 
Furthermore, a non-negligible portion of the incident solar radiation 
is absorbed by the atmosphere. 
Heat must not be confused with heat radiation.
The assumption that if gases emit 
heat radiation, then they will emit it  
only downwards, is rather obscure. 
The described mechanism of re-calibration 
to equilibrium has no physical basis.
The laws of cavity radiation do not apply 
to fluids and gases.
%
%
\subsubsection{Atmospheric greenhouse effect after Gra\ss l (1996)}
The former director of the 
World Meteorological Organization (WMO)
climate research program, 
Professor Hartmut Gra\ss l, 
states\TschSpace%
\cite{Grassl1996}:
%
\begin{quote}
\lq\lq
In so far as the gaseous hull [of the Earth] obstructs 
the propagation of solar energy down to the planet's surface 
less than the direct radiation of heat from the surface 
into space, the ground and the lower atmosphere must 
become warmer than \textit{without this atmosphere}, in order 
to re-radiate as much energy as received from the Sun.\rq\rq
\end{quote}
\textbf{Disproof:} 
This statement is vacuous, even in a literal sense.
One cannot compare the temperature of a planet's 
lower atmosphere with the situation where a planetary 
atmosphere does not exist at all. 
Furthermore, as shown in
Section~\ref{Sec:SunBlackBody}
the portion of the incoming infrared is larger 
than the portion of the incoming visible light.
Roughly speaking, we have a 50-50 relation.
Therefore the supposed warming from the bottom
must compare to an analogous warming from the top.
Even within the logics of Gra\ss l's oversimplified 
(and physically incorrect) conjecture one is left 
with a zero temperature gradient and thus a null effect.
%
%
\subsubsection{Atmospheric greenhouse effect after 
               Ahrens (2001)}
In his textbook 
\lq\lq Essentials in Meteorology:
       In Invitation to the Atmosphere\rq\rq\
the author Ahrens states\TschSpace%
%
\cite{Ahrens2001}:
%
\begin{quote}
\lq\lq 
The absorption characteristics of water vapor, 
${\rm CO}_2$,
and other gases such as methane and nitrous oxide
$\dots$
were, at one time, thought to be similar to the
glass of a florist's greenhouse. In a greenhouse, 
the glass allows visible radiation to come in, 
but inhibits to some degree the passage 
of outgoing infrared radiation. 
For this reason, 
the behavior of the water vapor and 
${\rm CO}_2$,
the atmosphere is popularly called the greenhouse
effect. However, studies have shown that the warm air
inside a greenhouse is probably caused more by the air's
inability to circulate and mix with the cooler outside air,
rather than by the entrapment of infrared energy. Because
of these findings, some scientists insist that the
greenhouse effect should be called the atmosphere effect.
To accommodate everyone, we will usually use the term
atmospheric greenhouse effect when describing the role
that water vapor and 
${\rm CO}_2$,
play in keeping the Earth''s
mean surface temperature higher than it otherwise
would be.\rq\rq
\end{quote}
\textbf{Disproof:}
The concept of the Earth's 
mean temperature is ill-defined.
Therefore the concept of a rise 
of a mean temperature 
is ill-defined as well.
%
%
\subsubsection{Atmospheric greenhouse effect after 
               Dictionary of Geophysics, Astrophysics,
               and Astronomy (2001)}
The Dictionary of Geophysics, Astrophysics,
and Astronomy says\TschSpace%
%
\cite{DictionaryGeophysicsAstronomyAstrophysics2001}:
%
\begin{quote}
\lq\lq 
Greenhouse Effect: The enhanced warming
of a planet's surface temperature caused by the
trapping of heat in the atmosphere by certain
types of gases 
(called greenhouse gases; primarily
carbon dioxide, water vapor, methane,
and chlorofluorocarbons). 
Visible light from the Sun 
passes through most atmospheres and is absorbed
by the body''s surface. The surface reradiates
this energy as longer-wavelength infrared
radiation (heat). If any of the greenhouse gases
are present in the body''s troposphere, the atmosphere
is transparent to the visible but opaque to
the infrared, and the infrared radiation will be
trapped close to the surface and will cause the
temperature close to the surface to be warmer
than it would be from solar heating alone.\rq\rq
\end{quote}
\textbf{Disproof:}
Infrared radiation is confused with heat.
It is not explained at all what is meant by 
\lq the infrared radiation will be trapped\rq\rq.
Is it a MASER, 
is it \lq\lq superinsulation\rq\rq,
i.e.\ vanishing thermal conductivity,
or is it simple thermalization?
%
%
\subsubsection{Atmospheric greenhouse effect after 
               Encyclopaedia of Astronomy and Astrophysics (2001)}
The Encyclopaedia of Astronomy and Astrophysics 
defines the greenhouse effect as follows\TschSpace%
%
\cite{EncyclopaediaAstronomyAndAstrophysics2001}:
%
\begin{quote}
\lq\lq
The greenhouse effect is the radiative influence exerted by
the atmosphere of a planet which causes the temperature
at the surface to rise above the value it would normally
reach if it were in direct equilibrium with sunlight 
(taking into account the planetary albedo). 
This effect stems from the fact 
that certain atmospheric gases have the ability
to transmit most of the solar radiation and to absorb
the infrared emission from the surface. 
The thermal (i.e.\ infrared) radiation 
intercepted by the atmosphere is then
partially re-emitted towards the surface, 
thus contributing additional heating of the surface.
Although the analogy is not entirely satisfactory in
terms of the physical processes involved, it is easy to
see the parallels between the greenhouse effect in the
atmosphere-surface system of a planet and a horticultural
greenhouse: the planetary atmosphere plays the role of the
glass cover that lets sunshine through to heat the soil 
while partly retaining the heat that escapes from the ground.
The analogy goes even further, since an atmosphere may
present opacity \lq windows\rq\ allowing infrared radiation
from the surface to escape, the equivalent of actual
windows that help regulate the temperature inside a
domestic greenhouse.\rq\rq
\end{quote}
\textbf{Disproof:}
The concept of the 
\lq\lq direct equilibrium with the sunlight\rq\ 
is physically wrong, 
as will be shown in detail in 
Section~\ref{Sec:RadiativeBalance}.
The description of the physics of a horticultural 
greenhouse is incorrect. Thus the analogy stinks. 
%
%
\subsubsection{Atmospheric greenhouse effect after 
               Encyclopaedia Britannica Online (2007)}
Encyclopaedia Britannica Online 
explains the greenhouse effect 
in the following way\TschSpace%
%
\cite{BritanicaOnline2007}:
%
\begin{quote}
\lq\lq
The atmosphere allows most of the visible light 
from the Sun to pass through and reach the 
Earth's surface. As the Earth's surface 
is heated by sunlight, it radiates part 
of this energy back toward space as 
infrared radiation. This radiation, 
unlike visible light, tends to be 
absorbed by the greenhouse gases 
in the atmosphere, raising its temperature. 
The heated atmosphere in turn radiates 
infrared radiation back toward the 
Earth's surface. (Despite its name, 
the greenhouse effect is different 
from the warming in a greenhouse, 
where panes of glass transmit visible 
sunlight but hold heat inside the 
building by trapping warmed air.) 
Without the heating caused by the 
greenhouse effect, the Earth's average 
surface temperature would be only about 
$-{\rm 18}\,^\circ{\rm C}$ ($0\,^\circ{\rm F}$).\rq\rq
\end{quote}
\textbf{Disproof:}
The concept of the Earth's average temperature
is a physically and mathematically ill-defined 
and therefore useless concept as will be shown 
in 
Section~\ref{Sec:RadiativeBalance}.
%
%
\subsubsection{Atmospheric greenhouse effect after Rahmstorf (2007)}
The renowned German climatologist Rahmstorf claims\TschSpace%
%
\cite{Rahmstorf2007}:
%
\begin{quote}
\lq\lq
To the solar radiation reaching Earth's surface 
$\dots$ 
the portion of the long-wave radiation is added, 
which is radiated by the molecules partly downward
and partly upward. Therefore more radiation arrives
down, and for reasons of compensation the surface 
must deliver more energy and thus has to be warmer
($+15\,^\circ{\rm C}$), 
in order to reach also there down again an equilibrium. 
A part of this heat is transported upward from the surface 
also by atmospheric convection. Without this natural 
greenhouse effect the Earth would have frozen life-hostilely 
and completely. 
The disturbance of the radiative balance 
[caused by the enrichment of the atmosphere with trace gases] 
must lead to a heating up of the Earth's surface, 
as it is actually observed.\rq\rq
\end{quote}
\textbf{Disproof:}
Obviously, reflection is confused with emission.
The concept of radiative balance is faulty. 
This will be explained in 
Section~\ref{Sec:RadiativeBalance}.
%
%
\subsubsection{Conclusion}
It is interesting to observe,
\begin{itemize}
\item that until today
      the \lq\lq atmospheric greenhouse effect\rq\rq\ 
      does not appear 
      \begin{itemize}
      \item in any fundamental work of thermodynamics,
      \item in any fundamental work of physical kinetics, 
      \item in any fundamental work of radiation theory;
      \end{itemize}
\item that the definitions given in the literature
      beyond straight physics are very different
      and, partly, \textbf{contradict} to each other.
\end{itemize}
%

\subsection{The conclusion of the US Department of Energy}
\begin{figure}[htbp]
\ifthenelse{\equal{IJMPB}{\TschStyle}}{%
\centerline{\psfig{file=PictureLarge_doe_.eps,width=3.00in}}
}{}%
\ifthenelse{\equal{arXiv}{\TschStyle}}{%
\centerline{\includegraphics[scale=1.00]{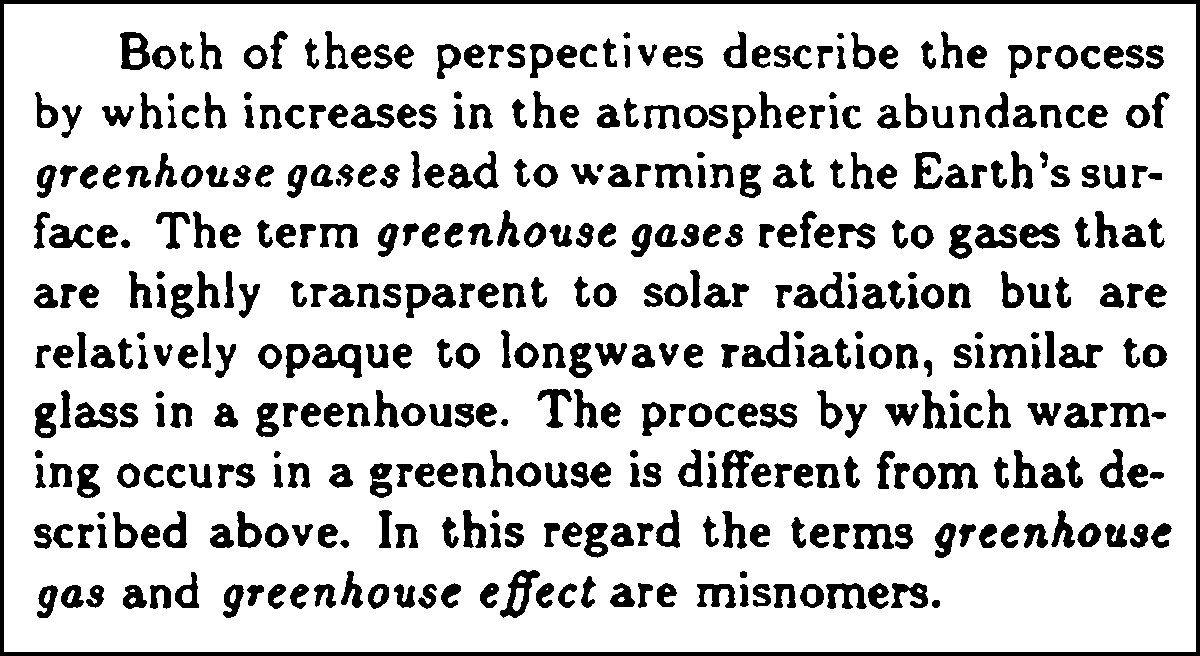}}
}{}%
\ifthenelse{\equal{TeX4ht}{\TschStyle}}{%
\centerline{\includegraphics[scale=1.00]{PictureLarge_doe_.png}}
}{}
\vspace*{8pt}
\caption{An excerpt from page 28 of the DOE report (1985).}
%
\label{Fig:Misnomer}
\end{figure}

All fictitious greenhouse effects have in common, 
that there is supposed to be one and only one
cause for them: 
An eventual rise in the concentration of ${\rm CO}_2$ 
in the atmosphere is supposed to lead 
to higher air temperatures near the ground.
For convenience,
in the context of this paper it is called 
\textit{the $\textit{CO}_2$-greenhouse effect}.%
\footnote{The nomenclature naturally extents to 
          other trace gases.}
Lee's 1973 result\TschSpace%
%
\cite{Lee1973}
%
that the warming phenomenon in a glass house 
does not compare to the supposed atmospheric 
greenhouse effect was confirmed 
in the 1985 report of the United States 
Department of Energy
\lq\lq Projecting the climatic effects
       of increasing carbon dioxide\rq\rq\TschSpace%
%
\cite{DOE1985}.
%
In this comprehensive pre-IPCC publication
MacCracken explicitly states that 
the terms \lq\lq greenhouse gas\rq\rq\
and \lq\lq greenhouse effect\rq\rq\ 
are misnomers\TschSpace%
%
\cite{DOE1985,MacCracken1985}.
%
A copy of the last paragraph 
of the corresponding section on page 28
in shown in 
Figure~\ref{Fig:Misnomer}.

The following should be emphasized:
\begin{itemize}
\item 
The warming phenomenon in a glass house and 
the supposed atmospheric greenhouse effects 
have the same participants,
but in the latter case the situation is reversed.
\item  
Methodically, there is a huge difference: 
For the physical greenhouse effect 
one can make measurements, look 
at the differences of the instruments readings
and observe the effect without any scientific 
explanation and such without any prejudice.
\end{itemize}
For the fictitious atmospheric greenhouse 
effect one cannot watch anything, 
and only calculations are compared 
with one another: 
Formerly extremely simple calculations, 
they got more and more intransparent.
Nowadays computer simulations are used, 
which virtually nobody can reproduce\TschSpace%
%
\cite{JIR2007}.
%

In the following the different
aspects of the physics 
underlying the atmospheric situation
are discussed in detail. 

\subsection{Absorption/Emission is not Reflection}
\label{Sec:Radiation}
%
%
\subsubsection{An inconvenient popularization of physics}
\label{AIPP}

\begin{figure}[hbtp]
\ifthenelse{\equal{IJMPB}{\TschStyle}}{%
\centerline{\psfig{file=PictureLarge_An_Inconvenient_Truth_.eps,width=4.00in}}
}{}%
\ifthenelse{\equal{arXiv}{\TschStyle}}{%
\centerline{\includegraphics[scale=4.00]{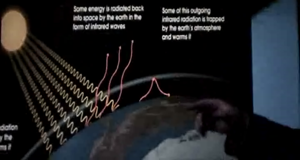}}
}{}%
\ifthenelse{\equal{TeX4ht}{\TschStyle}}{%
\centerline{\includegraphics[scale=1.00]{PictureLarge_An_Inconvenient_Truth_.png}}
}{}
\vspace*{8pt}
\caption{A very popular physical error illustrated
         in the movie \lq\lq An Inconvenient truth\rq\rq\
         by Davis Guggenheim featuring Al Gore (2006).}
\label{fig:AnInconvenientTruth}
\end{figure}

Figure~\ref{fig:AnInconvenientTruth}
is a screenshot from a controversial award-winning
\lq\lq documentary film\rq\rq\ 
about 
\lq\lq climate change\rq\rq, 
specifically \lq\lq global warming\rq\rq, 
starring Al Gore, the former United States 
Vice President, and directed by Davis Guggenheim\TschSpace%
%
\cite{AIT2006Book,AIT2006}.
%
This movie has been supported by managers and
policymakers around the world and has been shown 
in schools and in outside events, respectively.
Lewis wrote an interesting 
\lq\lq A Skeptic's Guide to An Inconvenient Truth\rq\rq\
evaluating Gore's work in detail\TschSpace%
%
\cite{Lewis2006}.
%

From the view of a trained physicist, 
Gore's movie is rather grotesque, 
since it is shockingly wrong. 
Every licensed radio amateur%
%
\footnote{\textit{Callsign of R.D.T.:} DK8HH}
%
knows that
what is depicted in 
Figure~\ref{fig:AnInconvenientTruth}
would be true only,
\begin{itemize}
\item if the radiation graphically
      represented here was long wave 
      or short wave radiation;
\item if the reflecting sphere was 
      a certain layer of the ionosphere\TschSpace%
%
\cite{Budden1966}.
%
\end{itemize}
Short waves 
(e.g.\ in the 20\,m/14\,MHz band)
are reflected 
by the F layer of the ionosphere
(located 120\,-\,400\,km above the Earth's surface)
enabling transatlantic connections (QSOs).
Things depend pretty much on the solar
activity, i.e.\ on the sun spot cycle,
as every old man (OM) knows well.
The reflective characteristics of the 
ionosphere diminish above about
${\rm 30}\,{\rm MHz}$.
In the very high frequency (VHF) bands 
(e.g.\ 2\,m/144\,MHz band)
one encounters the so called Sporadic-E clouds
(90\,-\,120\,km above the Earth's surface),
which still allow QSOs from Germany to Italy,
for example.
On the other hand at the extremely low 
frequencies (ELF, i.e.\ frequency range 3\,-\,30\,Hz)
the atmosphere of the Earth behaves as 
a cavity and one encounters the so called
Schumann resonances\TschSpace%
%
\cite{Schumann1952}.
%
These may be used 
to estimate a lower bound for the mass of the photon%
%
\footnote{As a teaching assistant at Hamburg University/DESY, R.D.T.\ 
           learned this from Professor Herwig Schopper.}
%
and, surprisingly, appear in the climate change discussion\TschSpace%
%
\cite{Fullekrug2006}.
%

However, the radio signal of Al Gore's cellular phone 
(within the centimeter range) does not travel around 
the world and so does not Bluetooth, Radar, microwave 
and infrared radiation (i.e.\ electromagnetic waves 
in the sub millimeter range). 

Ionosphere Radars typically work
in the $6\,{\rm m}$ Band,
i.e.\ at $50\,{\rm MHz}$.
Meteorological Radars work 
in the 0.1\,-\,20\,cm range 
(from 90\,GHz\ down to 1.5\,GHz),
those 
in the 3\,-\,10\,cm range
(from 10\,GHz\ down to 3\,GHz)
are used for wind finding and weather watch\TschSpace%
%
\cite{Atlas1990}.
%
It is obvious, that Al Gore confuses 
the ionosphere with the tropopause,
the region in the atmosphere, that
is the boundary between the troposphere 
and the stratosphere.
The latter one is located 
between 
$6\ {\rm km}$ (at the poles)
and 
$17\ {\rm km}$ (at the equator) 
above the surface of the Earth.%
\footnote{Some climatologists claim that there is a
          ${\rm CO}_2$ layer in the troposphere 
          that traps or reflects
          the infrared radiation coming from the ground.}

Furthermore, Al Gore confuses
\textit{absorption/emission} 
with \textit{reflection}.
Unfortunately, this is also done implicitly and
explicitly in many climatologic papers, often
by using the vaguely defined terms
\lq\lq re-emission\rq\rq,
\lq\lq re-radiation\rq\rq\
and
\lq\lq backradiation\rq\rq.
%
%
\subsubsection{Reflection}
When electromagnetic waves move 
from a medium of a given refractive index $n_1$ into 
a second medium with refractive index $n_2$, both 
reflection and refraction of the waves may occur\TschSpace%
%
\cite{Born1997}. 
%
In particular, 
when the jump of the refractive index 
occurs within a length of the order
of a wavelength, there will be a reflection.
The fraction of the intensity of incident 
electromagnetic wave that is reflected 
from the interface is given by the 
reflection coefficient $R$, 
the fraction refracted 
at the interface is given by the 
transmission coefficient $T$. 
The Fresnel equations, which are 
based on the assumption that the 
two materials are both dielectric, 
may be used to calculate 
the reflection coefficient $R$ and 
the transmission coefficient $T$ 
in a given situation.

In the case of a normal incidence %
the formula for the reflection 
coefficient is
\begin{equation}
R
=
\left(
\frac{n_2 - n_1}{n_2 + n_1}
\right)
^2
\end{equation}
In the case of strong absorption 
(large electrical conductivity $\sigma$)
simple formulas can be given for larger 
angles $\gamma$ of incidence, as well 
(Beer's formula): 
\begin{equation}
R_s
=
\frac{(n_2-n_1\cos\gamma)^2 + n_2^2\sigma^2}
     {(n_2+n_1\cos\gamma)^2 + n_2^2\sigma^2}
\end{equation}
and
\begin{equation}
R_p
=
\frac{(n_1-n_2\cos\gamma)^2 + n_2^2\sigma^2\cos^2\gamma}
     {(n_1+n_2\cos\gamma)^2 + n_2^2\sigma^2\cos^2\gamma}
\end{equation}
When the jump of the refractive index 
occurs within a length of the order
of a wavelength, there will be a reflection,
which is large at high absorption. 
In the case of gases 
this is only possible 
for radio waves of a comparatively 
long wave length in the ionosphere,
which has an electrical conductivity, 
at a diagonal angle of incidence. 
There is no reflection in the 
homogeneous absorbing range.
As already elucidated 
in Section~\ref{AIPP}
this has been well-known to radio 
amateurs ever since and affects
their activity e.g.\ in the 
15 m band, but never in the 
microwave bands. 
On the other hand,
most glasses absorb the infrared light 
almost completely at approximately
$1\,\mu{\rm m}$
and longer wavelength: 
therefore, the reflection 
of the infrared waves for normal 
glasses is very high.

For dielectric media, whose electrical conductivity is zero,
one cannot use Beer's formulas. This was a severe problem
in Maxwell's theory of light.
%
%
\subsubsection{Absorption and Emission}
If an area is in thermodynamical equilibrium 
with a field of radiation, 
the intensity 
${\rm E}_\nu$ 
(resp.\ ${\rm E}_\lambda$)
emitted by the unit solid angle
into a frequency unit 
(resp.\ a wavelength unit) 
is equal to the absorptance
${\rm A}_\nu$ 
(resp.\ ${\rm A}_\lambda$)
multiplied with 
a universal frequency function ${\rm B}_\nu(T)$ 
(resp.\ a wavelength function ${\rm B}_\lambda(T)$) 
of the absolute temperature $T$. 
One writes, respectively, 
\begin{eqnarray}
{\rm E}_\nu 
&=&
{\rm A}_\nu \cdot {\rm B}_\nu (T)
\\
{\rm E}_\lambda 
&=&
{\rm A}_\lambda \cdot {\rm B}_\lambda (T)
\end{eqnarray}
This is a theorem by \textit{Kirchhoff}. 
The function 
${\rm B}_\nu(T)$ 
(resp.\ ${\rm B}_\lambda(T)$)
is called the 
\textit{Kirchhoff-Planck-function.} 
It was already considered in Section~\ref{SBlaw}.

\begin{figure}[hbtp]
\ifthenelse{\equal{IJMPB}{\TschStyle}}{%
\centerline{\psfig{file=PictureLarge_cavity_.eps,width=1.50in}}
}{}%
\ifthenelse{\equal{arXiv}{\TschStyle}}{%
\centerline{\includegraphics[scale=1.00]{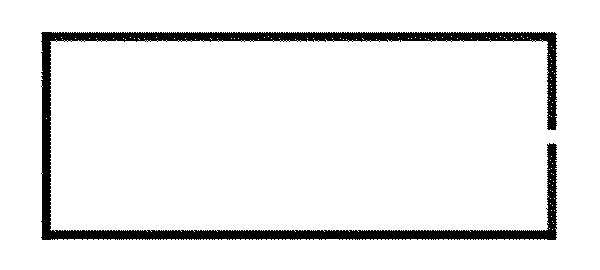}}
\vspace*{-16pt}
}{}%
\ifthenelse{\equal{TeX4ht}{\TschStyle}}{%
\centerline{\includegraphics[scale=1.00]{PictureLarge_cavity_.png}}
\vspace*{-16pt}
}{}
\vspace*{8pt}
\caption{A cavity realizing a perfect black body.}
\label{fig:cavity}
\end{figure}

The \textit{reflectance} is, respectively,
\begin{eqnarray}
{\rm R}_\nu &=& 1- {\rm A}_\nu \\
{\rm R}_\lambda &=& 1- {\rm A}_\lambda
\end{eqnarray}
and lies between zero and one,
like the \textit{absorptance} 
${\rm A}_\nu$. 
If ${\rm R}$ is equal to zero 
and ${\rm A}$ is equal 
to one, the body is called a 
perfect black body. 
The emissivity is largest 
for a perfect black body. 
The proposal to realize a 
perfect black body by using a cavity 
with a small radiating opening 
had already been made by Kirchhoff 
and is visualized 
in Figure~\ref{fig:cavity}.
For this reason, the emission of 
a black body for 
${\rm A}_\nu=1$ 
(resp.\ ${\rm A}_\lambda=1$)
is called \textit{cavity radiation}. 
The emitted energy comes from the walls, 
which are being held at a fixed temperature. 
If this is realized with a 
part of a body's surface, 
it will become clear, that 
these points of view will only be 
compatible, if the electromagnetic 
radiation is emitted and absorbed 
by an extremely thin surface layer.
For this reason, it is impossible 
to describe the volumes of gases 
with the model of black cavity 
radiation.
Since thermal radiation is electromagnetic 
radiation, this radiation would have to be 
caused by thermal motion in case of gases, 
which normally does not work effectively 
at room temperatures. At the temperatures 
of stars the situation is different:
The energy levels of the atoms are 
thermally excited by impacts.
%
%
\subsubsection{Re-emission}
In case of radiation transport calculations, 
Kirchhoff's law is \lq\lq generalized\rq\rq\ 
to the situation, in which the corresponding 
formula for the emission, or respectively, 
for the absorption 
(per unit length along the direction $ds$) 
is supposed to be applicable
\begin{equation}
\varepsilon_\nu ds 
=
\alpha_\nu ds
\cdot
{\rm B}_\nu (T)
\label{Eq:generalizedKirchhoff}
\end{equation}
The physical meaning of this 
\lq\lq generalization\rq\rq\ 
can be seen most easily, 
if the above mentioned Kirchhoff law 
is mathematically extracted out 
of this formula. For this,
one may introduce
\begin{eqnarray}
\varepsilon_\nu &=& {\rm E}_\nu \, \delta(s-s_0) 
\\
\alpha_\nu      &=& {\rm A}_\nu \, \delta(s-s_0)
\end{eqnarray}
with a $\delta$-density 
localized at the interface.
Physically, this means that all 
of the absorption and emission 
comes out of a thin superficial plane. 
Just like with the correct Kirchhoff law, 
use is made of the fact, 
that all absorbed radiation 
is emitted again, as otherwise 
the volume area would raise 
its temperature in thermal balance.

This assumption is called 
the assumption of  
\textit{Local Thermodynamical Equilibrium (LTE)}.
Re-emission does never mean reflection, 
but, rather, that the absorption 
\textit{does not cause any rise 
of temperature in the gas.}

An important physical difference 
to the correct Kirchhoff law lies
in the fact, that there is 
no formula for the absorption 
per linear unit analogous to 
\begin{equation}
{\rm R}_\nu  = 1 - {\rm A}_{\nu}
\end{equation}  

With $\rho$ being the density of the medium
one can define a 
\textit{absorption coefficient} $\kappa_\nu$
and an 
\textit{emission coefficient} $j_\nu$,
respectively, by setting 
\begin{eqnarray}
\alpha_\nu      &=& \kappa_\nu \, \rho 
\label{Eq:alphanu}
\\
\varepsilon_\nu &=& j_\nu \, \rho 
\label{Eq:varepsilonnu}
\end{eqnarray}
The ratio of the emission to the absorption coefficient
\begin{equation}
{\rm S}_\nu = \frac{j_\nu}{\kappa_\nu}
\label{Eq:sourcefunction}
\end{equation}
describes the re-emission of the radiation 
and is called the \textit{source function}.
%
%
\subsubsection{Two approaches of Radiative Transfer}
In a gas the radiation intensity 
of an area changes in the direction 
of the path element $ds$ according to
\begin{equation}
-
\frac {d{\rm I}_\nu} {ds}
=
\alpha_\nu{\rm I}_\nu
-
\varepsilon_\nu
\end{equation}
With the aid of the functions introduced in 
Equations
(\ref{Eq:alphanu}) - (\ref{Eq:sourcefunction}) 
this can be expressed as
\begin{equation}
\frac {{\rm 1}} {\kappa_\nu \varrho}
\frac {d{\rm I}_\nu} {ds}
=
{\rm I}_\nu 
-
{\rm S}_\nu
\label{Eq:RTE} 
\end{equation}
This equation is called the \textit{radiative transfer equation}.

Two completely different approaches
show that this emission function is 
not just determined by physical laws\TschSpace%
%
\cite{Chandrasekhar1960}:
%
\begin{enumerate}
\item
The usual one, i.e.\ the one in case of LTE, 
is given by the ansatz
\begin{equation}
{\rm S}_\nu ( x,y,z ; l,m,n )
=
{\rm B}_\nu ( {\rm T} ( x,y,z ; l,m,n ) )
\end{equation}
where the coordinates $(x,y,z)$ and the direction cosines $(l,m,n)$ define the point and
the direction to which $S_\nu$ and $B_\nu$ (resp.\ $T$) refer. 
This approach is justified 
with the aid of  
the Kirchhoff-Planck-function 
${\rm B}_\nu$
and the \lq\lq generalized\rq\rq\
Kirchhoff law introduced in 
Equation 
(\ref{Eq:generalizedKirchhoff}).
This assumption of 
\textit{Local Thermodynamical Equilibrium (LTE)} 
is ruled out by many scientists
even for the extremely hot atmospheres 
of stars. The reader is referred to
Chandrasekhar's classical book on
radiative transfer\TschSpace%
%
\cite{Chandrasekhar1960}.
%
LTE does only bear a certain significance 
for the radiation transport calculations, 
if the absorption coefficients were 
not dependent on the temperature, 
which is not the case at low temperatures. 
Nevertheless, in modern climate 
model computations, this approach is 
used unscrupulously\TschSpace%
%
\cite{DOE1985}.
%
\item
Another approach is the 
\textit{scattering atmosphere}
given by
\begin{equation}
{\rm S}_\nu
=
\frac{1}{4\pi}
\int_0^{\pi}
\int_0^{2\pi}
{\rm p} ( \vartheta, \varphi ; \vartheta', \varphi' )
\,
{\rm I}_\nu ( \vartheta', \varphi' )
\,
\sin \vartheta'
d \vartheta'
d \varphi'
\end{equation}
\end{enumerate}

These extremely different approaches show, 
that even the physically well-founded 
radiative transfer calculations are somewhat arbitrary. 
Formally, the radiative transfer equation
(\ref{Eq:RTE}) 
can be integrated leading to 
\begin{equation}
{\rm I}_\nu(s)
=
{\rm I}_\nu(0)
\exp (-\tau(s,0))
+
\int_0^s
{\rm S}_\nu(s')
\exp (-\tau(s,s'))
\kappa_\nu
\varrho
\,
ds'
\end{equation}
with the optical thickness
\begin{equation}
\tau(s,s') 
=
\int_{s'}^{s}
\kappa_\nu
\, 
\varrho
\,
ds''
\end{equation}
The integrations for the separate directions 
are independent of one another.
In particular, 
the ones up 
have nothing to do 
with the ones down.
It cannot be overemphasized,
that differential equations only allow 
the calculation of changes on the basis 
of known parameters. The initial values 
(or boundary conditions) cannot be derived 
from the differential equations to be solved.
In particular, this even holds for this simple
integral.

If one assumes that the temperature of a volume element
should be constant, one cannot calculate a rising 
temperature.

\subsection{The hypotheses of Fourier, Tyndall, and Arrhenius}
\subsubsection{The traditional works}
In their research and review papers
the climatologists refer 
to legendary publications
of Svante August Arrhenius 
(19 Feb.\ 1859 - 2 Oct.\ 1927),
a Nobel Prize winner for chemistry.
Arrhenius published one of the earliest, 
extremely simple calculations in 1896, 
which were immediately - and correctly - 
doubted and have been forgotten for many decades\TschSpace%
%
\cite{Arrhenius1896,Arrhenius1901,Arrhenius1906}.
%
It is a paper about the influence of carbonic acid 
in the air on the Earth's ground temperature.
In this quite long paper, Arrhenius put 
the hypothesis up for discussion, 
that the occurrences of warm and ice ages 
are supposed to be explainable by certain 
gases in the atmosphere, which absorb 
thermal radiation. 

In this context 
Arrhenius cited a 1824 publication by Fourier%
\footnote{There is a misprint in Arrhenius'
          work. The year of publication of
          Fourier's paper is 1824, not 1827
          as stated in many current papers,
          whose authors apparently did not
          read the original work of Fourier.
          It is questionable whether Arrhenius
          read the original paper.}
entitled 
\lq \lq 
M\'emoire sur les temp\'eratures 
du globe terrestre et des espaces plan\'etaires\rq\rq\TschSpace%
%
\cite{Fourier1824a,Fourier1824b}.
%

Arrhenius states incorrectly
that Fourier was the first,   
who claimed that 
the atmosphere works like a glass 
of a greenhouse 
as it lets the rays of the Sun through but keeps 
the so-called dark heat from the ground inside.

The English translation of the
relevant passage (p.\ 585) reads: 
\begin{quote}
We owe to the celebrated voyager M.\ de Saussure 
an experiment which appears very important 
in illuminating this question. It consists 
of exposing to the rays of the Sun a vase 
covered by one or more layers of well 
transparent glass, spaced at a certain distance. 
The interior of the vase is lined with 
a thick envelope of blackened cork, 
to receive and conserve heat. 
The heated air is sealed in all parts, 
either in the box or in each interval 
between plates. Thermometers placed 
in the vase and the intervals mark 
the degree of heat acquired in each place. 
This instrument has been exposed to the Sun 
near midday, and one saw, in diverse experiments, 
the thermometer of the vase reach 
70, 80, 100, 110 degrees and beyond 
(octogesimal division). Thermometers 
placed in the intervals acquired 
a lesser degree of heat, and which 
decreased from the depth of the box 
towards the outside.
\end{quote} 
Arrhenius work was also preceded by the
work of Tyndall who discovered that 
some gases absorb infrared radiation. 
He also suggested that changes 
in the concentration 
of the gases could bring climate change\TschSpace%
%
\cite{Tyndall1861,Tyndall1863a,Tyndall1863b,Tyndall1873a,Tyndall1873b}.
%
A faksimile of the front pages of
Fourier's and Arrhenius often cited
but apparently not really known papers
are shown in
Figure \ref{fig:Fourier1824}
and in 
Figure \ref{fig:Arrhenius1896},
respectively.
\ifthenelse{\equal{IJMPB}{\TschStyle}}{}{}%
\ifthenelse{\equal{arXiv}{\TschStyle}}{}{}%
\ifthenelse{\equal{TeX4ht}{\TschStyle}}{}{}
%
%
\ifthenelse{\equal{IJMPB}{\TschStyle}}{%
\begin{figure}[hbtp]
\centerline{\psfig{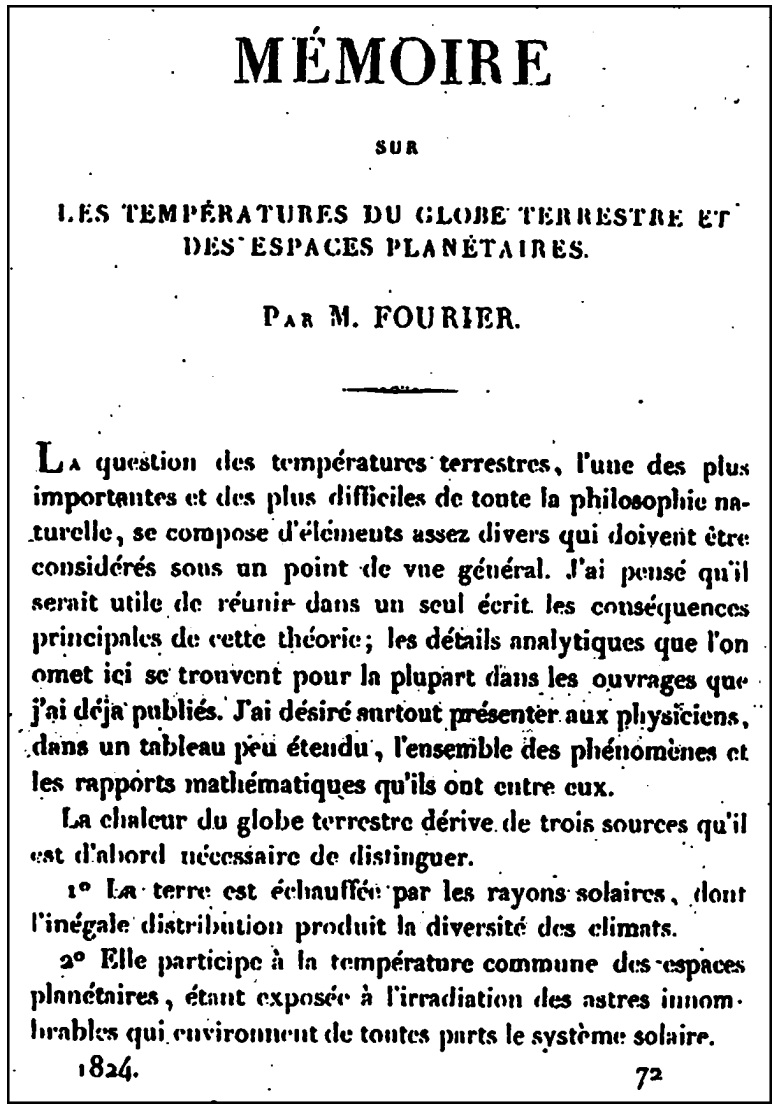}}
\vspace*{8pt}
\caption{The front page of Fourier's 1824 
         paper.}
\label{fig:Fourier1824}
\end{figure}
}{}%
\ifthenelse{\equal{arXiv}{\TschStyle}}{%
\ \vfill
\begin{figure}[hbtp]
\centerline{\includegraphics[scale=1.00]{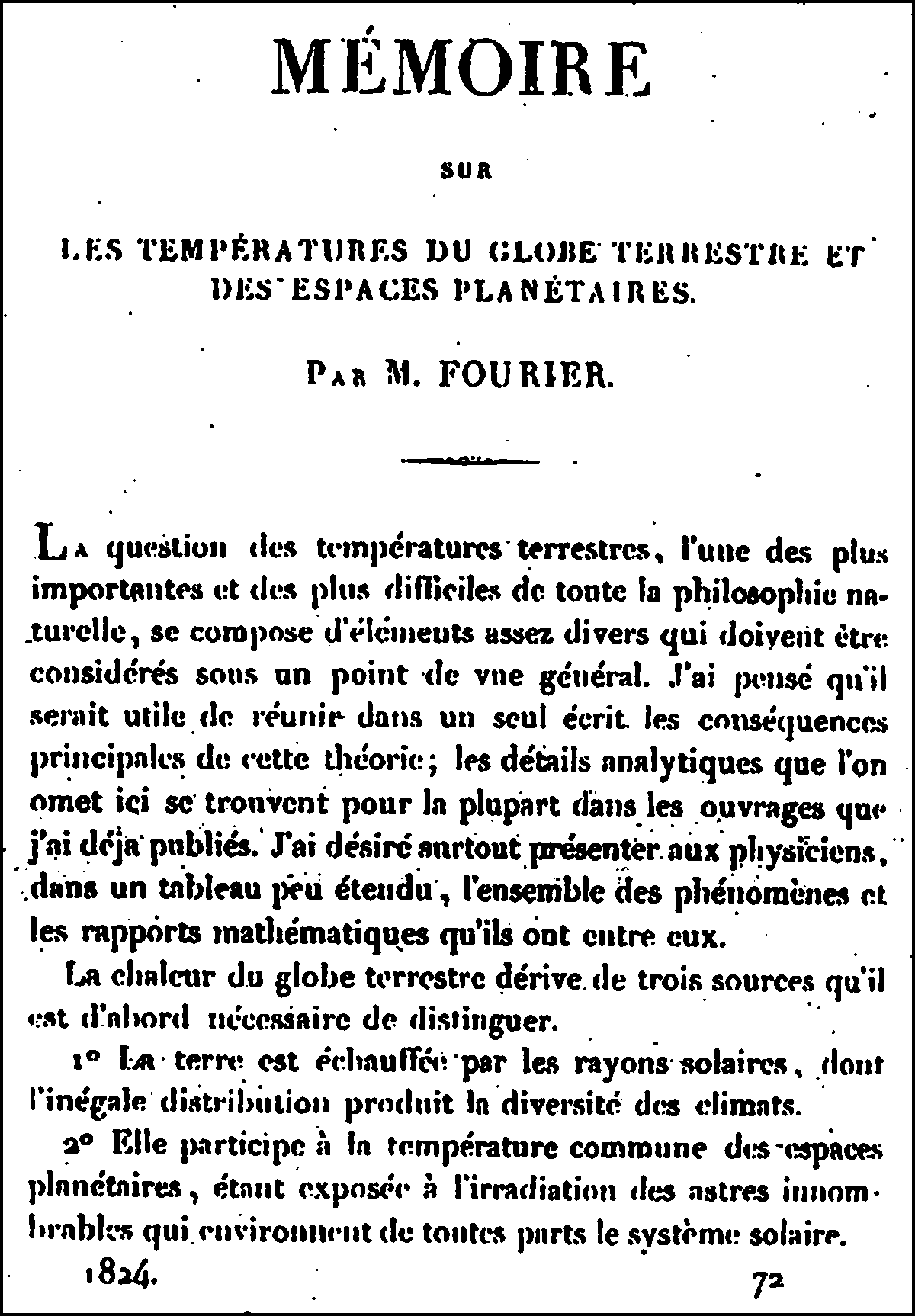}}
\vspace*{8pt}
\caption{The front page of Fourier's 1824 
         paper.}
\label{fig:Fourier1824}
\end{figure}
\ \vfill
}{}%
\ifthenelse{\equal{TeX4ht}{\TschStyle}}{%
\ \vfill
\begin{figure}[hbtp]
\centerline{\includegraphics[scale=1.00]{PictureLarge_Fourier1824_.png}}
\vspace*{8pt}
\caption{The front page of Fourier's 1824 
         paper.}
\label{fig:Fourier1824}
\end{figure}
\ \vfill
}{}
\ifthenelse{\equal{IJMPB}{\TschStyle}}{}{}%
\ifthenelse{\equal{arXiv}{\TschStyle}}{\newpage}{}%
\ifthenelse{\equal{TeX4ht}{\TschStyle}}{\newpage}{}
%
%
\ifthenelse{\equal{IJMPB}{\TschStyle}}{%
\begin{figure}[hbtp]
\centerline{\psfig{file=PictureLarge_Arrhenius1896_.eps,width=3.00in}}
\vspace*{8pt}
\caption{The front page of Arrhenius' 1896 
         paper.}
\label{fig:Arrhenius1896}
\end{figure}
}{}%
\ifthenelse{\equal{arXiv}{\TschStyle}}{%
\ \vfill
\begin{figure}[hbtp]
\centerline{\includegraphics[scale=1.00]{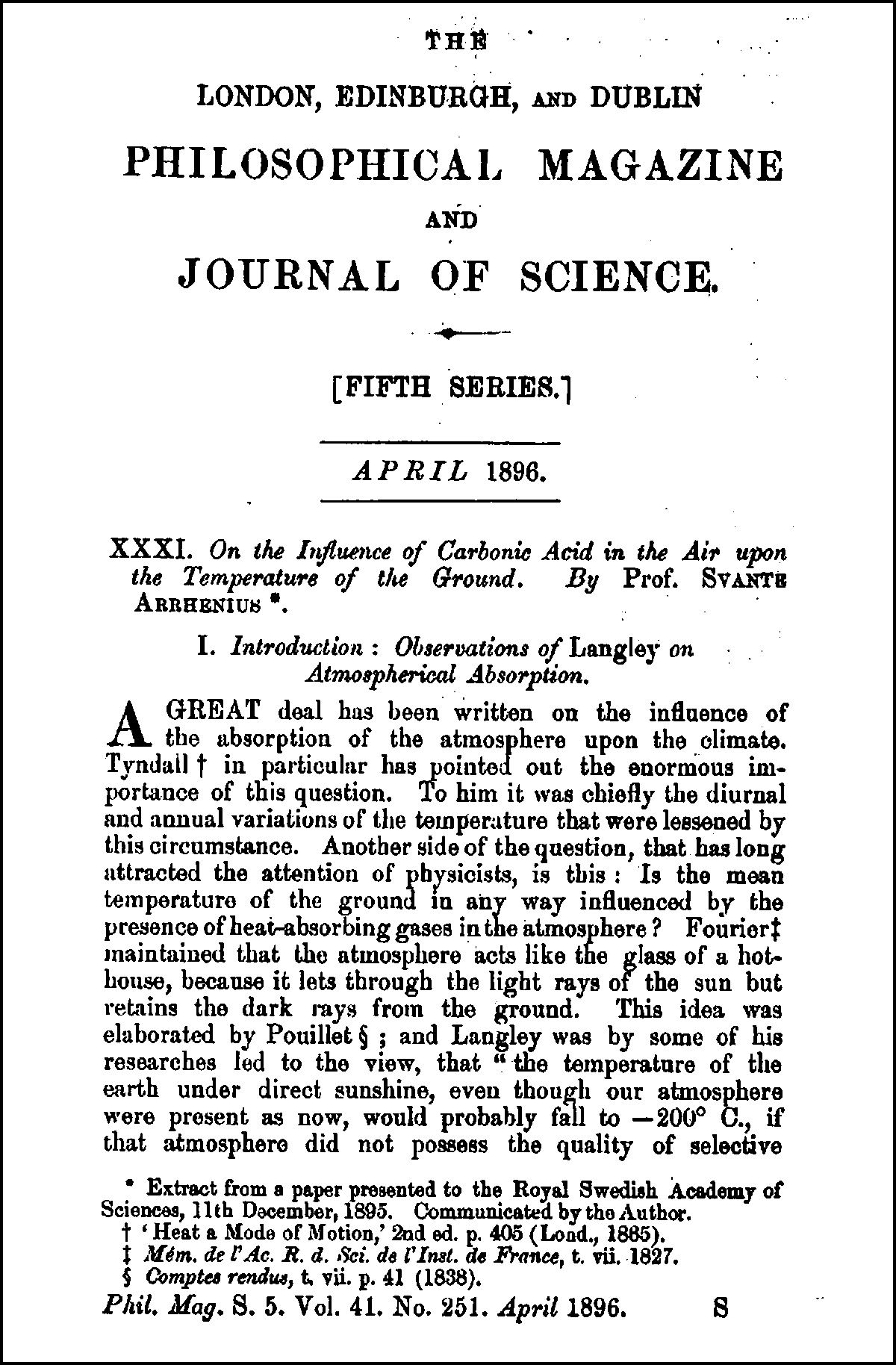}}
\vspace*{8pt}
\caption{The front page of Arrhenius' 1896 
         paper.}
\label{fig:Arrhenius1896}
\end{figure}
\ \vfill
}{}%
\ifthenelse{\equal{TeX4ht}{\TschStyle}}{%
\ \vfill
\begin{figure}[hbtp]
\centerline{\includegraphics[scale=1.00]{PictureLarge_Arrhenius1896_.png}}
\vspace*{8pt}
\caption{The front page of Arrhenius' 1896 
         paper.}
\label{fig:Arrhenius1896}
\end{figure}
\ \vfill
}{}
\ifthenelse{\equal{IJMPB}{\TschStyle}}{}{}%
\ifthenelse{\equal{arXiv}{\TschStyle}}{\newpage\noindent}{}%
\ifthenelse{\equal{TeX4ht}{\TschStyle}}{\newpage\noindent}{}%
%
%
In which fantastic way Arrhenius uses 
Stefan-Boltzmann's law to calculate 
this \lq\lq effect\rq\rq, 
can be seen better in another 
publication, in which he defends 
his ice age-hypothesis\TschSpace%
%
\cite{Arrhenius1906},
%
see 
Figures
\ref{fig:Arrhenius1906a},
\ref{fig:Arrhenius1906b},
and
\ref{fig:Arrhenius1906c}.
\begin{figure}[hbtp]
\ifthenelse{\equal{IJMPB}{\TschStyle}}{%
\centerline{\psfig{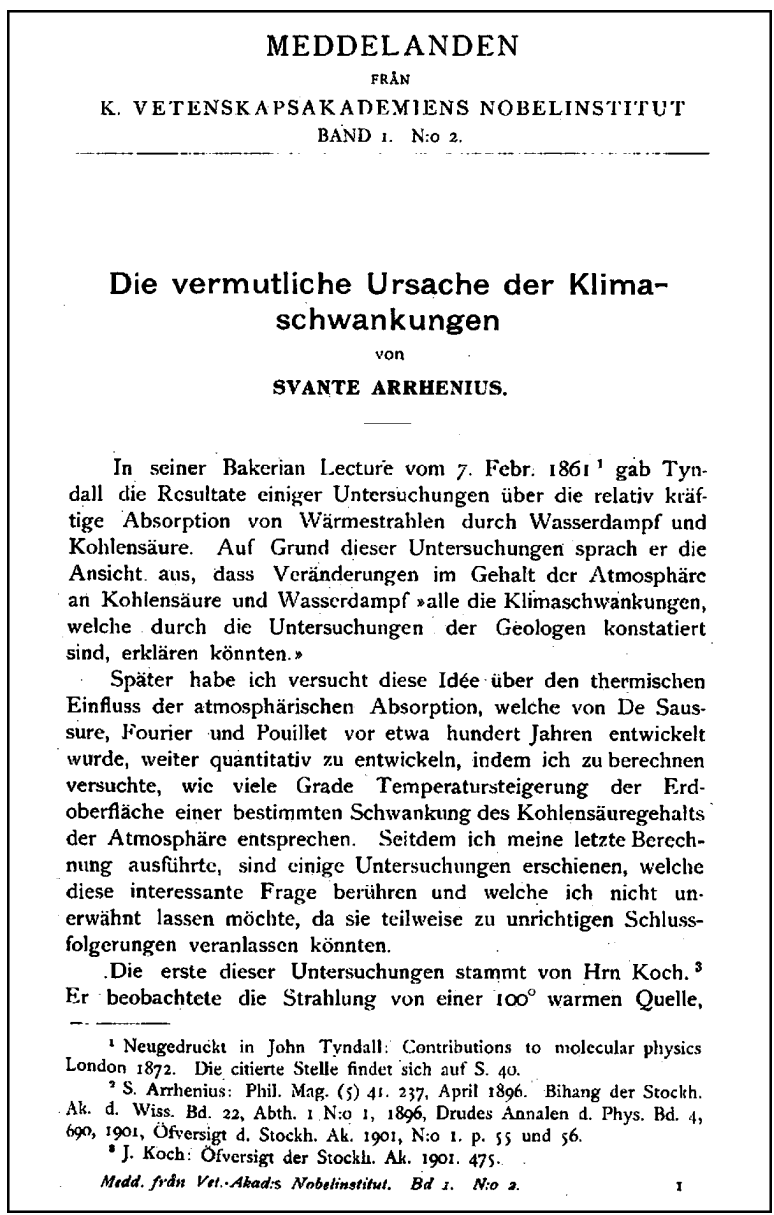}}
\vspace*{8pt}
\caption{Excerpt (a) of Arrhenius' 1906
         paper.}
\label{fig:Arrhenius1906a}
\end{figure}
}{}%
\ifthenelse{\equal{arXiv}{\TschStyle}}{%
\ \vfill
\vspace*{16pt}
\centerline{\includegraphics[scale=1.00]{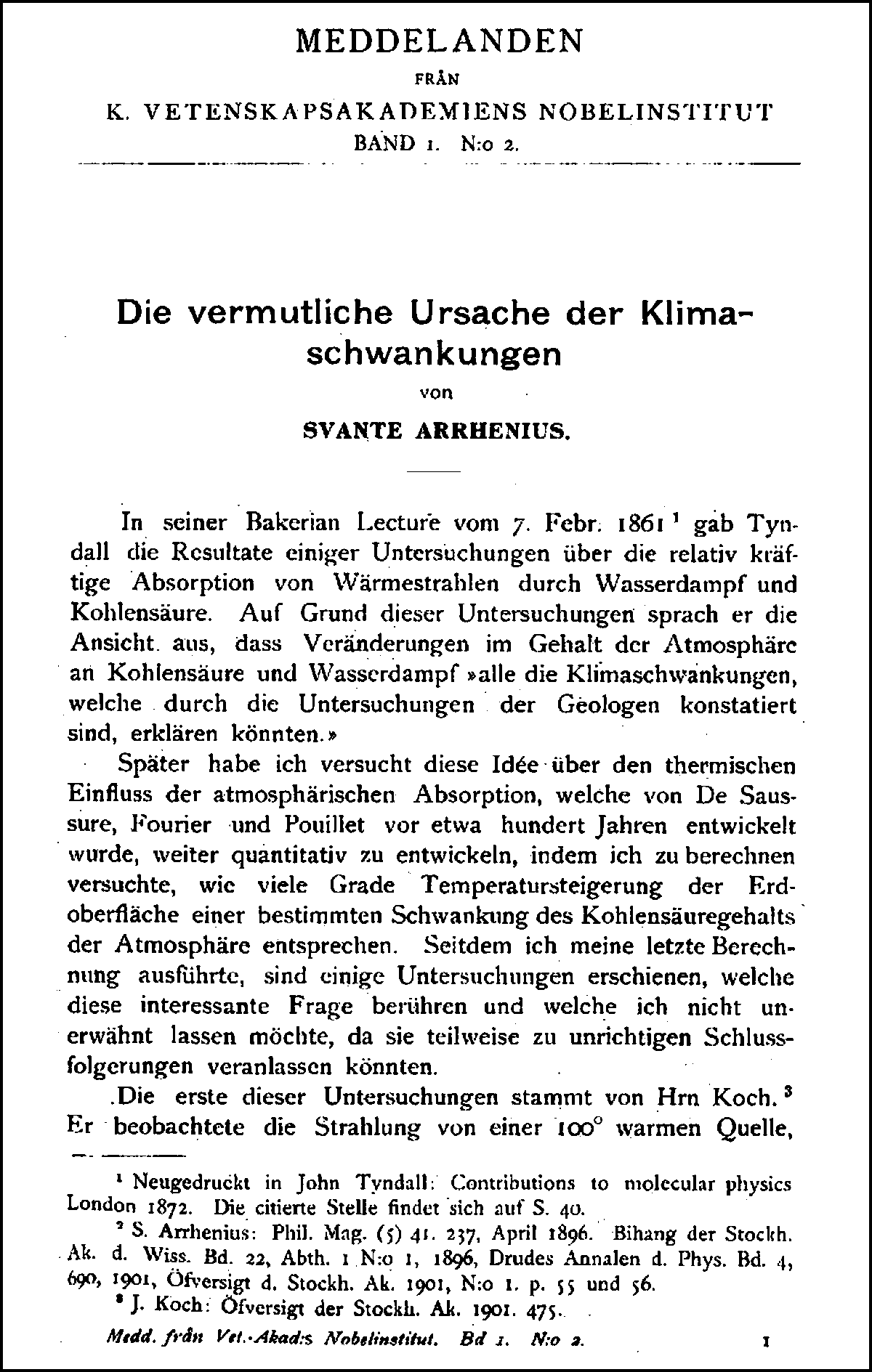}}
\vspace*{8pt}
\caption{Excerpt (a) of Arrhenius' 1906
         paper.}
\label{fig:Arrhenius1906a}
\end{figure}
\ \vfill
}{}%
\ifthenelse{\equal{TeX4ht}{\TschStyle}}{%
\ \vfill
\vspace*{16pt}
\centerline{\includegraphics[scale=1.00]{PictureLarge_Arrhenius1906a_.png}}
\vspace*{8pt}
\caption{Excerpt (a) of Arrhenius' 1906
         paper.}
\label{fig:Arrhenius1906a}
\end{figure}
\ \vfill
}{}
\ifthenelse{\equal{IJMPB}{\TschStyle}}{}{}%
\ifthenelse{\equal{arXiv}{\TschStyle}}{\newpage\noindent}{}%
\ifthenelse{\equal{TeX4ht}{\TschStyle}}{\newpage\noindent}{}%
%
%
%
First, Arrhenius estimates that $18.7\,\%$ 
of the Earth's infrared radiation would not 
be emitted into space because of its absorption 
by carbonic acid. 
This could be taken into account by reducing 
the Earth's effective radiation temperature 
$T_{\mbox{{\scriptsize\rm eff}}}$ 
to a reduced temperature
$T_{\mbox{{\scriptsize\rm reduced}}}$.
Arrhenius assumed
\begin{equation}
T_{\mbox{{\scriptsize\rm eff}}}
=
15\, ^\circ{\rm C} 
= 
288\,{\rm K}
\end{equation}
and, assuming the validity of the 
Stefan-Boltzmann law, made the ansatz
\begin{equation}
\frac{\sigma\cdot T_{\mbox{{\scriptsize\rm reduced}}}^4}
     {\sigma\cdot T_{\mbox{{\scriptsize\rm eff}}}^4}
=
\frac{(1-0.187)\cdot I_0}{I_0}
\end{equation}
yielding
\begin{equation}
T_{\mbox{{\scriptsize\rm reduced}}}
=
T_{\mbox{{\scriptsize\rm eff}}}
\cdot
\sqrt[4]{1-0.187} 
\end{equation}
and
\begin{equation}
T_{\mbox{\scriptsize\rm reduced}}
=
\sqrt[4]{0.813} \cdot 288 = 273.47
\end{equation}
which corresponds to a lowering 
of the Earth's temperature of 
$ 14.5\, ^\circ{\rm C}$. 

As one would probably not think 
that such an absurd claim is possible, 
a scan of this passage is displayed in 
Figures
\ref{fig:Arrhenius1906b}
and
\ref{fig:Arrhenius1906c}. 
\begin{figure}[hbtp]
\ifthenelse{\equal{IJMPB}{\TschStyle}}{%
\centerline{\psfig{file=PictureLarge_Arrhenius1906b_.eps,width=3.00in}}
\vspace*{8pt}
\caption{Excerpt (b) of Arrhenius' 1906
         paper.}
\label{fig:Arrhenius1906b}
\end{figure}
}{}%
\ifthenelse{\equal{arXiv}{\TschStyle}}{%
\vspace*{16pt}
\centerline{\includegraphics[scale=1.00]{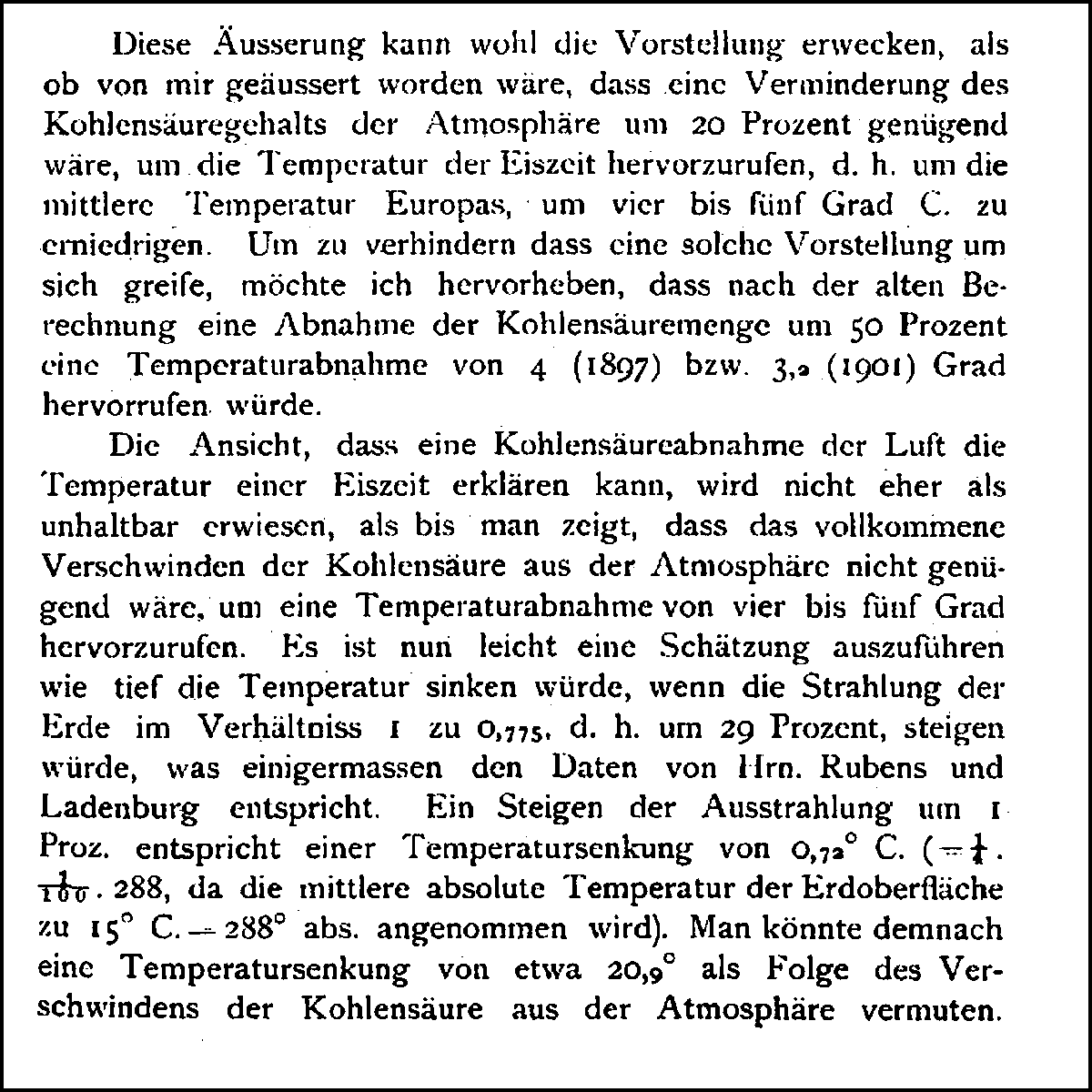}}
\vspace*{8pt}
\caption{Excerpt (b) of Arrhenius' 1906
         paper.}
\label{fig:Arrhenius1906b}
\end{figure}
}{}%
\ifthenelse{\equal{TeX4ht}{\TschStyle}}{%
\vspace*{16pt}
\centerline{\includegraphics[scale=1.00]{PictureLarge_Arrhenius1906b_.png}}
\vspace*{8pt}
\caption{Excerpt (b) of Arrhenius' 1906
         paper.}
\label{fig:Arrhenius1906b}
\end{figure}
}{}
\ifthenelse{\equal{IJMPB}{\TschStyle}}{}{}%
\ifthenelse{\equal{arXiv}{\TschStyle}}{\newpage\noindent}{}%
\ifthenelse{\equal{TeX4ht}{\TschStyle}}{\newpage\noindent}{}%
%
%
\begin{figure}[hbtp]
\ifthenelse{\equal{IJMPB}{\TschStyle}}{%
\centerline{\psfig{file=PictureLarge_Arrhenius1906c_.eps,width=3.00in}}
\vspace*{8pt}
\caption{Excerpt (c) of Arrhenius' 1906
         paper.}
\label{fig:Arrhenius1906c}
\end{figure}
}{}%
\ifthenelse{\equal{arXiv}{\TschStyle}}{%
\centerline{\includegraphics[scale=1.00]{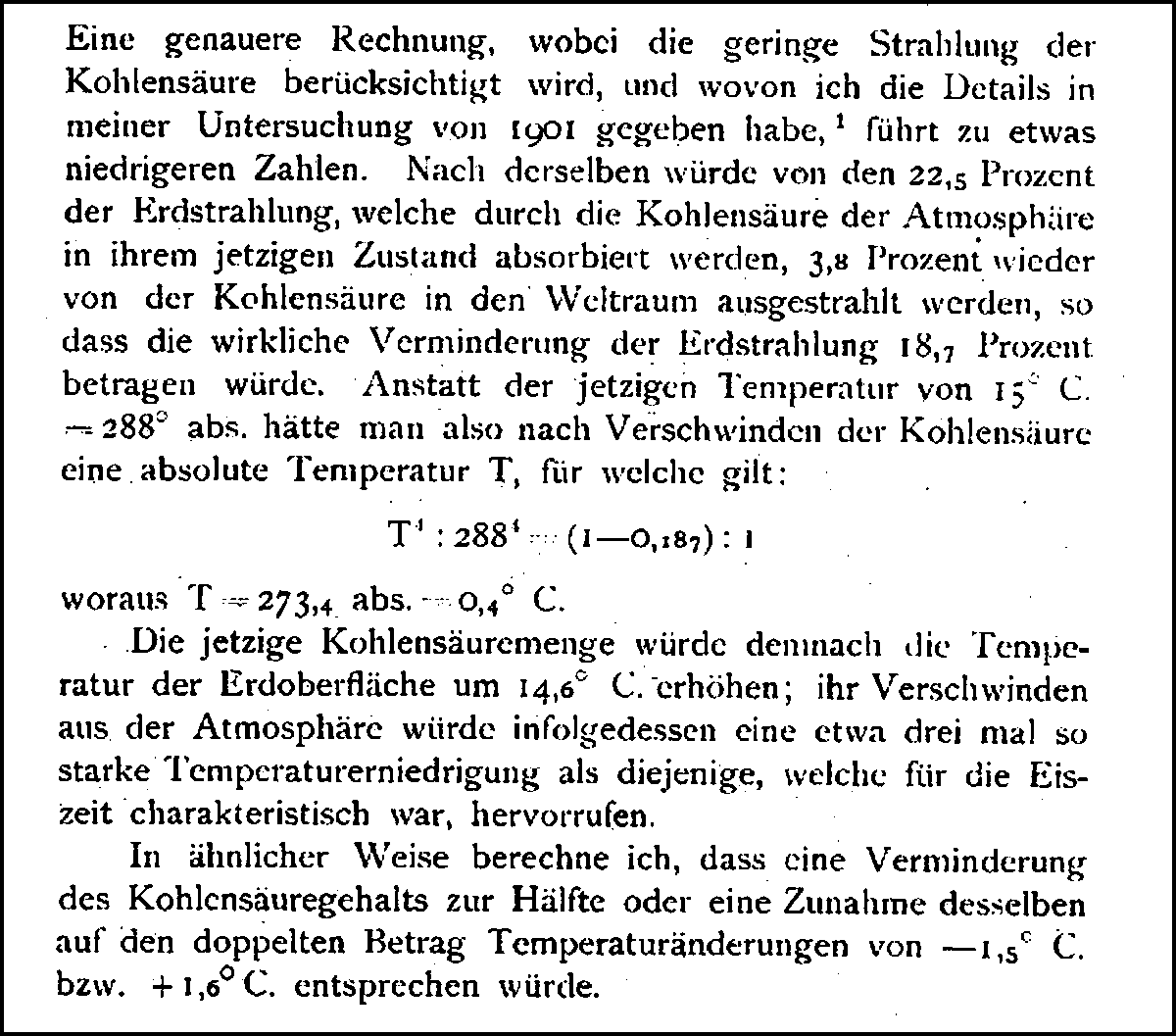}}
\vspace*{8pt}
\caption{Excerpt (c) of Arrhenius' 1906
         paper.}
\label{fig:Arrhenius1906c}
\end{figure}
}{}%
\ifthenelse{\equal{TeX4ht}{\TschStyle}}{%
\centerline{\includegraphics[scale=1.00]{PictureLarge_Arrhenius1906c_.png}}
\vspace*{8pt}
\caption{Excerpt (c) of Arrhenius' 1906
         paper.}
\label{fig:Arrhenius1906c}
\end{figure}
}{}

The English translation reads:
\begin{quote}
\lq\lq
This statement could lead to the impression, 
that I had claimed that a reduction 
of the concentration of carbonic acid 
in the atmosphere of
$20\,\%$
would be sufficient to cause ice-age temperatures, 
i.e.\ to lower the Europe's average temperature 
about four to five degrees
${\rm C}$.
To keep such an idea from spreading, 
I would like to point out that 
according to the old calculation a reduction 
of carbonic acid of
$50\,\%$
would cause the temperature to fall for 
$4$ (1897)
or, respectively, 
$3.2$ (1901)
degrees.
\textbf{
The opinion that a decrease of carbonic acid 
in the air can explain ice-age temperatures 
is not proved wrong until it is shown, 
that the total disappearance of carbonic 
acid from the atmosphere would not be sufficient 
to cause a lowering of temperatures 
about four to five degrees.}
It is now easy to estimate 
how low the temperature would fall, 
if the Earth's radiation rose in the ratio of
$1$ to $0.775$,
i.e.\ for
$29\,\%$,
which matches the data of 
Messrs.\ Rubens and Ladenburg. 
An increase of emissions of 
$1\,\%$
would be equivalent 
to a decrease of temperatures of 
$0.72\, ^\circ{\rm C}$, 
as the average absolute temperature 
of the Earth is taken to be 
$15\, ^\circ{\rm C} = 288 ^\circ{\rm C}$.
Therefore, one could estimate 
a lowering of the temperatures 
about $20,9\, ^\circ{\rm C}$ 
as a result of the disappearance 
of carbonic acid from the atmosphere. 
A more exact calculation, which takes 
into account the small amount of radiation 
of the carbonic acid and of which I have 
given details in my paper of 1901, 
leads to slightly lower numbers. 
According to this calculation,
$3.8\,\%$
out of the
$22.5\,\%$
of terrestrial radiation, 
which are being absorbed by the carbonic acid 
in the atmosphere at its current state, 
are emitted into space by the carbonic acid, 
so the real decrease of terrestrial radiation 
would be
$18.7\,\%$.
After the disappearance 
of the carbonic acid, instead of the current 
temperature of
$15\, ^\circ{\rm C} = 288\,{\rm K}$,
there would be an absolute temperature 
$T$, which is:
\begin{equation}
T^4 : 288^4 = (1-0,187) : 1
\end{equation}
being 
\begin{equation}
T = 273,4\,{\rm K} = 0,4\, ^\circ{\rm C}.
\end{equation}
The current amount of carbonic acid 
would therefore raise the temperature 
of the Earth's surface for
$14,6\, ^\circ{\rm C}$
its disappearance from the atmosphere 
would result in a lowering of temperatures 
about three times as strong as the one, 
which caused the ice ages. 
I calculate in a similar way, 
that a decrease in the concentration 
of carbonic acid by half or a doubling 
would be equivalent to changes 
of temperature of
$- 1,5\, ^\circ{\rm C}$
or
$+ 1,6\, ^\circ{\rm C}$
respectively.%
\rq\rq
\end{quote}
It is an interesting point that there is an
\textbf{inversion of the burden of proof}
in Arrhenius' paper, which is typeset in 
boldface here, because it winds its way as a red thread
through almost all contemporary papers 
on the influence of ${\rm CO}_2$ of 
the so-called global climate.

\subsubsection{Modern works of climatology}
\label{Sec:ModernWorks}
Callendar\TschSpace%
%
\cite{%
Callendar1938,%
Callendar1939,%
Callendar1940,%
Callendar1941,%
Callendar1949,%
Callendar1958,%
Callendar1961%
}
%
and Keeling\TschSpace%
%
\cite{%
Keeling1960,%
Keeling1973,%
Keeling1976,%
Keeling1978,%
Keeling1989,%
Keeling1996,%
Keeling1998%
}%
, 
the founders of the modern 
greenhouse hypothesis, 
recycled Arrhenius' 
\lq\lq 
discussion of yesterday and 
the day before yesterday\rq\rq%
\footnote{a phrase used by von\,Storch in
%
\ifthenelse{\equal{IJMPB}{\TschStyle}}
           {Ref.~\refcite{Stockholm2006}}
           {}%
\ifthenelse{\equal{arXiv}{\TschStyle}}
           {Ref.~\cite{Stockholm2006}}
           {} 
\ifthenelse{\equal{TeX4ht}{\TschStyle}}
           {Ref.~\cite{Stockholm2006}}
           {} 
%
}
by perpetuating the errors of the past 
and adding lots of new ones.

In the 70s and 80s two developments coincided:
A accelerating progress in computer technology and
an emergence of two contrary policy preferences,
one supporting the development of 
civil nuclear technology, the other 
supporting Green Political movements.
Suddenly the ${\rm CO}_2$ issue became
on-topic, and so did computer simulations
of the climate. The research results have 
been vague ever since: 
\begin{itemize}
\item 
In the 70s, computer simulations of the 
\lq\lq global climate\rq\rq\ predicted 
for a doubling of the ${\rm CO}_2$ concentration
a global temperature rise of about 
0.7\,-9.6\,K\TschSpace%
\cite{Schneider1975}.
\item 
Later, computer simulations pointed towards
a null effect%
\footnote{%
G.G.\ is indebted to the late science journalist 
Holger Heuseler for this valuable information\TschSpace%
%
\cite{Heuseler1996}.
}:
\begin{itemize}
\item
In the IPCC 1992 report,
computer simulations of the 
\lq\lq global climate\rq\rq\ predicted 
a global temperature rise of about 
0.27\,-\,0.82\,K per decade\TschSpace%
\cite{IPCC1992}.
\item 
In the IPCC 1995 report,
computer simulations of the 
\lq\lq global climate\rq\rq\ predicted 
a global temperature rise of about 
0.08\,-0.33\,K per decade\TschSpace%
\cite{IPCC1996}.
\end{itemize}
\item 
Two years ago (2005), computer simulations of the 
\lq\lq global climate\rq\rq\ predicted 
for a doubling of the ${\rm CO}_2$ concentration
a global temperature rise of about 
2\,-\,12\,K,
whereby six so-called scenarios have been
omitted that yield a \textit{global cooling}\TschSpace%
\cite{Stainforth2005}.
\end{itemize}
The state-of-the-art in climate modeling 1995 
is described in 
%
\ifthenelse{\equal{IJMPB}{\TschStyle}}
           {Ref.~\refcite{Cubasch1995}}
           {}%
\ifthenelse{\equal{arXiv}{\TschStyle}}
           {Ref.~\cite{Cubasch1995}}
           {}%
\ifthenelse{\equal{TeX4ht}{\TschStyle}}
           {Ref.~\cite{Cubasch1995}}
           {} 
%
in detail.
Today every home server is larger than 
a mainframe at that time and
every amateur can test and modify
the vintage code\TschSpace%
%
\cite{McGuffie2006}.
%
Of course, 
there exist no realistic solvable equations
for the weather parameters.
Meanwhile, 
\lq\lq computer models\rq\rq\
have been developed which run 
on almost every PC\TschSpace%
%
\cite{McGuffie2006,Stainforth2005}
%
or even in the internet
\cite{BBC2007}.
%

To derive a climate catastrophe from
these computer games and scare mankind to death
is a crime.
%

%
\subsection{The assumption of radiative balance}
\label{Sec:RadiativeBalance}
%
%
\subsubsection{Introduction}
Like the physical mechanism in glass houses 
the ${\rm CO}_2$-greenhouse effect is about 
a comparison of two different physical situations.
Unfortunately, the exact definition of the
atmospheric greenhouse effect changes 
from audience to audience, that is, 
there are many variations of the theme.
Nevertheless, one common aspect 
lies in the methodology that 
a fictitious model computation 
for a celestial body \textit{without} an atmosphere 
is compared to 
another fictitious model computation
for a celestial body \textit{with} an atmosphere. 
For instance, \lq\lq average\rq\rq\ temperatures
are calculated 
for an Earth \textit{without} an atmosphere 
and 
for an Earth \textit{with} an atmosphere.
Amusingly, there seem to exist no calculations 
for an Earth  \textit{without} oceans 
opposed to  calculations 
for an Earth \textit{with} oceans. 
However, in many studies, models for oceanic currents 
are included in the frameworks considered, 
and radiative \lq\lq transport\rq\rq\ 
calculations are incorporated too. 
Not all of these refinements can be 
discussed here in detail. The reader is referred to
%
%
\ifthenelse{\equal{IJMPB}{\TschStyle}}
           {Ref.~\refcite{McGuffie2006}}
           {}%
\ifthenelse{\equal{arXiv}{\TschStyle}}
           {Ref.~\cite{McGuffie2006}}
           {} 
\ifthenelse{\equal{TeX4ht}{\TschStyle}}
           {Ref.~\cite{McGuffie2006}}
           {} 
%
%
and further references therein.
Though there exists a huge family of 
generalizations, one common aspect is the assumption
of a radiative balance, which plays a central role
in the publications of the IPCC 
and, hence, in the public propaganda.
In the following it is proved that this assumption
is physically wrong.
%
%
\subsubsection{A note on \lq\lq radiation balance\rq\rq\ diagrams}
\begin{figure}[htbp]
\ifthenelse{\equal{IJMPB}{\TschStyle}}{%
\centerline{\psfig{file=PictureLarge_EarthsEnergyBalance_.eps,width=4.00in}}
}{}%
\ifthenelse{\equal{arXiv}{\TschStyle}}{%
\centerline{\includegraphics[scale=0.80]{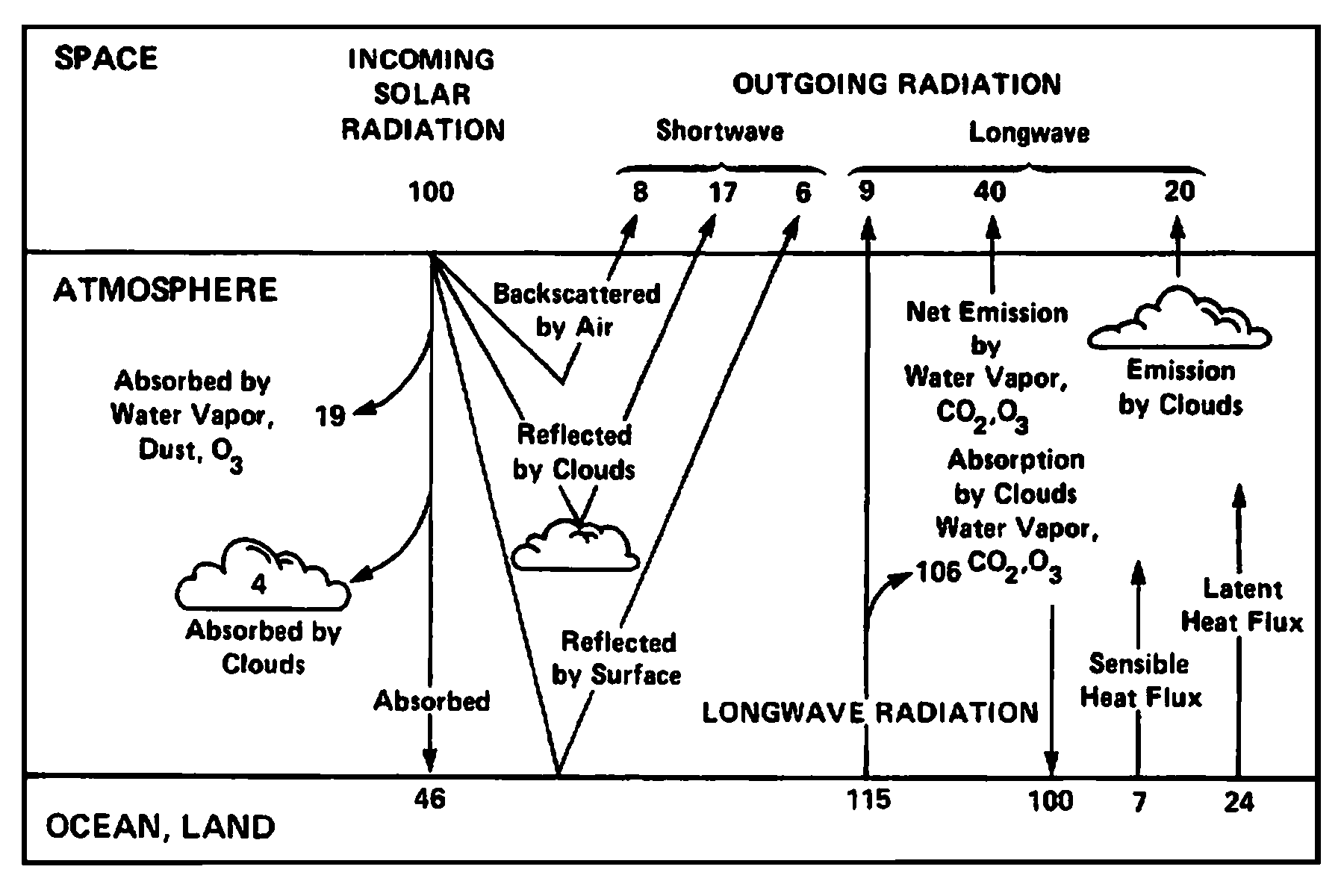}}
\vspace*{-8pt}
}{}%
\ifthenelse{\equal{TeX4ht}{\TschStyle}}{%
\centerline{\includegraphics[scale=0.80]{PictureLarge_EarthsEnergyBalance_.png}}
\vspace*{-8pt}
}{}
\vspace*{8pt}
\caption{A schematic diagram supposed to describe 
         the global average components of the Earth's 
         energy balance.       
         Diagrams of this kind \textbf{contradict to physics.}}
\label{Fig:Diagrams}
\end{figure}

From the definition given in Section~\ref{Sec:ISI}  
it is immediately evident that 
a radiation intensity 
$I_\nu$
is \textit{not} a current density 
that can be described by a vector field 
$\textbf{j}(\textbf{x},t)$.
That means that conservation laws
(continuity equations, 
balance equations, 
budget equations)
cannot be written down for intensities.
Unfortunately this is done in most
climatologic papers, 
\textbf{the cardinal error of global climatology},
that may have been overlooked so long due to
the oversimplification of the 
real world problem towards a quasi 
one-dimensional problem.
Hence the popular climatologic 
\lq\lq radiation balance\rq\rq\ diagrams 
describing quasi-one-dimensional situations
(cf.\ Figure~\ref{Fig:Diagrams}) 
are scientific misconduct since they do not properly represent 
the mathematical and physical fundamentals.

Diagrams of the type of Figure~\ref{Fig:Diagrams}
are the cornerstones of \lq\lq climatologic proofs\rq\rq\ 
of the supposed greenhouse effect in the atmosphere\TschSpace%
%
\cite{MacCracken1985}.
They are highly suggestive, because they bear some 
similarity to Kirchhoff rules of electrotechnics,
in particular to the node rule describing the conservation
of charge\TschSpace%
%
\cite{Paul2001}.
%
Unfortunately, in the literature on global 
climatology it is not explained, what the
arrows  in \lq\lq radiation balance\rq\rq\ 
diagrams mean physically. It is easily 
verified that within the frame of physics 
they cannot mean anything.

Climatologic radiation balance diagrams are nonsense,
since they  
\begin{enumerate}
\item cannot represent radiation intensities,
      the most natural interpretation of the arrows
      depicted in Figure~\ref{Fig:Diagrams}, 
      as already explained in 
      Section~\ref{Sec:ISI} 
      and 
      Section~\ref{Sec:Conclusion} ;
\item cannot represent sourceless fluxes,  
      i.e.\ a divergence free vector fields
      in three dimensions, since a vanishing 
      three-dimensional divergence still allows
      that a portion of the field goes sidewards;  
\item do not fit in the framework of Feynman diagrams,
      which represent mathematical expressions clearly
      defined in quantum field theory\TschSpace%
%
\cite{Itzykson1980}.
%
\item do not fit in the standard language
      of system theory or system engineering\TschSpace%
%
\cite{SysML2007}.
%
\end{enumerate}
Kirchhoff-type node rules only hold in cases, 
where there is a conserved quantity and the 
underlying space may be described by a topological 
space that is a one-dimensional manifold almost 
everywhere, the singularities being the 
network nodes,
i.e. in conventional electric circuitry
%
\cite{Paul2001},
%
in mesoscopic networks\TschSpace%
%
\cite{Balachandran1992},
%
and, 
for electromagnetic waves, 
in waveguide networks%
\footnote{%
The second and the third type are beautifully related 
by the correspondence of the v.\,Klitzing resistance 
${\rm R}_{\rm vK} \approx 25,813\ {\rm k}\Omega$
with the characteristic impedance
${\rm Z}_0 \approx 376,73\ \Omega$
via the Sommerfeld fine structure constant
$\alpha={\rm Z}_0/2{\rm R}_{\rm vK} \approx 1/137,036\,$\TschSpace%
%
\cite{Tscheuschner1998}.
%
}%
\TschSpace%
%
\cite{Montgomery1948,Marcuvitz1986}.
%
However, although Kirchhoff's mesh analysis 
may be successfully applied to microwave networks, 
the details are highly involved 
and will break down if dissipation is allowed\TschSpace%
%
\cite{Montgomery1948,Marcuvitz1986}.
%

Clearly, neither the cryptoclimate of a glass house 
nor the atmosphere of the Earth's does compare 
to a waveguide network e.g.\ feeding the
acceleration cavities of a particle accelerator.
Therefore, the climatologic radiation balance  
diagrams are inappropriate and misleading,
even when they are supposed to describe 
averaged quantities. 
%
%
\subsubsection{The case of purely radiative balance}
If only thermal radiation was possible for 
the heat transfer of a radiation-exposed body 
one would use Stefan-Boltzmann's law
\begin{equation}
{\rm S}(T) = \sigma T^4
\label{Eq:epsilonsigmaTto4}
\end{equation}
to calculate the ground temperature determined by this balance. The irradiance ${\rm S}$ has dimensions of 
a power density and $\sigma$ is the Stefan-Boltzmann 
constant given by 
\begin{equation}
\sigma 
=
\frac{2 \pi^5 k^4}{15 c^2 h^3}
=
5.67040ß0 \cdot 10^{-8} 
\frac{{\rm W}}{{\rm m}^2{\rm K}^4}
\end{equation}
For example, the energy flux density of 
a black body at room temperature $300\,{\rm K}$
is approximately
\begin{equation}
{\rm S}(\,T\!=\!300\,{\rm K}\,) = 459\ {\rm W}/{\rm m}^2
\end{equation}

One word of caution is needed here:
As already emphasized in Section~\ref{Sec:Conclusion}
the constant $\sigma$ appearing in the $T^4$ law
is \textit{not} a universal constant of physics.
Furthermore, a gray radiator 
must be described by a temperature dependent 
$\sigma(T)$ spoiling the $T^4$ law. 
Rigorously speaking, for real objects Equation (\ref{Eq:epsilonsigmaTto4}) 
is invalid.
Therefore all crude approximations 
relying on $T^4$ expressions need 
to be taken with great care.
In fact, though popular in global
climatology, they prove nothing!  

In the balance equation
\begin{equation}
\sigma
\cdot
T_{\mbox{{\scriptsize\rm Earth's ground}}}^4 
=
\sigma
\cdot
T_{\mbox{{\scriptsize\rm Sun}}}^4
\cdot 
\frac{{\rm R}^2_{\mbox{\scriptsize\rm Sun}}}
     {{\rm R}^2_{\mbox{\scriptsize\rm Earth's orbit}}}
\end{equation}
one may insert 
a general phenomenological normalization factor
$\epsilon$
at the right side, leaving room
for a fine tuning and inclusion of geometric
factors.%
\footnote{%
The factor 
$\varepsilon$
is related to 
the albedo 
$A$
of the Earth
describing 
her reflectivity:
$A = 1 - \varepsilon$.
In the earlier literature 
one often finds 
$A=0.5$ 
for the Earth,
in current publications 
$A=0.3$.
The latter value 
is used here.%
}
Thus one may write
\begin{equation}
\sigma \cdot T^4_{\mbox{{\scriptsize\rm Earth's ground}}}
=
\epsilon \cdot \sigma \cdot 5780^4 \cdot \frac{1}{46225}
=
\epsilon \cdot 1368\ {\rm W}/{\rm m}^2
=
\epsilon \cdot {\rm s}
\end{equation}
which yields
\begin{equation}
T_{\mbox{{\scriptsize\rm Earth's ground}}}
=
\sqrt[4]{\epsilon} \cdot \frac{\,\,5780}{\,\,\sqrt{215}}\ {\rm K} 
=
\sqrt[4]{\epsilon} \cdot 394.2\ {\rm K}
\label{Eq:TE75}
\end{equation}
${\rm s}$ is the solar constant.
With the aid of 
Equation~(\ref{Eq:TE75}) 
one calculates the values 
displayed in 
Table~\ref{temps}.

\ifthenelse{\equal{IJMPB}{\TschStyle}}{%
\begin{table}[htbp] 
\tbl{Effective temperatures
     $T_{\mbox{{\scriptsize\rm ground}}}$
     in dependence of the parameter $\epsilon$.}
{     
\begin{tabular}{@{}crr@{}} 
\Hline 
\\[-1.8ex] 
$\epsilon$  
       &  $T_{\mbox{{\scriptsize\rm ground}}}\,[{\rm K}]$ 
       &  $T_{\mbox{{\scriptsize\rm ground}}}\,[^\circ{\rm C}]$ 
\\[0.8ex]
\hline
\\[-1.8ex] 
$1.00$            & $394.2$ & $121.2$ \\
$0.70$            & $360.6$ & $ 87.6$ \\
$0.62$            & $349.8$ & $ 76.8$ 
\\[0.8ex]
\hline
\\[-1.8ex] 
\end{tabular}
}
\label{temps}
\end{table}
}{}
\ifthenelse{\equal{arXivOrTeX4ht}{\TschStyles}}{%
\begin{table}[htbp] 
{
\begin{center}
\vspace*{0.5cm}     
\begin{tabular}{|c|r|r|} 
\hline 
$\epsilon$  &  $T_{\mbox{{\scriptsize\rm Earth's ground}}}\,[{\rm K}]$ 
            &  $T_{\mbox{{\scriptsize\rm Earth's ground}}}\,[^\circ{\rm C}]$ 
\\
\hline
$1.00$            & $394.2$ & $121.2$ \\
$0.70$            & $360.6$ & $ 87.6$ \\
$0.62$            & $349.8$ & $ 76.8$ 
\\
\hline
\end{tabular}
\end{center}
}
\caption{Effective temperatures
         $T_{\mbox{{\scriptsize\rm Earth's ground}}}$
         in dependence of the phenomenological
         normalization parameter $\epsilon$.}
\label{temps}
\vspace*{0.5cm}
\end{table}
}{} 

Only the temperature measured in the Sun inside 
the car bears some similarity 
with the three ones calculated here. 
Therefore, the radiation balance does 
not determine the temperature outside the car!
In contrast to this, Table~\ref{avertemps} 
displays the \lq\lq average effective\rq\rq\ temperatures
of the ground, which according to climatological consensus
are used to \lq\lq explain\rq\rq\ the atmospheric
greenhouse effect.
The factor of a quarter is introduced 
by \lq\lq distributing\rq\rq\ 
the incoming solar radiation 
seeing a cross section $\sigma_{\mbox{{\scriptsize\rm Earth}}}$
over the global surface $\Omega_{\mbox{{\scriptsize\rm Earth}}}$
\begin{equation}
\frac{ \sigma_{\mbox{{\scriptsize\rm Earth}}} }
     { \Omega_{\mbox{{\scriptsize\rm Earth}}} }
=
\frac{ \,\,\pi \cdot {\rm R}^2_{\mbox{\scriptsize\rm Earth}} }  
     {    4\pi \cdot {\rm R}^2_{\mbox{\scriptsize\rm Earth}} }
=
\frac{1}{4}    
\label{Eq:quarter}     
\end{equation}
\ifthenelse{\equal{IJMPB}{\TschStyle}}{%
\begin{table}[htbp] 
\tbl{Effective \lq\lq average\rq\rq\ temperatures
     $T_{\mbox{{\scriptsize\rm ground}}}$
     in dependence of the parameter 
     $\epsilon$.}
{     
\begin{tabular}{@{}crr@{}} 
\Hline 
\\[-1.8ex] 
$\epsilon$ 
       &  $T_{\mbox{{\scriptsize\rm ground}}}\,[{\rm K}]$ 
       &  $T_{\mbox{{\scriptsize\rm ground}}}\,[^\circ{\rm C}]$ 
\\[0.8ex] 
\hline 
\\[-1.8ex] 
$0.25 \cdot 1.00$ & $278.7$ & $  5.7$ \\
$0.25 \cdot 0.70$ & $255.0$ & $-18.0$ \\ 
$0.25 \cdot 0.62$ & $247.4$ & $-25.6$ 
\\[0.8ex] 
\hline
\\[-1.8ex] 
\end{tabular}
}
\label{avertemps}
\end{table}
}{}
\ifthenelse{\equal{arXivOrTeX4ht}{\TschStyles}}{%
\begin{table}[htbp] 
{
\begin{center}
\vspace*{0.5cm}     
\begin{tabular}{|c|r|r|} 
\hline 
$\epsilon$  &  $T_{\mbox{{\scriptsize\rm Earth's ground}}}\,[{\rm K}]$ 
            &  $T_{\mbox{{\scriptsize\rm Earth's ground}}}\,[^\circ{\rm C}]$ 
\\ 
\hline 
$0.25 \cdot 1.00$ & $278.7$ & $  5.7$ \\
$0.25 \cdot 0.70$ & $255.0$ & $-18.0$ \\ 
$0.25 \cdot 0.62$ & $247.4$ & $-25.6$ \\
\hline
\end{tabular}
\end{center}
}
\caption{Effective \lq\lq average\rq\rq\ temperatures
         $T_{\mbox{{\scriptsize\rm ground}}}$
         in dependence of the phenomenological 
         normalization parameter $\epsilon$
         incorporating a geometric factor of $0.25$.}
\label{avertemps}
\vspace*{0.5cm}
\end{table}
}{}

The fictitious natural greenhouse effect 
is the difference between the 
\lq\lq average effective\rq\rq\ temperature 
of 
$-18\,^\circ{\rm C}$
and the
Earth's 
\lq\lq observed\rq\rq\
average temperature
of
$+15\,^\circ{\rm C}$.

In summary, 
the factor 0.7 will enter the equations 
if one assumes that 
a grey body absorber 
is a black body radiator,
contrary to the laws of physics.
Other choices are possible,
the result is arbitrary.
Evidently, such an average value 
has no physical meaning at all.
This will be elucidated in the following 
subsection.
%
%
\subsubsection{The average temperature of a radiation-exposed globe}
\label{Sec:AverageTemperature}
\begin{figure}[htbp]
\ifthenelse{\equal{IJMPB}{\TschStyle}}{%
\centerline{\psfig{file=PictureLarge_Earthball_.eps,width=4.00in}}
}{}%
\ifthenelse{\equal{arXiv}{\TschStyle}}{%
\centerline{\includegraphics[scale=1.00]{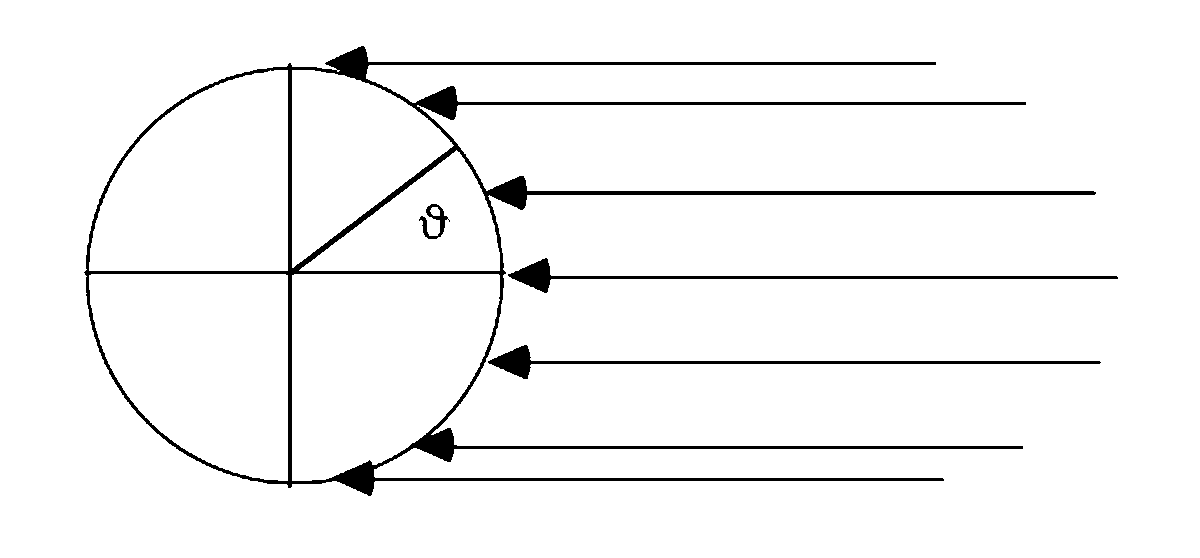}}
}{}%
\ifthenelse{\equal{TeX4ht}{\TschStyle}}{%
\centerline{\includegraphics[scale=1.00]{PictureLarge_Earthball_.png}}
}{}
\vspace*{8pt}
\caption{A radiation exposed static globe.}
\label{globe}
\end{figure}
For a radiation exposed static globe 
(cf.\ Figure~\ref{globe})
the corresponding balance equation 
must contain a geometric factor and 
reads therefore
\begin{equation}
\sigma \cdot T^4
=
\left\{\begin{array}{cl}
\epsilon \cdot {\rm S} \cdot \cos\, \vartheta 
=
\epsilon \cdot \sigma \cdot 5780^4 / 215^2 \cdot \cos \vartheta
&
\mbox{{\rm \ \ \ if\ \ }}
\phantom{\pi/}0\le\vartheta\leq\pi/2\phantom{}
\\
0
& 
\mbox{{\rm \ \ \ if\ \ }}
\phantom{}\pi/2\le\vartheta\leq\pi\phantom{/2}
\end{array}\right.
\end{equation}
It is obvious that one gets 
the effective temperatures
if the right side is divided by $\sigma$.

This in turn will determine the formerly mentioned 
\lq\lq average\rq\rq\ effective temperatures
over the global surface.
\begin{eqnarray}
T^4_{\mbox{\scriptsize\rm eff}} 
&=&
\frac{1}{4\pi} \int\!\!\!\int_{\mbox{\scriptsize\rm surface}} T^4
\,d\Omega
\nonumber \\
&=&
\frac{1}{4\pi} \int_{0}^{2\pi} \int_{0}^{\pi} T^4 
\sin\vartheta\,d\vartheta\,d\varphi
\nonumber \\
&=&
\frac{1}{4\pi} \int_{0}^{2\pi} \int_{1}^{-1} T^4 
d(-\cos\vartheta)\,d\varphi
\nonumber \\
&=&
\frac{1}{4\pi} \int_{0}^{2\pi} \int_{-1}^{1} T^4 
d(\cos\vartheta)\,d\varphi
\end{eqnarray}
Defining 
\begin{equation}
\mu = \cos\vartheta
\end{equation}
one gets 
\begin{eqnarray}
T^4_{\mbox{\scriptsize\rm eff}} 
&=&
\frac{1}{4\pi} \int_{0}^{2\pi} \int_{-1}^{1} T^4 \, d\mu \, d\varphi
\nonumber \\
&=&
\frac{1}{4\pi} \int_{0}^{2\pi} \int_{0}^{1} \epsilon \cdot 
\frac{{\rm S}}{\sigma} \cdot \mu \, d\mu \, d\varphi
\nonumber \\
&=&
\frac{1}{2} \cdot \epsilon \cdot \frac{{\rm S}}{\sigma} \cdot \int_{0}^{1} \, \mu \, d\mu
\nonumber \\
&=&
\frac{1}{2} \cdot \epsilon \cdot \frac{{\rm S}}{\sigma} \cdot 
\left( \left. \frac{\mu^2}{2} \right|_0^1 \right)
\nonumber \\
&=&
\frac{1}{4} \cdot \epsilon \cdot \frac{S}{\sigma}
\nonumber \\
&=&
\frac{1}{4} \cdot \epsilon \cdot (394.2)^4\ {\rm K}^4
\end{eqnarray}  
This is the correct derivation of the factor quarter
appearing in Equation (\ref{Eq:quarter}). 
Drawing the fourth root 
out of the resulting expression   
\begin{eqnarray}
T_{\mbox{\scriptsize\rm eff}} 
&=&
\sqrt[4]
{ 
\frac{\epsilon}{4} \cdot \frac{{\rm S}}{\sigma} 
}
\nonumber \\
&=&
\sqrt[4]
{ 
\frac{\epsilon}{4}
} 
\cdot 394.2\ {\rm K} 
\nonumber \\
&=&
(1 / \sqrt{2})
\cdot
\sqrt[4]{\epsilon}
\cdot 394.2\ {\rm K}
\nonumber \\
&=&
0.707
\cdot
\sqrt[4]{\epsilon}
\cdot 
394.2\ {\rm K}
\end{eqnarray}
Such a calculation, 
though standard in global climatology,
is plainly wrong. 
Namely, if one wants to calculate 
the average temperature, 
one has to draw the fourth root first and 
then determine the average, though:
\begin{eqnarray}
T_{\mbox{\scriptsize\rm phys}} 
&=&
\frac{1}{4\pi} \int_{0}^{2\pi} \int_{-1}^{1} T \, d\mu \, d\varphi
\nonumber \\
&=&
\frac{1}{4\pi} \int_{0}^{2\pi} \int_{0}^{1} 
\sqrt[4]
{
\epsilon 
\cdot 
\frac{{\rm S}}{\sigma} \cdot \mu
} 
\,\,\,d\mu \, d\varphi
\nonumber \\
&=&
\frac{1}{2}
\cdot 
\sqrt[4]
{
\epsilon 
\cdot 
\frac{{\rm S}}{\sigma}
} 
\cdot
\int_{0}^{1} 
\sqrt[4]{\mu}
\,\,\,
d\mu
\nonumber \\
&=&
\frac{1}{2} 
\cdot 
\sqrt[4]
{
\epsilon 
\cdot 
\frac{{\rm S}}{\sigma}
} 
\cdot
\left( \left. \frac{\mu^{5/4}}{5/4} \right|_0^1 \right)
\nonumber \\
&=&
\frac{1}{2}
\cdot 
\sqrt[4]
{
\epsilon 
\cdot 
\frac{{\rm S}}{\sigma}
} 
\cdot
\frac{4}{5}
\nonumber \\
&=&
\frac{2}{5}
\cdot 
\sqrt[4]
{
\epsilon 
\cdot 
\frac{{\rm S}}{\sigma}
} 
\end{eqnarray}
\noindent%
finally yielding 
\begin{eqnarray}
T_{\mbox{\scriptsize\rm phys}} 
&=&
\frac{2}{5}
\cdot
\sqrt[4]{\epsilon}
\cdot
394.2\ {\rm K}
\nonumber \\
&=&
0.4
\cdot
\sqrt[4] {\epsilon}
\cdot
394.2\ {\rm K}
\label{Eq:Tphys}
\end{eqnarray}
Now the averaged temperatures 
$T_{\mbox{\scriptsize\rm phys}}$ 
are considerably lower 
than the absolute temperature's fourth 
root of the averaged fourth power
(cf.\ Table~\ref{phystemps}).
\ifthenelse{\equal{IJMPB}{\TschStyle}}{%
\begin{table}[htbp] 
\tbl{Two kinds of \lq\lq average\rq\rq\ temperatures
     $T_{\mbox{{\scriptsize\rm eff}}}$
     and 
     $T_{\mbox{{\scriptsize\rm phys}}}$ 
     in dependence of the parameter 
     $\epsilon$
     compared.}
{ 
\begin{tabular}{@{}crr@{}} 
\Hline 
\\[-1.8ex]
\phantom{xxx} 
$\epsilon$ 
\phantom{xxx} 
&  
\phantom{xxx}
$T_{\mbox{{\scriptsize\rm eff}}}\,[^\circ{\rm C}]$
\phantom{xxx} 
&  
\phantom{xxx}
$T_{\mbox{{\scriptsize\rm phys}}}\,[^\circ{\rm C}]$
\phantom{xxx} 
\\[0.8ex] 
\hline 
\\[-1.8ex] 
\phantom{xxx} $ 1.00$ \phantom{xxx} 
& 
\phantom{xxx} $  5.7$ \phantom{xxx} 
& 
\phantom{xxx} $ -115$ \phantom{xxx} 
\\
\phantom{xxx} $ 0.70$ \phantom{xxx} 
& 
\phantom{xxx} $-18.0$ \phantom{xxx} 
& 
\phantom{xxx} $ -129$ \phantom{xxx} 
\\
\phantom{xxx} $ 0.62$ \phantom{xxx} 
& 
\phantom{xxx} $-25.6$ \phantom{xxx}  
& 
\phantom{xxx} $ -133$ \phantom{xxx}  
\\[0.8ex] 
\hline
\\[-1.8ex] 
\end{tabular}
}
\label{phystemps}
\end{table}
}{}
\ifthenelse{\equal{arXivOrTeX4ht}{\TschStyles}}{%
\begin{table}[htbp] 
{
\begin{center}
\vspace*{0.5cm} 
\begin{tabular}{|c|r|r|} 
\hline 
\phantom{xxx} 
$\epsilon$ 
\phantom{xxx} 
&  
\phantom{xxx}
$T_{\mbox{{\scriptsize\rm eff}}}\,[^\circ{\rm C}]$
\phantom{xxx} 
&  
\phantom{xxx}
$T_{\mbox{{\scriptsize\rm phys}}}\,[^\circ{\rm C}]$
\phantom{xxx} 
\\ 
\hline
\phantom{xxx} $ 1.00$ \phantom{xxx} 
& 
\phantom{xxx} $  5.7$ \phantom{xxx} 
& 
\phantom{xxx} $ -115$ \phantom{xxx} 
\\
\phantom{xxx} $ 0.70$ \phantom{xxx} 
& 
\phantom{xxx} $-18.0$ \phantom{xxx} 
& 
\phantom{xxx} $ -129$ \phantom{xxx} 
\\
\phantom{xxx} $ 0.62$ \phantom{xxx} 
& 
\phantom{xxx} $-25.6$ \phantom{xxx}  
& 
\phantom{xxx} $ -133$ \phantom{xxx}  
\\ 
\hline
\end{tabular}
\end{center}
}
\caption{Two kinds of \lq\lq average\rq\rq\ temperatures
         $T_{\mbox{{\scriptsize\rm eff}}}$
         and 
         $T_{\mbox{{\scriptsize\rm phys}}}$ 
         in dependence of the emissivity parameter 
         $\epsilon$
         compared.}
\label{phystemps}
\vspace*{0.5cm}
\end{table}
}{}

This is no accident but a general inequality
\begin{equation}
\langle T \rangle
=
\int_X T\,dW 
\leq
\sqrt[4]{\int_X T^4\,dW}
=
\sqrt[4]{\langle T^4\rangle}
\end{equation} 
for a non-negative measurable function $T$
and an probability measure $W$. 
It is a consequence of
H\"older's inequality\TschSpace%
%
\cite{Hoelder1889,Hardy1934,Beckenbach1961,Kuptsov2001}
%
\begin{equation}
\int_X fg\,dW
\le
\left\{
\int_X f^p\,dW
\right\}
^{1/p}
\cdot
\left\{
\int_X g^q\,dW
\right\}
^{1/q}
\end{equation}
for a probability measure
$W$
and  
for two non-negative measurable functions
$f$, $g$ and non-negative integers $p$, $q$
obeying
\begin{equation}
\frac{1}{p}
+
\frac{1}{q}
=
1
\end{equation}
In the case discussed here
one has
\begin{equation}
p=4,
\phantom{x} 
q=4/3,
\phantom{x}
g(x) \equiv 1
\end{equation}
and 
\begin{equation}
f = T
\end{equation}
%
%
\subsubsection{Non-existence of the natural greenhouse effect}
According to the consensus among global climatologists
one takes the $-18^\circ{\rm C}$ computed from
the $T^4$ average and compares it to the
fictitious Earth's average temperature of $+15\,^\circ{\rm C}$.
The difference of  $33\,^\circ{\rm C}$ is attributed to the
\textit{natural greenhouse effect}. 
As seen in Equation (\ref{Eq:Tphys}) a correct averaging yields a temperature
of $-129\,^\circ{\rm C}$. Evidently, something must be 
fundamentally wrong here.

In global climatology temperatures are computed from given
radiation intensities, and this exchanges cause and effect. 
The current \textit{local} temperatures determine the radiation
intensities and not  \textit{vice versa}. If the soil is warmed up by
the solar radiation many different local processes are triggered,
which depend on the local movement of the air, rain,
evaporation, moistness, and on the local ground conditions
as water, ice, rock, sand, forests, meadows, etc.\
One square meter of a meadow does not know anything of
the rest of the Earth's surface, which determine the
global mean value. 
Thus, the radiation is \textit{locally} 
determined by the \textit{local}
temperature.
Neither is there a global radiation balance, 
nor a global radiation budget, even in the mean-field limit.

While it is incorrect to determine a temperature from 
a given radiation intensity, one is allowed to compute 
an effective radiation temperature
$T_{\mbox{\scriptsize\rm eff rad}}$ 
from $T^4$ averages
representing a mean radiation emitted from the Earth
and to compare it with an assumed Earth's average 
temperature
$T_{\mbox{\scriptsize\rm mean}}$
H\"older's inequality says that the 
former is always larger than the latter
\begin{equation}
T_{\mbox{\scriptsize\rm eff rad}} 
> 
T_{\mbox{\scriptsize\rm mean}}
\end{equation}
provided sample selection 
and averaging (probability space)
remain the same. 

For example,
if $n$ weather stations distributed around 
the globe measure $n$ temperature values
$T_1$, \dots $T_n$, an \textit{empirical} 
mean temperature will be defined as
\begin{equation}
T_{\mbox{\scriptsize\rm mean}}
=
\frac{1}{n}
\sum_{i=1}^{n}
T_i
\end{equation}
For the corresponding black body radiation intensity 
one can approximately set
\begin{equation}
S_{\mbox{\scriptsize\rm mean}}
=
\frac{1}{n}
\sum_{i=1}^{n}
\sigma \, T_i^4
=:
\sigma \, 
T_{\mbox{\scriptsize\rm eff rad}}^4
\end{equation}
defining an \textit{effective} radiation temperature
\begin{equation}
T_{\mbox{\scriptsize\rm eff rad}}
=
\sqrt{
\frac{1}{\sigma} 
S_{\mbox{\scriptsize\rm mean}}
} 
\end{equation}
One gets immediately
\begin{equation}
T_{\mbox{\scriptsize\rm eff rad}} 
=
\sqrt[4]{
\frac{1}{n}
\sum_{i=1}^{n}
T_i^4
}
\label{Eq:SqrtSumT4}
\end{equation} 
H\"older's inequality shows that one always has 
\begin{equation}
T_{\mbox{\scriptsize\rm eff rad}}
>
T_{\mbox{\scriptsize\rm mean}}  
\end{equation}
%
%
\subsubsection{A numerical example}
From 
Equation~(\ref{Eq:SqrtSumT4}) 
one can construct numerical examples 
where e.g.\ a few high local temperatures 
spoil an average built from a large 
collection of low temperatures.
A more realistic distribution 
is listed in Table~\ref{Table:numerical}.
The effective radiation temperature
$T_{\mbox{\scriptsize\rm eff rad}}$ 
is slightly higher than the 
average
$T_{\mbox{\scriptsize\rm mean}}$
of the measured temperatures.
According to H\"older's inequality 
this will always be the case.
\ifthenelse{\equal{IJMPB}{\TschStyle}}{%
\begin{table}[htbp] 
\tbl{An example for a measured temperature distribution
     from which its associated effective radiation temperature
     is computed.}
{
\begin{tabular}{@{}cccccc@{}} \Hline 
\\[-1.8ex] 
Weather         & Instruments             
                & Absolute          
                & 4th     
                & 4th Root of    
                & 4th Root of    
\\
Station         & Reading                 
                & Temperature       
                & Power      
                & 4th Power Mean 
                & 4th Power Mean 
\\
                & $T_i$ [$^\circ{\rm C}]$ 
                & $T_i$ [${\rm K}]$ 
                & $T_i^4$    
                & $T_{\mbox{\scriptsize\rm eff rad}}$ 
                  [${\rm K}]$ 
                & $T_{\mbox{\scriptsize\rm eff rad}}$
                  [$^\circ{\rm C}]$               
\\[0.8ex] 
\hline          
\\[-1.8ex] 
1               & \phantom{X}0.00         
                & 273.15            
                & 5566789756 
                &                   
                &                   
\\
2               & \phantom{}10.00         
                & 283.15            
                & 6427857849 
                &                   
                &                   
\\
3               & \phantom{}10.00         
                & 283.15            
                & 6427857849 
                &                   
                &                   
\\
4               & \phantom{}20.00         
                & 293.15            
                & 7385154648 
                &                   
                &                   
\\
5               & \phantom{}20.00         
                & 293.15            
                & 7385154648 
                &                   
                &                   
\\
6               & \phantom{}30.00         
                & 303.15            
                & 8445595755 
                &                   
                &                   
\\
Average         & \phantom{}15.00         
                & 288.15            
                & 6939901750 
                & 288,63            
                & 15.48             
\\[0.8ex]
\hline 
\\[-1.8ex]
\end{tabular}
}
\label{Table:numerical}
\end{table}
}{}
\ifthenelse{\equal{arXivOrTeX4ht}{\TschStyles}}{%
\begin{table}[htbp]
{
\begin{center}
\vspace*{0.5cm} 
\begin{tabular}{|c|c|c|c|c|c|} 
\hline 
Weather         & Instruments             
                & Absolute          
                & 4th     
                & 4th Root of    
                & 4th Root of    
\\
Station         & Reading                 
                & Temperature       
                & Power      
                & 4th Power Mean 
                & 4th Power Mean 
\\
                & $T_i$ [$^\circ{\rm C}]$ 
                & $T_i$ [${\rm K}]$ 
                & $T_i^4$    
                & $T_{\mbox{\scriptsize\rm eff rad}}$ 
                  [${\rm K}]$
                & $T_{\mbox{\scriptsize\rm eff rad}}$
                  [$^\circ{\rm C}]$               
\\ 
\hline
1               & \phantom{X}0.00         
                & 273.15            
                & 5566789756 
                &                   
                &                   
\\
2               & \phantom{}10.00         
                & 283.15            
                & 6427857849 
                &                   
                &                   
\\
3               & \phantom{}10.00         
                & 283.15            
                & 6427857849 
                &                   
                &                   
\\
4               & \phantom{}20.00         
                & 293.15            
                & 7385154648 
                &                   
                &                   
\\
5               & \phantom{}20.00         
                & 293.15            
                & 7385154648 
                &                   
                &                   
\\
6               & \phantom{}30.00         
                & 303.15            
                & 8445595755 
                &                   
                &                   
\\
Mean            & \phantom{}15.00         
                & 288.15            
                & 6939901750 
                & 288,63            
                & 15.48  
\\                           
\hline
\end{tabular}
\end{center}
}
\caption{An example for a measured temperature distribution
         from which its associated effective radiation temperature
         is computed. The latter one corresponds to the fourth root 
         of the fourth power mean.  }
\label{Table:numerical}
\vspace*{0.5cm}
\end{table}
}{}

Thus there is no longer any room 
for a \textit{natural greenhouse effect}, 
both mathematically and physically:
\begin{itemize}
\item Departing from the 
      \textit{physically incorrect} 
      assumption of radiative balance a
      \textit{mathematically correct}
      calculation of the average temperature
      lets the difference temperature
      that defines the natural greenhouse effect
      explode.
\item Departing from the 
      \textit{mathematically correct}
      averages of 
      \textit{physically correct} 
      temperatures (i.e.\ measured temperatures)
      the corresponding effective radiation temperature
      will be \textit{always higher} than the  
      average of the measured temperatures. 
\end{itemize}
%
%
\subsubsection{Non-existence of a global temperature}
In the preceding sections mathematical and physical arguments have been presented
that the notion of a global  temperature is meaningless.
Recently, 
Essex, McKitrick, and Andresen showed 
%
\cite{Essex2007}:
%
\begin{quote}
\lq\lq
that there is no physically meaningful 
global temperature for the Earth in the 
context of the issue of global warming. 
While it is always possible to construct 
statistics for any given set of
local temperature data, an infinite range 
of such statistics is mathematically permissible
if physical principles provide no explicit basis 
for choosing among them. Distinct and
equally valid statistical rules can and 
do show opposite trends when applied to the
results of computations from physical models 
and real data in the atmosphere. A given
temperature field can be interpreted 
as both \lq warming\rq\ and \lq cooling\rq\ 
simultaneously, making the concept of warming 
in the context of the issue of global warming 
physically ill-posed.\rq\rq
\end{quote}
Regardless of any ambiguities, a global mean temperature 
could only emerge out of many local temperatures. Without knowledge
of any science everybody can see, how such a changing average 
near-ground temperature is constructed: 
There is more or less sunshine on the ground 
due to the distribution of clouds.
This determines a field of local 
near-ground temperatures, 
which in turn determines the change
of the distribution of clouds and, hence, 
the change of the temperature average, which 
is evidently independent of the carbon dioxide 
concentration.
Mathematically, an evolution of a temperature distribution 
may be phenomenologically described by a differential
equation. The averages are computed  afterwards 
from the solution of this equation. However, one
cannot write down a differential equation directly 
for averages.
%
%
\subsubsection{The rotating globe}
Since the time when Fourier formulated
the heat conduction equation,
a non-linear boundary condition
describing radiative transfer
of a globe with a Sun-side and
a dark side has never belonged 
to the family of solvable heat conduction problems, 
even in the case of a non-rotating globe.

Regardless of solvability, one can
write down the corresponding equations
as well as their boundary conditions.  
\begin{figure}[hbtp]
\ifthenelse{\equal{IJMPB}{\TschStyle}}{%
\centerline{\psfig{file=PictureLarge_Earthball_rotating_.eps,width=4.00in}}
}{}%
\ifthenelse{\equal{arXiv}{\TschStyle}}{%
\centerline{\includegraphics[scale=1.00]{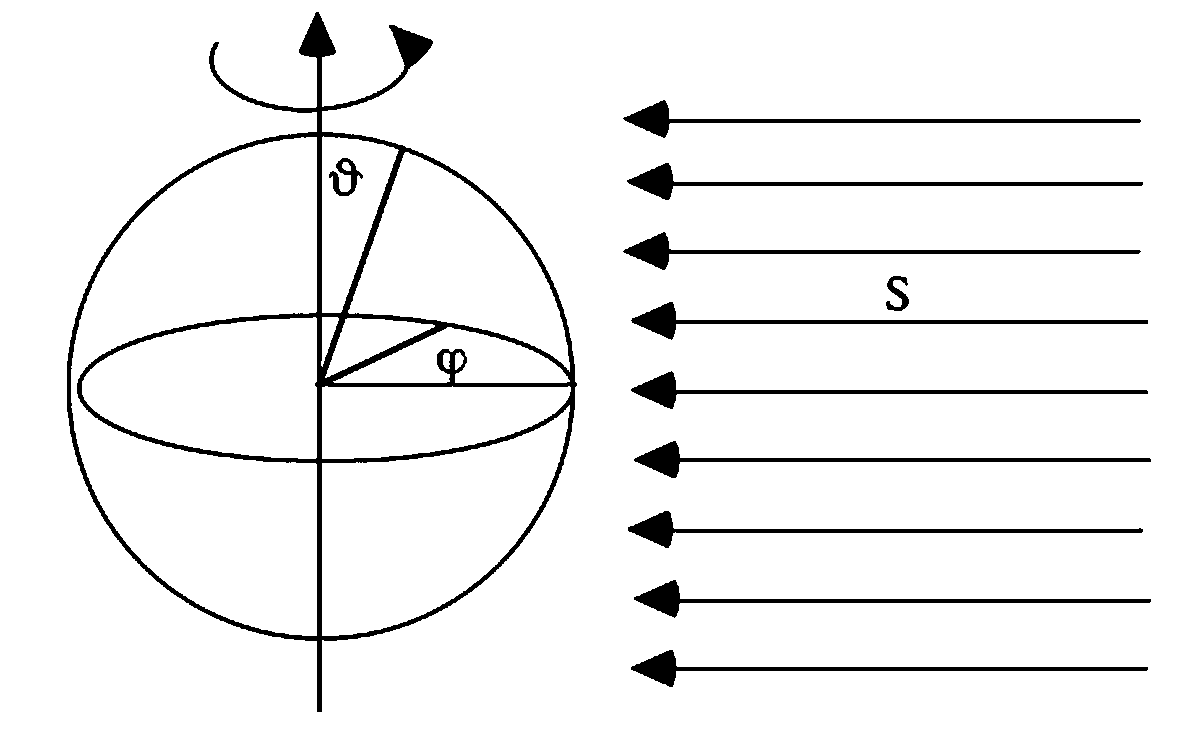}}
}{}%
\ifthenelse{\equal{TeX4ht}{\TschStyle}}{%
\centerline{\includegraphics[scale=1.00]{PictureLarge_Earthball_rotating_.png}}
}{}
\vspace*{8pt}
\caption{The rotating globe.}
\label{fig:rotatingglobe}
\end{figure}
If a rotating globe 
(Fig.~\ref{fig:rotatingglobe}) 
was exposed to radiation and only 
radiative heat transfer to its environment
was possible, the initial problem 
of the heat conduction equation 
would have to be solved with the 
following boundary condition
\begin{equation}
- \lambda\,
\frac{\partial T}{\partial\textbf{n}}
=
\left\{\begin{array}{lcc}
\sigma T^4 - {\rm S} \cdot \sin\vartheta \cos(\varphi-\omega_d t) 
&
\mbox{{\rm \ \ \ if\ }}
&
- \pi/2 \leq \varphi - \omega_d t \leq \pi/2
\\
\sigma T^4
& 
\mbox{{\rm \ \ \ if\ }}
&
\phantom{xx}
\pi/2 \leq \varphi - \omega_d t \leq 3\pi/2
\end{array}\right.
\end{equation}
where
\begin{equation}
\frac{\partial}{\partial\textbf{n}}
=
\textbf{n} \cdot \mbox{\boldmath$\nabla$}
\end{equation}
denotes the usual normal derivative
at the surface of the sphere
and $\omega_d$ the angular frequency 
associated with the day-night cycle.
By defining an appropriate geometry factor
\begin{equation}
\zeta(\vartheta,\varphi, \omega_d, t) = \sin\vartheta \cos(\varphi-\omega_d t) 
\end{equation}
and the corresponding Sun side area 
\begin{equation}
A 
= 
\lbrace 
(\varphi,\vartheta) 
\,|\,  
\zeta(\vartheta,\varphi,\omega_d, t)
\geq 0 
\rbrace
\end{equation} 
one can rewrite the expression as  
\begin{equation}
- \lambda\,
\frac{\partial T}{\partial\textbf{n}}
=
\left\{\begin{array}{lcc}
\sigma T^4 - {\rm S} \cdot \zeta(\vartheta,\varphi, \omega_d, t) 
&
\mbox{{\rm \ \ \ if\ }}
&
(\varphi,\vartheta) \in A
\\
\sigma T^4
& 
\mbox{{\rm \ \ \ if\ }}
&
(\varphi,\vartheta) \not\in A
\end{array}\right.
\end{equation}
%
%
\subsubsection{The obliquely rotating globe}
The result obtained above may be generalized 
to the case of an obliquely rotating globe. 
\begin{figure}[hbtp]
\ifthenelse{\equal{IJMPB}{\TschStyle}}{%
\centerline{\psfig{file=PictureLarge_Earthball_rotating_oblique_.eps,width=4.00in}}
}{}%
\ifthenelse{\equal{arXiv}{\TschStyle}}{%
\centerline{\includegraphics[scale=1.00]{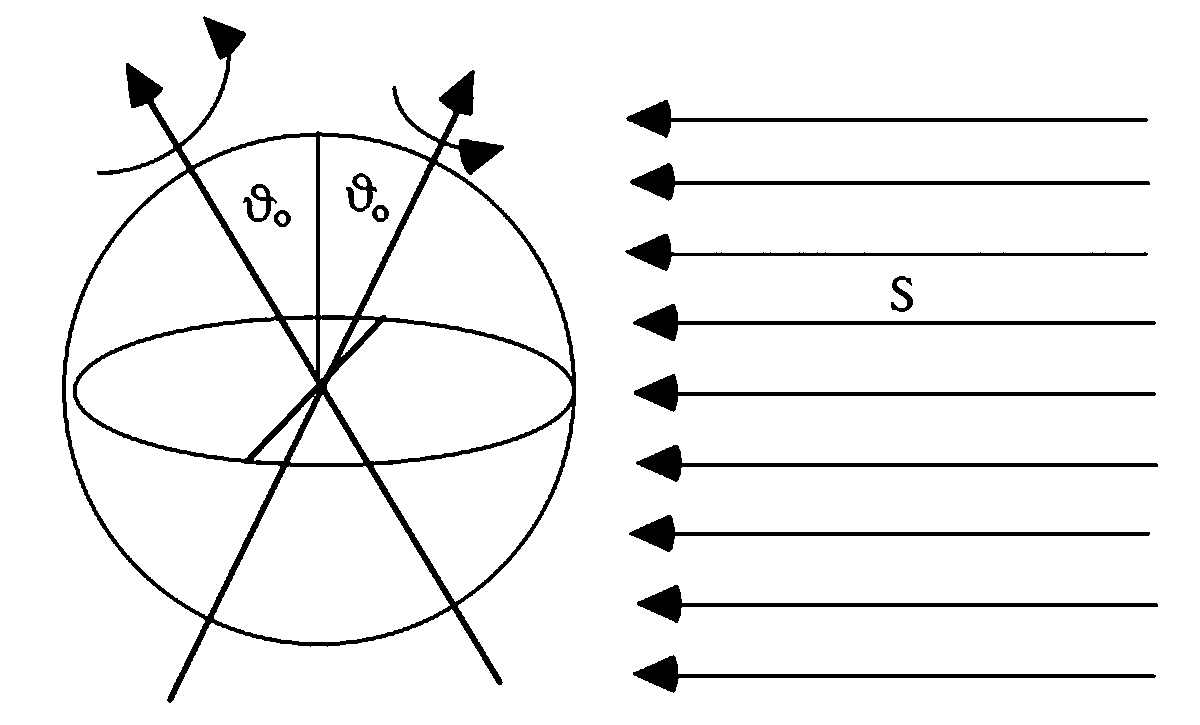}}
}{}%
\ifthenelse{\equal{TeX4ht}{\TschStyle}}{%
\centerline{\includegraphics[scale=1.00]{PictureLarge_Earthball_rotating_oblique_.png}}
}{}
\vspace*{8pt}
\caption{An obliquely rotating globe.}
\label{fig:obliquelyrotatingglobe}
\end{figure}

\noindent%
For an obliquely rotating globe 
(Fig.~\ref{fig:obliquelyrotatingglobe})
one has
\begin{equation}
- \lambda\,
\frac{\partial T}{\partial\textbf{n}}
=
\left\{\begin{array}{lcc}
\sigma T^4 - {\rm S} \cdot \xi(\vartheta_0,\vartheta,\varphi,\omega_y,\omega_d,t) 
&
\mbox{{\rm \ \ \ if\ }}
&
(\varphi,\vartheta) \in A
\\
\sigma T^4
& 
\mbox{{\rm \ \ \ if\ }}
&
(\varphi,\vartheta) \not\in A
\end{array}\right.
\end{equation}
where $\partial/\partial\textbf{n}$ 
denotes the usual normal derivative
on the surface of the sphere 
and $\omega_y$, $\omega_d$ the angular frequencies
with the year cycle and the day-night cycle,
respectively.%
\footnote{Here sidereal time is used\TschSpace%
\cite{%
DictionaryGeophysicsAstronomyAstrophysics2001,%
EncyclopaediaAstronomyAndAstrophysics2001%
}.
} 
The geometry factor now reads
\begin{eqnarray}
\xi(\vartheta_0,\vartheta,\varphi,\omega_y,\omega_d,t)
&=&  
\,\,\,\,\,\,
[
\,\,\,\,\, 
\sin(\omega_y t) \cos(\omega_d t) + \cos(\omega_y t)\sin(\omega_d t)\cos \vartheta_0 
]
\, 
\sin\vartheta\cos\varphi 
\nonumber \\
&&
+
\,
[ 
-
\sin(\omega_y t) \sin(\omega_d t) + \cos(\omega_y t)\cos(\omega_d t)\cos \vartheta_0 
]
\, 
\sin\vartheta\sin\varphi
\nonumber \\
&&
-
\,
[
\,
\cos(\omega_y t)\sin\vartheta_0
\,
]
\,
\cos\vartheta
\end{eqnarray}
and the expression for the sun-side surface
is given by
\begin{equation}
A 
= 
\lbrace 
(\varphi,\vartheta) 
\,|\,  
\xi(\vartheta_0,\vartheta,\varphi,\omega_y,\omega_d,t)
\geq 0
\rbrace 
\end{equation} 
Already the first unrealistic problem 
will be too much for any computer.
The latter more realistic model 
cannot be tackled at all.
The reasons for this is not only 
the extremely different frequencies
$\omega_y$ and $\omega_d$ but also
a very non-physical feature 
which affects the numeric as well: 
According to a famous law formulated by
Wiener, 
almost all particles in this 
mathematical model which cause 
the diffusion, move on paths 
at infinitely high speeds\TschSpace%
%
\cite{Bauer1964,Bauer2002}.

Rough estimates indicate that even these
oversimplified problems cannot be tackled
with any computer.
Taking a sphere with dimensions of the Earth
it will be impossible to solve this problem 
numerically even in the far future.
Not only the computer would work ages,
before a \lq\lq balanced\rq\rq\ temperature 
distribution would be reached,
but also the correct initial temperature
distributions could not be determined
at all.
%
%
\subsubsection{The radiating bulk}
The physical situation of a radiating 
volume where the radiation density
\begin{equation}
{\rm S}(T) = \sigma T^4
\end{equation}
emitted through the surface shell 
originates from the volume's heat content, 
cannot be realized easily, if at all.
However, it is interesting to study 
such a toy model in order to get a feeling about 
radiative equilibration processes which
are assumed to take place within a
reasonable time interval.
  
With disregard to the balancing processes 
inside, one gets the differential equation
\begin{equation}
{\rm V}  
\varrho\,\,
c_{\rm v}
\frac{dT}{dt} 
=
- \Omega \, \sigma T^4
\end{equation}
with 
${\rm V}$ denoting the volume,
$\varrho$ the density,
$c_{\rm v}$ the isochoric specific heat,
$\Omega$ the surface of the body.
By defining 
\begin{equation}
\eta = \frac{\Omega}{{\rm V}}
\end{equation}
the above equation can be rewritten as
\begin{equation}
\frac{dT}{dt} 
=
-
\frac{\eta\,\sigma}{\varrho\,c_{\rm v}}
\cdot T^4
\end{equation}
For a cube with an edge length of $a$
one has $\eta = 6/a$,
for a globe with radius $r$
one has $\eta = 3/r$ instead.
For bodies with unit volumes
$\eta = 6$ or $\eta = 4.8$,
respectively. 

\begin{figure}[hbtp]
\ifthenelse{\equal{IJMPB}{\TschStyle}}{%
\centerline{\psfig{file=PictureLarge_cooling_.eps,width=3.00in}}
}{}%
\ifthenelse{\equal{arXiv}{\TschStyle}}{%
\centerline{\includegraphics[scale=1.00]{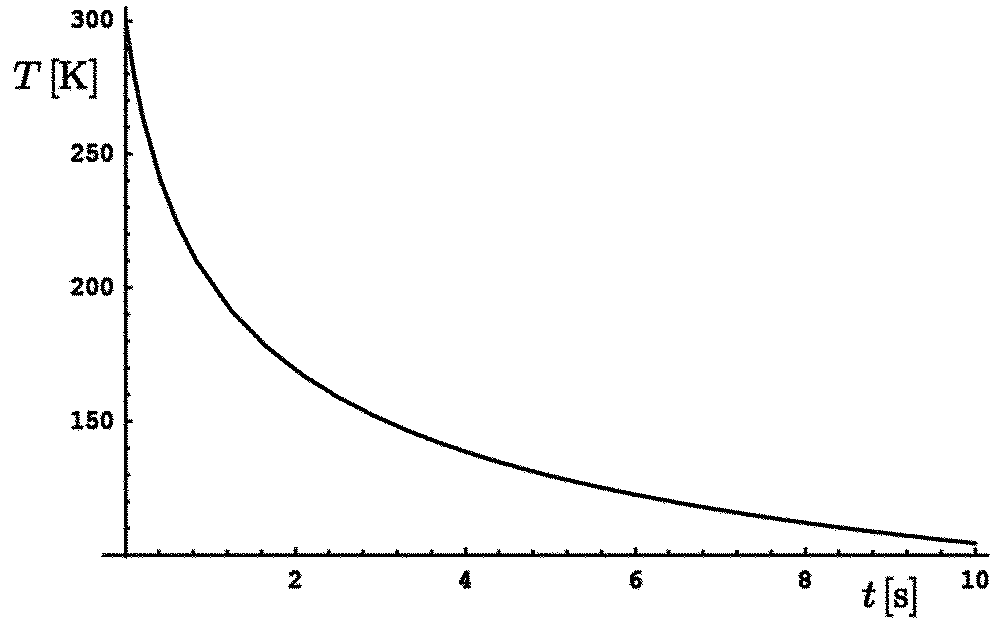}}
\vspace*{-8pt}
}{}%
\ifthenelse{\equal{TeX4ht}{\TschStyle}}{%
\centerline{\includegraphics[scale=1.00]{PictureLarge_cooling_.png}}
\vspace*{-8pt}
}{}
\vspace*{8pt}
\caption{The cooling curve for a 
         radiating standard cube.}
\label{Fig:CubeCooling}         
\end{figure}

The differential equation is easily solvable.
The solution reads
\begin{equation}
T(t) 
=
T_0
/
{
\sqrt[3]
{
1
+
\frac{3\,\eta\,\sigma T_0^3}
     {\varrho\,c_{\rm v}}
\, t
}
}
\end{equation}
At an initial temperature of 
$300\,{\rm K}$ 
with the values of 
$\varrho$ and $c_{\rm v}$ 
for air, one gets one half of the temperature value 
within three seconds for the standard cube
(cf.\ Figure~\ref{Fig:CubeCooling})
For iron the 
isochoric thermal diffusivity 
\begin{equation}
a_{\rm v}=\varrho\,c_{\rm v}
\end{equation}
is about 3000 times 
higher than for air, 
the half time for the temperature 
decrease is approximately three hours.
For air, even if only one of the cube's planes 
were allowed to radiate, one would get 
a fall in temperatures of seventy degrees 
within the first three seconds, 
and almost 290 degrees within ten hours 
- a totally unrealistic cooling process.
 
Hence, this simple assessment will prove
that one has to be extremely careful, 
if the radiation laws for black-body 
radiation, where the energy comes 
from the heated walls of the cavity, 
are to be used for gases, where the 
emitted electromagnetic radiation 
should originate from the movements 
of the gas molecules 
(cf.\ Section~\ref{Sec:Radiation}).
%
%
\subsubsection{The comprehensive work of Schack}
Professor Alfred Schack, 
the author of a standard textbook 
on industrial heat transfer\TschSpace%
%
\cite{AlfredSchackBook}, 
%
was the first scientist who pointed out 
in the twenties of the past century
that the infrared light absorbing fire gas components 
\textit{carbon dioxide} (${\rm CO}_2$) 
and 
\textit{water vapor} (${\rm H}_2{\rm O}$) 
may be responsible 
for a higher heat transfer in the combustion 
chamber \textit{at high burning temperatures} 
through \textit{an increased emission in the infrared}.
He estimated the emissions by measuring 
the spectral absorption capacity of 
carbon dioxide and water vapor.
  
In the year 1972 Schack published a paper in
Physikalische Bl\"atter entitled 
\lq\lq The influence of the carbon dioxide content
of the air on the world's climate\rq\rq.
With his article he got involved in the 
climate discussion and emphasized
the important role of water vapor\TschSpace%
%
\cite{AlfredSchack1972}. 
%

Firstly, Schack estimated the mass 
of the consumed fossil fuels up
\begin{equation}
m_{ 
\mbox{\scriptsize\rm burned}
}
=
5 \cdot 10^{12}\,{\rm kg}
=
5 \, {\rm GtC}
\end{equation}
\textit{per anno}. 
Since $1\,{\rm kg}$ produces 
$10\,{\rm m}^3$ waste gas 
with $15\,\%$ ${\rm CO}_2$,
a volume of 
\begin{equation}
V_{
\mbox{\scriptsize{${\rm CO}_2$}}
}
= 
7.5 \cdot 10^{12}\,{\rm m}^3 
\end{equation}
is blown into the Earth's atmosphere,
whose total volume under normal conditions 
($0\,^\circ{\rm C}$ and 760\,{\rm mm}\,{\rm Hg})
is  
\begin{equation}
V_{ 
\mbox{\scriptsize\rm atmosphere} 
} 
= 
4 \cdot 10^{18}\,{\rm m}^3 
\end{equation}
It follows immediately that the increase 
of the ${\rm CO}_2$ concentration
is approximately 
$1.9\cdot 10^{-6}$
\textit{per anno}. 
About one half is absorbed by the oceans, 
such that the increase of ${\rm CO}_2$
is reduced to
\begin{equation}
\frac
{
\Delta
V_{
\mbox{\scriptsize{${\rm CO}_2$}}
}
}
{
V_{
\mbox{\scriptsize{${\rm CO}_2$}}
}
}
=
0.95\cdot 10^{-6}
\end{equation}
\textit{per anno.}

With the \lq\lq current\rq\rq\ (1972) atmospheric 
${\rm CO}_2$ 
volume concentration of 
\begin{equation}
0.03\,\% = 300 \cdot 10^{-6}
\end{equation}
and an relative annual increase of
\begin{equation}
0.32\,\% 
=
\frac
{0.95 \cdot 10^{-6}}
{300 \cdot 10^{-6}}
\end{equation}
the ${\rm CO}_2$ concentration in the atmosphere 
would rise by one third of current
concentration within 100 years, 
supposed the fossil fuel consumption
will remain constant.

Schack then shows that ${\rm CO}_2$ 
would absorb only one seventh of the ground's 
heat radiation at most, if the water vapor had 
not already absorbed the infrared light 
in most situations. Furthermore, 
a doubling of the ${\rm CO}_2$-content 
in the air would only halve the 
radiation's characteristic absorption length, 
that is, the radiation would be absorbed 
at a length of $5\,{\rm km}$ instead of 
at a length of $10\,{\rm km}$, for example.

Schack discussed the ${\rm CO}_2$ contribution
only under the aspect that ${\rm CO}_2$ acts
as an absorbent medium.
He did not get the absurd idea 
to heat the radiating warmer ground 
with the radiation absorbed and re-radiated
by the gas. 

In a comment on an article by 
the science journalist Rudzinski\TschSpace%
%
\cite{Rudzinski1976}
%
the climatologist Oeschger 
objected against Schack's analysis 
of the influence of the 
${\rm CO}_2$ concentration 
on the climate that Schack had not calculated 
thoroughly  enough\TschSpace%
%
\cite{Oeschger1976}.
%
In particular, he referred to radiation transport calculations.
However, such calculations have formerly been performed  
only for the atmospheres of stars, because the processes 
in planetary atmospheres are far too complicated 
for such simple models. The goal of astrophysical 
radiation transport calculations is to calculate as many 
absorption lines as possible with one boundary density 
distribution and one temperature dependency with respect 
to the height with Saha's equation
and many other additional hypotheses\TschSpace%
%
\cite{Unsoeld1955}.
%
However, the boundary density of the radiation intensity 
cannot be derived from these calculations. 

One should emphasize that Schack was the first 
scientist to take into account the selective 
emission by the infrared light absorbing 
fire-gases for combustion chambers. 
Therefore one is driven to the verge of irritation 
when global climatologists blame 
him for not calculating complicatedly enough, 
simply because he saw the primitive 
physical concepts behind the equations for the 
radiation transfer. 
%

\subsection{Thermal conductivity versus radiative transfer} 
%
%
\subsubsection{The heat equation}
In many climatological texts it seems 
to be implicated that 
thermal radiation does not need to be taken 
into account when dealing with 
heat conduction, which is incorrect\TschSpace%
\cite{Weise1966}.
Rather, always the entire heat flow 
density $\textbf{q}$ must be taken into account.
This is given by the equation
\begin{equation}
\textbf{q} = - \lambda \cdot \mbox{\textbf{grad}} \, T
\end{equation}
in terms of the gradient 
of the temperature $T$.
It is inadmissible to separate 
the radiation transfer from 
the heat conduction, when
balances are computed.

\begin{figure}[htbp]
\ifthenelse{\equal{IJMPB}{\TschStyle}}{%
\centerline{\psfig{file=PictureLarge_heat_conduction_.eps,width=2.00in}}
}{}
\ifthenelse{\equal{arXiv}{\TschStyle}}{%
\centerline{\includegraphics[scale=1.00]{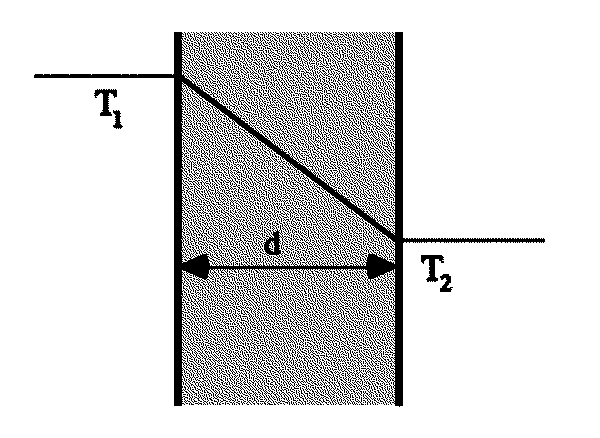}}
\vspace*{-16pt}
}{}
\ifthenelse{\equal{TeX4ht}{\TschStyle}}{%
\centerline{\includegraphics[scale=1.00]{PictureLarge_heat_conduction_.png}}
\vspace*{-16pt}
}{}
\vspace*{8pt}
\caption{A simple heat transport problem.}
\label{Fig:heattransport}
\end{figure}

In the following, a quasi one-dimensional
experimental situation for the determination 
of the thermal conductivity is considered
(Fig.~\ref{Fig:heattransport}).
With $F$ being the cross section,
$d$ the distance between the two walls,
and $Q$ being the heat per time 
transported
from 1 to 2, such that,
\begin{equation}
q_{x} 
= 
\frac{Q}{F} 
\end{equation}
we have
\begin{equation}
Q
= 
F \cdot q_x 
= 
- \lambda  \cdot F \cdot \frac{\partial T}{\partial x}
=
- \lambda \cdot F \cdot \frac{T_2 - T_1}{d}
=
\lambda \cdot F \cdot \frac{T_1 - T_2}{d}
\end{equation}
in case of a stationary temperature distribution. 

$Q$ is produced and measured 
for the stationary situation 
by Joule heat 
(i.e.\ electric heat) 
at the higher temperature. 
The heat transfer by radiation 
cannot be separated from the heat 
transfer of kinetic energy. 
Of course, one tries to avoid 
the heat convection by 
the experimental arrangement. 
Hence any effects 
of the thermal radiation 
(long wave atmospheric radiation to Earth) 
are  simply contained in the stationary 
temperatures and the measured Joule heat.

In the non-stationary case the divergence
of the heat flow no longer vanishes,
and we have for constant thermal conductivity 
\begin{equation}
\mbox{\textbf{div}} \, \textbf{q} 
= 
- \lambda \cdot 
\mbox{\textbf{div}} \, \mbox{\textbf{grad}} \, T
=
- \lambda \cdot
\Delta T
=
- \varrho\,c_{\rm v} \cdot \frac{\partial T}{\partial t}
\label{Eq:preheatequation}
\end{equation}
where 
$\Delta T$ 
is the Laplacean of the temperature 
and 
$\varrho\,c_{\rm v}$
the specific heat of unit volume. We finally obtain

\begin{equation}
\frac{\partial T}{\partial t}
=
\frac{\lambda}{\varrho\,c_{\rm v}}\,
\Delta T
\end{equation}
It is important to note, 
that the thermal conductivity 
is divided by 
$\varrho\,c_{\rm v}$,
which means that 
the isochoric thermal diffusivity
\begin{equation}
a_{\rm v} = \frac{\lambda}{\varrho\,c_{\rm v}}
\end{equation}
of gases and metals 
can be of the the same order of magnitude, 
even if 
the thermal conductivities $\lambda$ 
are completely different. 

Unfortunately, the work on even
the simplest examples of heat conduction
problems needs techniques of mathematical
physics, which are far beyond the 
undergraduate level.
Because a concise treatment of the 
partial differential equations 
lies even outside the scope of this paper,
the following statements should 
suffice: Under certain circumstances 
it is possible to calculate the 
space-time dependent temperature distribution 
with given initial values and boundary conditions. 
If the temperature changes have 
the characteristic length 
$L_{\mbox{\scriptsize\rm char}}$, 
the characteristic time for 
the heat compensation process is
\begin{equation}
\frac{
1
}
{
t_{\mbox{\scriptsize\rm char}}
}
=
\frac{ 
\lambda 
}
{ 
\varrho \, c_{\rm v} 
}
\cdot
\frac{
1
}
{
{\rm L_{\mbox{\scriptsize\rm char}}^2}
}
\end{equation}
If the radius of the Moon were used 
as the characteristic length and typical values 
for the other variables, the 
relaxation time would be equivalent 
\textit{to many times the age of the universe}.
Therefore, an average ground temperature 
(over hundreds of years) is 
\textbf{no indicator at all} 
that the total irradiated solar 
energy is emitted. If there were 
a difference, it would be 
impossible to measure it, due to the 
large relaxation times. 
At long relaxation times, 
the heat flow from the Earth's 
core is an important factor 
for the long term reactions 
of the average ground temperature; 
after all, according to certain hypotheses 
the surfaces of the planetary bodies 
are supposed to have been very hot 
and to have cooled down. These 
temperature changes can never 
be separated experimentally 
from those, which were caused 
by solar radiation.
%
%
\subsubsection{Heat transfer across and near interfaces}
In the real world things become even more complex 
through the existence of interfaces, namely 
\begin{itemize}
\item solid-gas interfaces
\item solid-liquid interfaces
\item liquid-gas interfaces
\end{itemize}
for which a general theory of heat transport does not exist yet.
The mechanisms of air cooling and water cooling and the influence 
of radiation have been studied in engineering thermodynamics\TschSpace%
%
\cite{AlfredSchackBook,Kreith1999,Baukal2000}
%
and are of practical interest e.g.\ in solar collectors, 
fire research, chemistry, nuclear engineering, electronic 
cooling, and in constructing reliable computer hardware\TschSpace%
%
\cite{Safran1994,Bouali2006}.
%
Obviously, they are of utmost importance
in geophysics and atmospheric physics as well. 
Since they add an additional degree of complexity 
to the problem discussed
here, they are not discussed further in this context.    
%
%
\subsubsection{In the kitchen: Physics-obsessed housewife versus IPCC}
In Section~\ref{Sec:WaterPot} 
it was indicated how simple it is 
to falsify 
the atmospheric greenhouse hypotheses,
namely by observing a water pot
on the stove:
Without water filled in, the bottom of the pot
will soon become glowing red.
However, with water filled in, 
the bottom of the pot will be substantially 
colder.

In particular, such an experiment can be performed on
a glass-ceramic stove. The role of the Sun is
played by the electrical heating coils or by 
infrared halogen lamps that are used as heating 
elements. Glas-ceramic has a very low 
heat conduction coefficient, but lets infrared 
radiation pass very well. The dihydrogen monoxide in the pot,
which not only plays the role of the 
\lq\lq greenhouse gas\rq\rq\
but also realizes a very dense phase 
of such a magic substance,
absorbs the infrared extremely well. 
Nevertheless,
there is no additional \lq\lq backwarming\rq\rq\ effect of the
bottom of the pot. In the opposite, the ground becomes colder.

There are countless similar experiments that immediately show that the atmospheric
greenhouse picture is absolutely ridiculous 
from an educated physicist's point of view
or 
from the perspective
of a well-trained salesman offering
high performance tinted glass that reduces
solar heat gain mainly in the infrared\TschSpace%
%
\cite{EfficientWindowsOrg}:
%
\begin{quote}
\lq\lq 
Daylight and view are two of the fundamental 
attributes of a window. Unfortunately, windows 
are also the source of significant solar heat 
gain during times when it is unwanted. 
Traditional solutions to reducing solar 
heat gain such as tinted glazing or shades 
mean that the amount of light is reduced 
as well. New glazings with low-solar-gain 
Low-E (spectrally selective) coatings 
can provide better solar heat gain 
reduction than tinted glass, with 
a minimal loss of visible light. 
This also means that views can be 
clearer and unobstructed.\rq\rq
\end{quote}

Ironically, this works already in the case of 
dihydrogen monoxide.
Such experiments can be performed easily 
on every overhead projector, showing that 
the absorption of the infrared portion of the 
incoming radiation by water is a non-neglible
and leads to a drop of the temperature of
the illuminated surface dressed by an infrared 
absorbing layer that is transparent to visible
light. 
%

%
\subsection{The laws of thermodynamics}
\label{Sec:Thermodynamics}
%
%
\subsubsection{Introduction}
At the time of Fourier's publication\TschSpace%
%
\cite{Fourier1824a,Fourier1824b}
%
the two fundamental laws of 
classical thermodynamics were 
not known.
For each law 
two equivalent versions as 
formulated by Rudolf Clausius
(January 2, 1822 - August 24, 1888),
the founder of axiomatic thermodynamics,
are given by\TschSpace%
%
\cite{Clausius1887a,Clausius1887b}:
%
%
\begin{itemize}
\item 
\textbf{\textit{First law of thermodynamics:}}
\begin{itemize}
\item
\textit{%
In all cases, when work is transformed 
into heat, an amount of heat in proportion 
to the produced work is used up, and vice versa, 
the same amount of heat can be produced by the 
consumption of an equal amount of work.}
\item
\textit{%
Work can be transformed into heat and vice versa, 
where the amount of one is in proportion to 
the amount of the other.}
\end{itemize}
This is a definition of the
\textit{mechanical heat equivalent}.
\end{itemize}
%
%
\begin{itemize}
\item
\textbf{\textit{Second law of thermodynamics:}}
\begin{itemize}
\item
\textit{%
Heat cannot move itself from 
a cooler body into a warmer one.}
\item
\textit{%
A heat transfer from a cooler body into a warmer 
one cannot happen without compensation.}
\end{itemize}
A fictitious heat engine which works 
in this way is called a 
\textit{perpetuum mobile of the second kind}.
\end{itemize}
%
%
Clausius examines thoroughly, that the second law 
is relevant for radiation as well, even if 
image formations with mirrors and lenses are taken 
into account\TschSpace%
%
\cite{Clausius1887a,Clausius1887b}.
%
%
\subsubsection{Diagrams}
\begin{figure}[hbtp]
\ifthenelse{\equal{IJMPB}{\TschStyle}}{%
\centerline{\psfig{file=PictureLarge_machine_steam_engine_.eps,width=2.00in}}
}{}%
\ifthenelse{\equal{arXiv}{\TschStyle}}{%
\centerline{\includegraphics[scale=1.00]{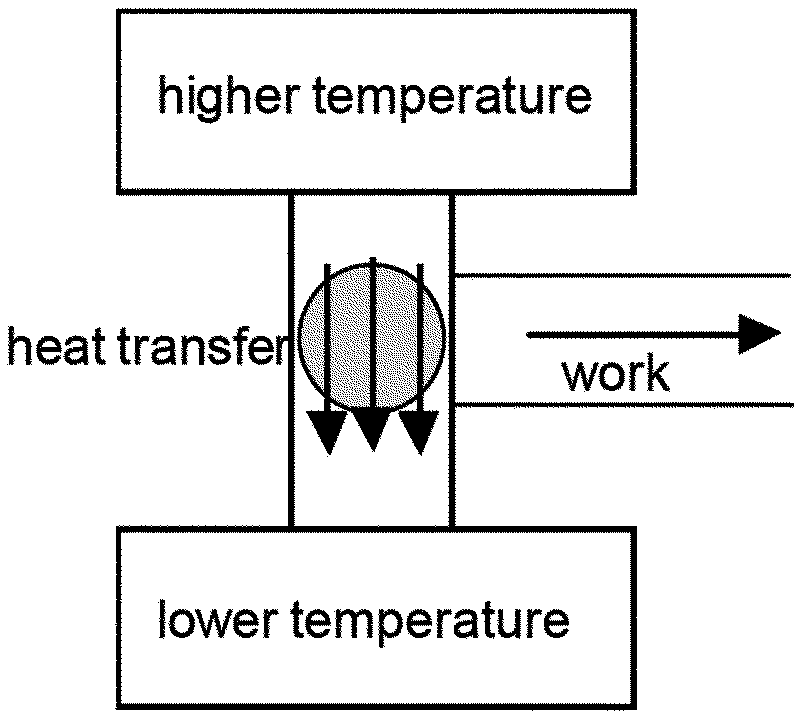}}
}{}%
\ifthenelse{\equal{TeX4ht}{\TschStyle}}{%
\centerline{\includegraphics[scale=1.00]{PictureLarge_machine_steam_engine_.png}}
}{}
\vspace*{8pt}
\caption{A steam engine works
         transforming heat 
         into mechanical energy.}
\label{Fig:SteamEngine}         
\end{figure}

\begin{figure}[hbtp]
\ifthenelse{\equal{IJMPB}{\TschStyle}}{%
\centerline{\psfig{file=PictureLarge_machine_heat_pump_.eps,width=2.00in}}
}{}%
\ifthenelse{\equal{arXiv}{\TschStyle}}{%
\centerline{\includegraphics[scale=1.00]{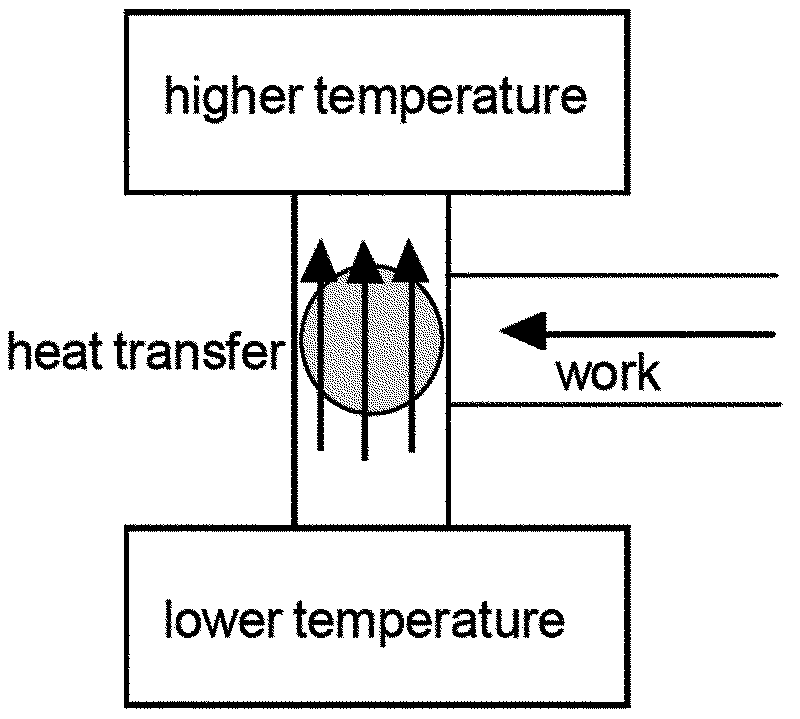}}
}{}%
\ifthenelse{\equal{TeX4ht}{\TschStyle}}{%
\centerline{\includegraphics[scale=1.00]{PictureLarge_machine_heat_pump_.png}}
}{}
\vspace*{8pt}
\caption{A heat pump (e.g.\ a refrigerator)
         works, because an external work is 
         applied.}
\label{Fig:HeatPump}  
\end{figure}

\begin{figure}[hbtp]
\ifthenelse{\equal{IJMPB}{\TschStyle}}{%
\centerline{\psfig{file=PictureLarge_machine_pm2_.eps,width=4.00in}}
}{}%
\ifthenelse{\equal{arXiv}{\TschStyle}}{%
\centerline{\includegraphics[scale=1.00]{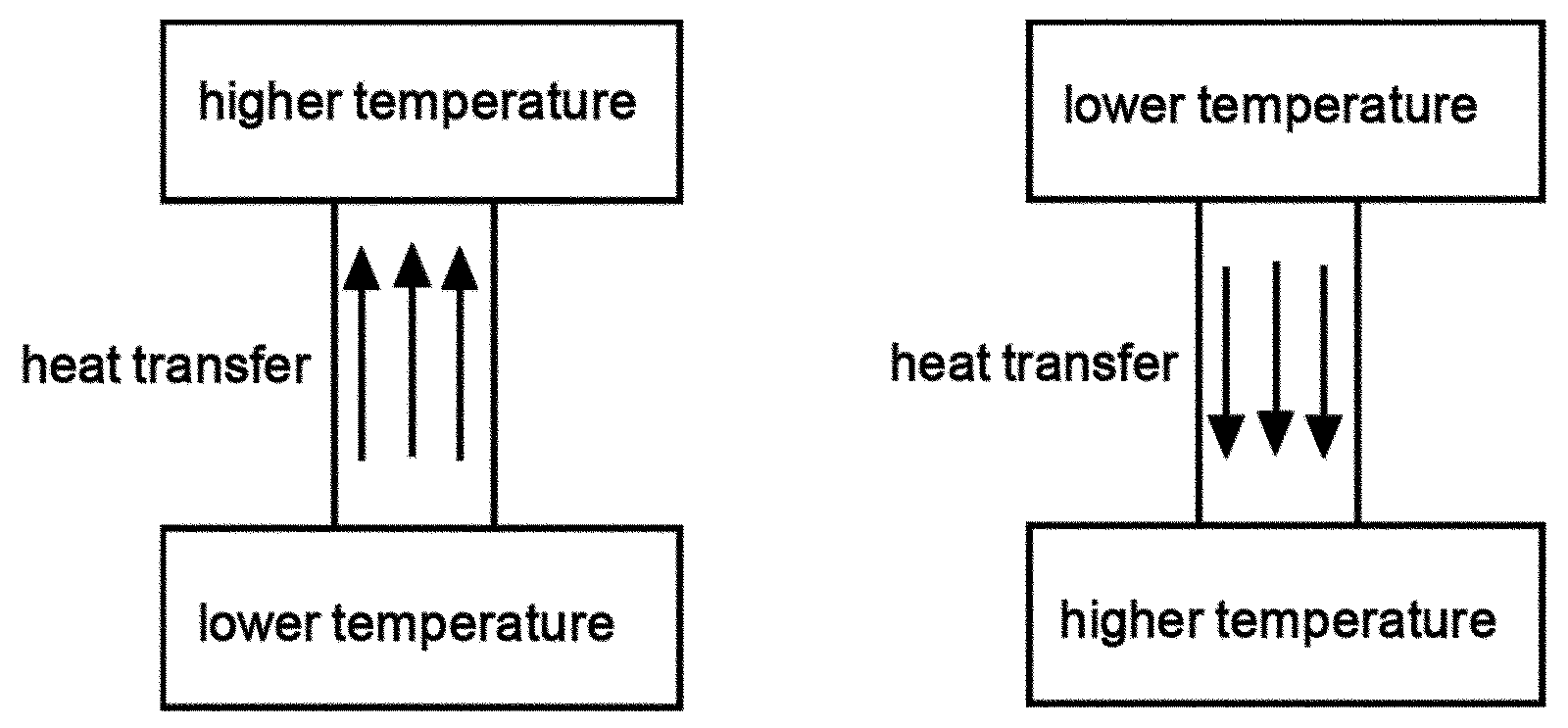}}
}{}%
\ifthenelse{\equal{TeX4ht}{\TschStyle}}{%
\centerline{\includegraphics[scale=1.00]{PictureLarge_machine_pm2_.png}}
}{}
\vspace*{8pt}
\caption{Any machine which transfers heat 
         from a low temperature reservoir 
         to a high temperature reservoir
         without external work applied
         cannot exist:
         A \textit{perpetuum mobile of the second kind}
         is impossible.}
\label{Fig:PerpetuumMobile2}          
\end{figure}

It is quite useful to clarify 
the second law of thermodynamics 
with (self-explaining) diagrams.
\begin{itemize}
\item
A steam engine works
transforming heat 
into mechanical energy,
whereby heat is transferred
from the warmth to the cold
(see\ Figure~\ref{Fig:SteamEngine}).
\end{itemize}
\begin{itemize}
\item
A heat pump (e.g.\ a refrigerator)
works, because an external work is 
applied,
whereby heat is transferred
from the the cold to the warmth
(see\ Figure~\ref{Fig:HeatPump}).
\end{itemize}
\begin{itemize}
\item
In a perpetuum mobile of the second
kind heat is transferred from the 
cold to the warmth without external 
work applied 
(see\ Figure~\ref{Fig:PerpetuumMobile2}).
\end{itemize}
%
%
\subsubsection{A paradox}
\begin{figure}[hbtp]
\ifthenelse{\equal{IJMPB}{\TschStyle}}{
\centerline{\psfig{file=PictureLarge_machine_pm2_climate_model_.eps,width=2.50in}}
}{}%
\ifthenelse{\equal{arXiv}{\TschStyle}}{
\centerline{\includegraphics[scale=1.00]{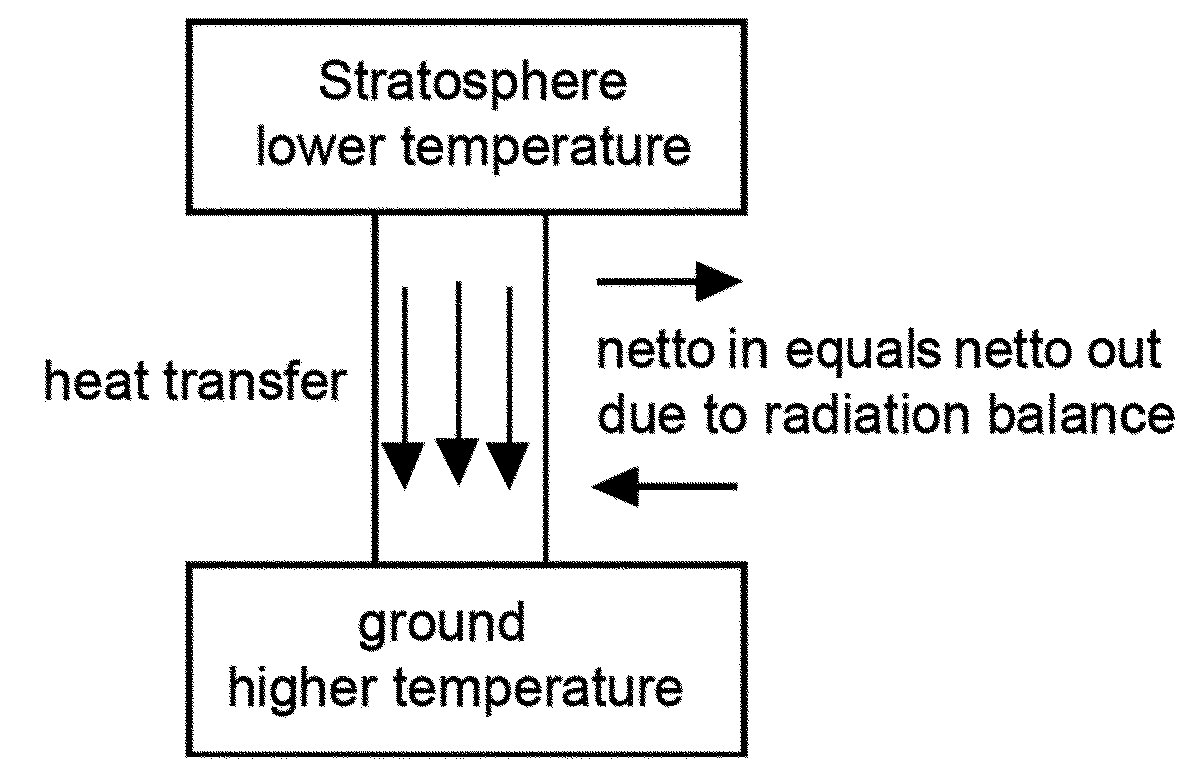}}
}{}%
\ifthenelse{\equal{TeX4ht}{\TschStyle}}{
\centerline{\includegraphics[scale=1.00]{PictureLarge_machine_pm2_climate_model_.png}}
}{}
\vspace*{8pt}
\caption{A machine which transfers heat 
         from a low temperature reservoir (e.g.\ stratosphere) 
         to a high temperature reservoir (e.g.\ atmosphere)
         without external work applied,
         cannot exist - even if it is radiatively coupled to 
         an environment, to which it is radiatively balanced.
         A modern climate model is supposed to be 
         such a variant of a perpetuum mobile of the second kind.}
\label{Fig:PerpetuumMobile3}         
\end{figure}

The use of
\textit{a perpetuum mobile of the second kind} 
can be found in many modern pseudo-explanations 
of the ${\rm CO}_2$-greenhouse effect
(see\ Figure~\ref{Fig:PerpetuumMobile3}). 
Even prominent physicists have relied 
on this argumentation. 
One example was the hypothesis of Stichel
already discussed in Section~\ref{Sec:Stichel1995}\TschSpace%
%
\cite{Stichel1995}.
%

The renowned German climatologist 
Rahmstorf has claimed that the greenhouse 
effect does not contradict the second law of thermodynamics\TschSpace%
%
\cite{Rahmstorf2007}:
%
\begin{quote}
\lq\lq
Some \lq sceptics\rq\ state that the greenhouse effect 
cannot work since (according to the second law of 
thermodynamics) no radiative energy can be transferred 
from a colder body (the atmosphere) to a warmer one 
(the surface). However, the second law is not violated 
by the greenhouse effect, of course, since, during 
the radiative exchange, in both directions the net 
energy flows from the warmth to the cold.%
\rq\rq\
\end{quote}
Rahmstorf's reference to the second law of thermodynamics 
is plainly wrong. The second law is a statement about heat, 
not about energy.  
Furthermore the author introduces an obscure notion of 
\lq\lq net energy flow\rq\rq. The relevant quantity 
is the \lq\lq net heat flow\rq\rq, which, of course, 
is the sum of the upward and the downward heat flow
within a fixed system, here the atmospheric system.
It is inadmissible to apply the second law for 
the upward and downward heat separately 
redefining the thermodynamic system 
on the fly.

A similar confusion is currently 
seen in the German version of Wikipedia\TschSpace%
%
\cite{Wikipedia2007}:
%
\begin{quote}
\lq\lq 
Some have problems with the energy that is radiated 
by the greenhouse gases towards the surface of the 
Earth ($150\,{\rm W}/{\rm m}^2$ - as shown above) 
because this \mbox{energy} flows from a colder body 
(approx.\ $-40\,^\circ{\rm C}$) 
to a warmer one 
(Earth's ground approx.\ $+15\,^\circ{\rm C}$) 
apparently violating the second law of thermodynamics. 
This is a wrong interpretation, since it ignores the radiation 
of the Sun (even 6000 K). With respect to the total 
balance the second law is obeyed indeed.%
\rq\rq
\end{quote}
Obviously, the authors are confusing energy with heat. 
Furthermore, the system in question here is the 
atmospheric system of the Earth including the Earth's 
ground. Since this system is 
\textbf{assumed to be in radiative balance} 
with its environment, and any other forms of 
energy and mass exchange with its environment are 
strictly prohibited, it defines a system 
in the sense of thermodynamics for which the second 
law holds strictly.

The difference among heat, energy 
and work is crucial for the understanding 
of thermodynamics. The second law is a
statement about this difference.
%
%
\subsubsection{Possible resolution of the paradox}
It may be due to the following approximation 
that something is possible in climate models, 
which contradicts the second law of thermodynamics. 
In the field theoretical description of 
irreversible thermodynamics, the second 
law is found in the statement, that the heat 
flow density and the gradient of the temperature
point into opposite directions 
\begin{equation}
\textbf{q} = 
-
\mbox{\large\boldmath$\lambda$}
\cdot 
\textbf{grad} \,T
\end{equation}
In this formula, the heat conduction   
necessarily is a positive definite
tensor. 
In climate models it is customary
to neglect the thermal conductivity
of the atmosphere, which means 
to set it to zero\TschSpace%
%
\cite{Hansen1983}.
%
\begin{equation}
\mbox{\large\boldmath$\lambda$}
= 
\mbox{\large\boldmath$0$}
\end{equation}
This could explain, why the numerical 
simulations could produce small effects 
in contradiction to the second law 
of thermodynamics.
To set the heat conduction to zero 
would not be a real violation 
of the second law of thermodynamics as it corresponds to an
approximation of an ideal system:
In spite of the temperature differences 
no heat flow could move from a warmer area 
to a colder one. It would be in accordance 
to the second law, if there were 
no temperature rise.
In the past, the \lq\lq predictions\rq\rq\ 
of the climate models were pointing 
sometimes in this direction, 
as was shown in detail in 
Section~\ref{Sec:ModernWorks}.
%

\newpage%
\section{Physical Foundations of Climate Science}
%
\subsection{Introduction}
A fundamental theory of the weather and its local averages, 
the climates, must be founded on a reasonable physical theory.
Under the premise that such a theory has
already been formulated there are still two basic problems left 
unresolved, namely 
\begin{itemize}
\item 
the embedding of the purely physical 
theory in a much more wider framework including the 
chemical and biological interactions within the 
geophysical realm, 
\item 
the correct physical account of a possible 
non-trivial radiative effect, which must go far 
beyond the famous black body approach, which is 
suggestive but does not apply to gases.
\end{itemize}
A review of the issues of chemistry and biology 
such as the carbon cycle lies outside 
the perspective of this paper, but it must not be neglected. %
In his criticism of global warming studies
by means of computer models the eminent theoretical
physicist Freeman J.\ Dyson stated\TschSpace%
%
\cite{Dyson2005}:
%
%
\begin{quote}
\lq\lq
The models solve the equations of fluid dynamics, 
and they do a very good job of describing the fluid 
motions of the atmosphere and the oceans. 
They do a very poor job of describing the clouds, 
the dust, the chemistry and the biology of fields 
and farms and forests. 
They do not begin to describe the real world 
that we live in. 
The real world is muddy and messy and 
full of things that we do not yet understand. 
It is much easier for a scientist to sit 
in an air-conditioned building and 
run computer models, than to put on 
winter clothes and measure what is 
really happening outside in the swamps 
and the clouds. 
That is why the climate model experts 
end up believing in their own models.\rq\rq 
\end{quote}
However, it can be shown that even within the borders of 
theoretical physics with or without radiation 
things are extremely complex so that one very 
quickly arrives at a point where verifiable 
predictions no longer can be made. Making such 
predictions nevertheless may be interpreted 
as an escape out of the department of sciences, 
not to say as a scientific fraud.

In the following the conservation laws 
of magnetohydrodynamics are reviewed. 
It is generally accepted that 
a Navier-Stokes-type approach
or a simplified magnetohydrodynamics 
provides the backbone to climatological 
computer simulations\TschSpace%
%
\cite{McGuffie2006,Scaife2007,Chorin1993}.
%
In these frameworks neither the radiative budget 
equations can be derived, nor is it possible 
to integrate radiative interactions in a consistent way. 
Therefore it would conceptually be necessary to go into 
the microscopic regime, which is described by 
non-equilibrium multi-species quantum electrodynamics 
of particles incorporating bound states with internal 
degrees of freedom, whereby the rich structure and 
coexistence of phases have to be taken into account 
in the discussion of natural situations. From these 
only formally sketchable microscopic \textit{ab initio} 
approaches there is no path known that leads to a family 
of more realistic phenomenological climate models\TschSpace%
%
\cite{Zichichi2007}.
%

\subsection{The conservation laws of magnetohydrodynamics}
\label{Sec:TheEquations}
\subsubsection{Overview}
The core of a climate model 
must be a set of equations 
describing the equations of fluid flow, 
namely the Navier-Stokes equations\TschSpace%
%
\cite{Scaife2007,Chorin1993}.
%
The Navier-Stokes equations are nonlinear 
partial differential equations, which, in general,  
are impossible to solve analytically.
In very special cases numerical methods lead to 
useful results, but there is no systematics for 
the general case. In addition, the Navier-Stokes
approach has to be extended to multi-component
problems, which does not simplify the analysis. 

Climate modelers often do not accept that
\lq\lq climate models are too complex and 
uncertain to provide useful projections 
of climate change\rq\rq\TschSpace%
\cite{Mitchell2007}.
Rather, they claim that 
\lq\lq current models enable [them] to attribute 
the causes of past climate change and predict the 
main features of the future climate with a high 
degree of confidence\rq\rq\TschSpace%
\cite{Mitchell2007}.
Evidently, this claim 
(not specifying the observables subject to the prediction)
contradicts to what is well-known 
from theoretical meteorology, namely 
that the predictability of the weather forecast 
models is (and must be) rather limited 
(i.e.\ limited to a few days)\TschSpace%
\cite{Zdunkowski2003}.

The non-solvability of Navier-Stokes-type 
equations is related (but not restricted) 
to the chaotic character of turbulence. 
But this is not the only reason 
why the climate modeling cannot be 
built on a solid ground. 
Equally importantly,
even the full set of equations providing 
a proper model of the atmospheric system
(not to say atmospheric-oceanographic system)
are not known (and never will) to a full extent. 
All models used for \lq\lq simulation\rq\rq\ 
are (and have to be) oversimplified. 
However, in general a set of oversimplified nonlinear 
partial differential equations exhibits 
a totally different behavior than 
a more realistic, more complex 
system. Because there exists no strategy 
for a stepwise refinement 
within the spirit of the renormalization (semi-)group, 
one cannot make any useful predictions.
The real world is too complex to be represented 
properly by a feasible system of equations ready 
for processing\TschSpace%
%
\cite{Zichichi2007}.
%
The only safe statement that can be made is that 
the dynamics of the weather is probably governed 
by a generalized Navier-Stokes-type dynamics. 

Evidently, the electromagnetic interactions
have to be included, leading straightly 
to the discipline of Magnetohydrodynamics (MHD)\TschSpace%
%
\cite{Davidson2001,Gerlich1970,Shu1992a,Shu1992b}.
%
This may be regarded as a set of equations 
expressing all the essential 
physics of a fluid, gas and/or plasma.

In the following these essential equations are reviewed.
The purpose is twofold:
\begin{itemize}
\item Firstly, it should be made a survey 
      of what budget relations really exist
      in the case of atmospheric physical systems.
\item Secondly, the question should be discussed 
      at what point the supposed greenhouse 
      mechanism does enter the equations and where
      the carbon dioxide concentration appears.
\end{itemize}
Unfortunately, the latter aspect seems 
to be obfuscated in the mainstream 
approaches of climatology.

\subsubsection{Electric charge conservation}
As usual, electric charge conservation is described
by the continuity equation
\begin{equation}
\frac{\partial\varrho_e}{\partial t}
+
\mbox{\boldmath$\nabla$} \cdot \textbf{j}
=
0
\label{Eq:ECC}
\end{equation}
where $\varrho_e$ is the electrical (excess) charge density
and $\textbf{j}$ is the electrical (external) current density.
\subsubsection{Mass conservation}
The conservation of mass is described by another
sort of continuity equation
\begin{equation}
\frac{\partial\varrho}{\partial t}
+
\mbox{\boldmath$\nabla$} \cdot ( \varrho\,\textbf{v} )
=
0
\label{Eq:MC}
\end{equation}
where $\varrho$ is the mass density
and $\varrho\,\textbf{v}$ is the
density of the mass current.
\subsubsection{Maxwell's equations}
The electromagnetic fields are described 
by Maxwell's field equations that read
\begin{eqnarray}
%
%
\mbox{\boldmath$\nabla$} \cdot \textbf{D}
&=&
\varrho_e
\\
%
%
\mbox{\boldmath$\nabla$} \times \textbf{E}
&=&
- \, \frac{\partial\textbf{B}}{\partial t}
\\
%
%
\\
\mbox{\boldmath$\nabla$} \cdot \textbf{B}
&=&
0
\\
%
%
\mbox{\boldmath$\nabla$} \times \textbf{H}
&=&
\textbf{j}
+
\frac{\partial\textbf{D}}{\partial t}
%
%
\label{Eq:MXE}
\end{eqnarray}
where the standard notation is used.
They have to be supplemented by
the material equations
\begin{eqnarray}
\textbf{D}
&=&
\varepsilon\,\varepsilon_0\, 
\textbf{E}
\\
\textbf{B}
&=&
\mu\,\mu_0\, 
\textbf{H}
\label{Eq:MTE}
\end{eqnarray}
where $\varepsilon$ and $\mu$ are assumed 
to be constant in space and time, 
an assumption that was already made by Maxwell.

\subsubsection{Ohm's law for moving media}
Electric transport is described 
by Ohm's law for moving media
\begin{eqnarray}
\textbf{j} - \varrho_e \textbf{v}
&=&
\mbox{\Large\boldmath$\sigma$} 
\, 
( \textbf{E} + \textbf{v} \times \textbf{B} )
\label{Eq:OLS}
\end{eqnarray}
with
$\mbox{\Large\boldmath$\sigma$}$ 
being the electrical conductivity tensor. Expressed in terms
of the resistivity tensor
$\mbox{\Large\boldmath$\rho$}$ 
this reads
\begin{eqnarray}
\mbox{\Large\boldmath$\rho$}
\, 
( \textbf{j} - \varrho_e \textbf{v} )
&=&
\textbf{E} + \textbf{v} \times \textbf{B}
\label{Eq:OLMM}
\end{eqnarray}
\subsubsection{Momentum balance equation}
Conservation of momentum is described by
a momentum balance equation, also known 
as Navier-Stokes equation,
\begin{equation}
\frac{\partial}{\partial t}
( \varrho\,\textbf{v} )
+
\mbox{\boldmath$\nabla$}
\cdot 
( \varrho\,\textbf{v}\otimes\textbf{v} )
=
- 
\mbox{\boldmath$\nabla$} p
-
\varrho \, \mbox{\boldmath$\nabla$} \Phi
+
\varrho_e \textbf{E}
+
\textbf{j} \times \textbf{B}
+
\mbox{\boldmath$\nabla$} \cdot \textbf{R}
+
\textbf{F}_{\mbox{\scriptsize\rm ext}}
\label{Eq:MBE}
\end{equation}
where 
$\textbf{v}$ is the velocity vector field,
$p$ the pressure field,
$\Phi$ the gravitational potential,
$\textbf{R}$ the friction tensor,
and 
$\textbf{F}_{\mbox{\scriptsize\rm ext}}$
are the external force densities, 
which could describe the Coriolis and
centrifugal accelerations.
\subsubsection{Total energy balance equation}
The conservation of energy is described by
\begin{eqnarray}
\frac{\partial}{\partial t}
\left(
\frac{ \varrho }{2}
|\textbf{v}|^2
+
\frac{1}{2} 
\,
\textbf{H} \cdot \textbf{B}
+
\frac{1}{2} \textbf{E} \cdot \textbf{D}
+
\varrho \, \Phi
+
\varrho \, u
\right)
+
\phantom{xxxxxxxxxxxxxxxxxxxx}
&&
\nonumber \\
+
\mbox{\boldmath$\nabla$}
\cdot
\left(
\frac{ \varrho }{2}
|\textbf{v}|^2 \, \textbf{v}
+
\textbf{E} \times \textbf{H}
+
\varrho \, \Phi \, \textbf{v}
+
\varrho \, u \, \textbf{v}
+ 
p\,\textbf{v}
-
\textbf{v} \cdot \textbf{R}
+
\mbox{\large\boldmath$\lambda$}
\cdot 
\mbox{\boldmath$\nabla$} T
\right)
=&&
\nonumber \\
=
\varrho\,
\frac{\partial \Phi}{\partial t}
+
\textbf{F}_{\mbox{\scriptsize\rm ext}}
\cdot
\textbf{v}
+
\textbf{Q}
\,\,
&&
\label{Eq:TEBE}
\end{eqnarray} 
where 
$u$ is the density of the internal energy,
$T$ is the temperature field, 
and
$\mbox{\large\boldmath$\lambda$}$
the thermal conductivity tensor,
respectively. Furthermore a term
\textbf{Q} 
has been added which could describe
a heat density source or sink distribution.
\subsubsection{Poynting's theorem} 
From Maxwell's equation with 
space-time independent $\varepsilon$
and $\mu$ one obtains the relation
\begin{equation}
\frac{ \partial }{ \partial t}
\left(
\frac{1}{2}
\,
\textbf{H} \cdot \textbf{B}
+
\frac{1}{2}
\,
\textbf{E} \cdot \textbf{D}
\right)
+
\mbox{\boldmath$\nabla$}
\cdot
( \textbf{E} \times \textbf{H} )
=
- \,
\textbf{j} \cdot \textbf{E}
\label{Eq:PT} 
\end{equation}
This relation is a balance equation.
The Pointing vector field 
$\textbf{E} \times \textbf{H}$
may be interpreted as an energy current density 
of the electromagnetic field.%
\subsubsection{Consequences of the conservation laws}
Multiplying Ohm's law for moving media 
(Equation \ref{Eq:OLMM})
with 
$( \textbf{j} - \varrho_e \, \textbf{v} )$ 
one gets
\begin{eqnarray}
( \textbf{j} - \varrho_e \textbf{v} )
\, 
\mbox{\Large\boldmath$\rho$} 
\, 
( \textbf{j} - \varrho_e \textbf{v} )
&=&
\textbf{j} \cdot \textbf{E} 
+ 
\textbf{j} \cdot ( \textbf{v} \times \textbf{B} )
-
\varrho_e \, \textbf{v} \cdot \textbf{E}
\nonumber\\
&=&
\textbf{j} \cdot \textbf{E} 
- 
\textbf{v} \cdot ( \textbf{j} \times \textbf{B} )
-
\varrho_e \, \textbf{v} \cdot \textbf{E}
\label{Eq:CCL1}
\end{eqnarray}
which may be rewritten as
\begin{equation}
\textbf{j} \cdot \textbf{E}
= 
( \textbf{j} - \varrho_e \textbf{v} )
\, 
\mbox{\Large\boldmath$\rho$} 
\, 
( \textbf{j} - \varrho_e \textbf{v} )
+
\textbf{v} \cdot ( \textbf{j} \times \textbf{B} )
+
\varrho_e \, \textbf{v} \cdot \textbf{E}
\label{Eq:CCL2}
\end{equation}
Inserting this into Poynting's theorem 
(Equation~\ref{Eq:PT}) 
one obtains
\begin{eqnarray}
\frac{ \partial }{ \partial t}
\left(
\frac{1}{2}
\,
\textbf{H} \cdot \textbf{B}
+
\frac{1}{2}
\,
\textbf{E} \cdot \textbf{D}
\right)
+
\mbox{\boldmath$\nabla$}
\cdot
( \textbf{E} \times \textbf{H} )
=
\phantom{xxxxxxxxxxxxxxxxxxxx}
&&
\nonumber\\
=
- \,
( \textbf{j} - \varrho_e \textbf{v} )
\, 
\mbox{\Large\boldmath$\rho$}  
\, 
( \textbf{j} - \varrho_e \textbf{v} )
-
\textbf{v} \cdot
(
\varrho_e \, \textbf{E}
+
\textbf{j} \times \textbf{B}
) 
&&
\label{Eq:CCL3}
\end{eqnarray}
On the other hand, if one applies
the scalar product with $\textbf{v}$
on the momentum balance equation
(\ref{Eq:MBE})
one gets
\begin{eqnarray}
\frac{\partial}{\partial t}
\left( 
\frac{\varrho}{2} \, |\textbf{v}|^2 
\right)
+
\mbox{\boldmath$\nabla$} \cdot 
\left( \frac{\varrho}{2} \, |\textbf{v}|^2 \, \textbf{v} \right)
=
\phantom{xxxxxxxxxxxxxxxxxxxxxxxxxxxxxxxx}
&&
\nonumber\\
=
- 
\textbf{v} \cdot \mbox{\boldmath$\nabla$} p
-
\varrho \, \textbf{v} \cdot \mbox{\boldmath$\nabla$} \Phi
+
\textbf{v} \cdot
( 
\varrho_e \textbf{E}
+
\textbf{j} \times \textbf{B}
)
+
\textbf{v} \cdot
( 
\mbox{\boldmath$\nabla$} \cdot \textbf{R}
)
+
\textbf{v} \cdot 
\textbf{F}_{\mbox{\scriptsize\rm ext}}
\,\,
&&
\label{Eq:CCL4}
\end{eqnarray}
Replacing 
$
\textbf{v} \cdot
( 
\varrho_e \textbf{E}
+
\textbf{j} \times \textbf{B}
)
$
with 
Equation (\ref{Eq:CCL3}) and doing 
some elementary manipulations one 
finally obtains
\begin{eqnarray}
\frac{\partial}{\partial t}
\left(
\frac{\varrho}{2}
|\textbf{v}|^2
+
\frac{1}{2}
\,
\textbf{H} \cdot \textbf{B}
+
\frac{1}{2}
\textbf{E} \cdot \textbf{D}
+
\varrho \, \Phi
\right)
+
\phantom{xxxxxxxxxxxxxxxxxxxxxxxxx}
&&
\nonumber \\
+
\mbox{\boldmath$\nabla$}
\cdot
\left(
\frac{\varrho}{2}
|\textbf{v}|^2
\textbf{v}
+
\textbf{E} \times \textbf{H}
-
\textbf{v} \cdot \textbf{R}
+ 
p\,\textbf{v}
+
\varrho\,\Phi\,\textbf{v}
\right)
=
\phantom{xxxxxxxx}
&&
\nonumber\\ 
=
p\,\mbox{\boldmath$\nabla$}\cdot\textbf{v}
+
\varrho\,\frac{\partial\Phi}{\partial t}
-
{\rm Tr} ( (\mbox{\boldmath$\nabla$}\otimes\textbf{v})\cdot\textbf{R})
- \,
( \textbf{j} - \varrho_e \textbf{v} )
\, 
\mbox{\Large\boldmath$\rho$} 
\, 
( \textbf{j} - \varrho_e \textbf{v} )
+
\textbf{F}_{\mbox{\scriptsize\rm ext}}
\cdot 
\textbf{v} 
\,\,
\label{Eq:WithQ}
\end{eqnarray}
Hence, this relation is a consequence of the 
fundamental equations of magnetohydrodynamics.    
The heat density source term $\textbf{Q}$,
the internal energy density $u$, 
and the divergence of the heat current 
density $\textbf{q}$ are missing here.
\subsubsection{General heat equation}
With
\begin{equation}
du
=
\frac{p}{\varrho^2}
\,
d\varrho
+
T
\,
ds
\end{equation}
for reversible processes one can substitute the density of 
the internal energy $u$ by the density of the entropy $s$.

With the aid of
Equations
(\ref{Eq:TEBE}) 
and 
(\ref{Eq:WithQ})
one derives a differential equation  
for the entropy density $s$:
\begin{eqnarray}
&&
\frac{ \partial(\varrho\,s) }{ \partial t }
+
\mbox{\boldmath$\nabla$} \cdot (\varrho\,s\,\textbf{v})
=
\nonumber \\
&&
\phantom{xxxxxxxxxx}  
=
\frac{1}{T}
\,
{\rm Tr} ( ( \mbox{\boldmath$\nabla$} \otimes \textbf{v} ) \cdot \textbf{R} )
+
\frac{1}{T}
\,
( \textbf{j} - \varrho_e \textbf{v} )
\,
\mbox{\Large\boldmath$\rho$}
\,
( \textbf{j} - \varrho_e \textbf{v} )
\nonumber \\
&& 
\phantom{xxxxxxxxxxxxxxxxxx} 
- \, \frac{1}{T}
\,
\mbox{\boldmath$\nabla$} 
\cdot 
(
\mbox{\large\boldmath$\lambda$}
\cdot 
\mbox{\boldmath$\nabla$} T )
+
\frac{\textbf{Q}}{T}
\label{Eq:GHE}
\end{eqnarray}
This is the generalized form 
of the heat equation.

Only with artificial heat densities 
\textbf{Q} 
in Equations 
(\ref{Eq:GHE})
and
(\ref{Eq:TEBE})
one can incorporate a hypothetical
warming by radiation. There is no term that depends on the carbon dioxide concentration.
\subsubsection{Discussion}
The equations discussed above 
comprise a system of one-fluid equations only. 
One can (and must) write down many-fluid equations 
and, in addition, the averaged equations describing 
the turbulence.
To get a realistic model of the real world,
the above equations \textit{must} be generalized
to take into account
\begin{itemize}
\item the dependency of all relevant coefficients 
      on space and time;
\item the presence and coexistence 
      of various species of fluids and gases;  
\item the inhomogenities of the media,
      the mixture and separation of phases.
\end{itemize}
In principle such a generalization will be feasable, if one
cuts the domains of definition  into pieces and treats the 
equations by a method of patches. 
Thus the final degree of complexity may be much 
larger than originally expected arriving at a system 
of thousands of phenomenological equations 
defining non-linear three-dimensional dynamics and heat transfer.

It cannot be overemphasized that 
even if these equations are simplified
considerably, one cannot determine 
numerical solutions, 
even for small space regions 
and 
even for small time intervals.
This situation will not change 
in the next 1000 years regardless 
of the progress made in computer hardware.
Therefore, global climatologists may continue 
to write updated research grant proposals
demanding next-generation supercomputers
\textit{ad infinitum}.
As the extremely simplified one-fluid 
equations are unsolvable, 
the many-fluid equations 
would be more unsolvable,
the equations that include 
the averaged equations describing 
the turbulence 
would be still more unsolvable,
if \lq\lq unsolvable\rq\rq\ had a comparative.

Regardless of the chosen level of complexity,
these equations are supposed to be the backbone
of climate simulations, or, in other words,  
the foundation of models of nature.
But even this is not true: 
In computer simulations heat conduction and friction
are compületely neglected,
since they are mathematically 
described by 
second order partial derivatives
that cannot be represented on
grids with wide meshes. 
Hence, the computer simulations 
of global climatology 
\textbf{are not based on physical laws.}

The same holds for the speculations about the influence 
of carbon dioxide:
\begin{itemize}
\item Although the electromagnetic field is included
      in the MHD-type global climatologic equations, 
      there are no terms 
      that correspond to the absorption of electromagnetic
      radiation.
\item It is hard if not impossible to find the point
      in the MHD-type global climatologic equations, 
      where the concentration 
      of carbon dioxide enters the game. 
\item It is impossible to include the 
      radiative transfer equation
      (\ref{Eq:RTE})
      into the MHD-type climatologic equations.
\item Apparently, there is no reference in the literature,
      where the carbon dioxide concentration is implemented
      in the MHD-type climatologic equations.       
\end{itemize}
Hence, one is left with the possibility 
to include a hypothetical warming by radiation by hand  in terms of artificial heat densities $\textbf{Q}$ 
in Equation~(\ref{Eq:GHE}). But this would 
be equivalent to imposing 
the \lq\lq political correctly\rq\rq\ 
requested anthropogenic rise 
of the temperature even from the beginning 
just saving an additional trivial calculation.  

In case of partial differential equations
more than the equations themselves  
the boundary conditions determine 
the solutions. 
There are so 
many different transfer phenomena,
radiative transfer,
heat transfer,
momentum transfer,
mass transfer,
energy transfer,
etc.\
and 
many types of interfaces,
static or moving,
between
solids,
fluids, 
gases,
plasmas, etc.\
for which there does not exist
an applicable theory, such that 
one even cannot write down the
boundary conditions\TschSpace%
%
\cite{Safran1994,Bouali2006}.
%

In the \lq\lq approximated\rq\rq\ discretized
equations artificial unphysical boundary conditions
are introduced, in order to prevent running the system
into unphysical states. Such a \lq\lq calculation\rq\rq,
which yields an arbitrary result, is no calculation
in the sense of physics, and hence, in the sense of science.
There is no reason to believe
that global climatologists do not 
know these fundamental scientific facts. 
Nevertheless, in their summaries for
policymakers, global climatologists
claim that they can compute the
influence of carbon dioxide on 
the climates.
%

\subsection{Science and Global Climate Modelling}
%
%
\subsubsection{Science and the Problem of Demarcation}
Science refers to any system of objective knowledge,
in particular knowledge based on the scientific method
as well as an organized body of knowledge gained through 
research\TschSpace%
%
\cite{WikipediaScience2007,REP2007}.
%

There are essentially three categories of sciences, namely
\begin{itemize}
\item formal sciences (mathematics),
\item natural sciences (physics, chemistry, biology)
\item social sciences
\end{itemize}
In natural sciences one has to distinguish 
between 
\begin{itemize}
\item \textit{a theory:}
      a logically self-consistent framework
      for describing the behavior of certain 
      natural phenomena based on fundamental 
      principles; 
\item \textit{a model:}
      a similar but weaker concept than a theory, 
      describing only certain aspects of natural 
      phenomena typically based on some simplified
      working hypothesis;
\item \textit{a law of nature:}
      a scientific generalization based on 
      a sufficiently large number of empirical
      observations that it is taken as fully
      verified;
\item \textit{a hypothesis:}
      a contention that has been neither proved 
      nor yet ruled out by experiment or falsified
      by contradiction to established laws of nature.
\end{itemize}
A \textit{consensus}, exactly speaking 
\textit{a consensus about a hypothesis} 
is a notion which lies outside natural science,
since it is completely irrelevant for 
objective truth of a physical law:
\begin{quote}
\textit{Scientific consens(us) is scientific nonsense}.
\end{quote}
The \textit{problem of demarcation} is how and where to draw
lines around science, i.e.\ to distinguish science
from religion, from pseudoscience, i.e.\ fraudulent
systems that are dressed up as science, and non-science
in general\TschSpace%
%
\cite{WikipediaDemarcationProblem2007,REP2007}.
%

In the philosophy of science several approaches 
to the definition of science are discussed\TschSpace%
%
\cite{WikipediaScience2007,REP2007}:
%
\begin{itemize}
\item 
\textit{empirism%
\footnote{%
also \textit{logical positivism} or \textit{verificationism}
}
(Vienna Circle):}
only statements of empirical observations are meaningful,
i.e.\ if a theory is verifiable, then it will be scientific;
\item 
\textit{falsificationism (Popper):}
if a theory is falsifiable, then it will be scientific;  
\item 
\textit{paradigm shift (Kuhn):}
within the process of normal science anomalies 
are created which lead eventually to a crisis 
finally creating a new paradigm; the acceptance of 
a new paradigm by the scientific community 
indicates a new demarcation between science
and pseudoscience;
\item 
\textit{democratic and anarchist 
approach to science (Feyerabend):}
science is not an autonomous form of reasoning
but inseparable from the larger body of human 
thought and inquiry: \lq\lq Anything goes\rq\rq.
\end{itemize}
Superficially, the last point provides a nice argument 
\textit{for} computer modelers in the framework
of global climatology. 
However, it is highly questionable whether 
this fits into the frame of physics.
Svozil remarked that Feyerabend's understanding of physics 
was superficial\TschSpace%
%
\cite{Svozil2004}.
%
Svozil emphasizes: 
\begin{quote}
\lq\lq Quite generally, partly due to the complexity 
       of the formalism and the new challenges of 
       their findings, which left philosophy proper 
       at a loss, physicists have attempted to develop 
       their own meaning of their subject.\rq\rq
\end{quote}       
Physics provides a fundament for engineering and, hence, for production and modern economics.
Thus the citizen is left with the alternative (in the sense of a choice between two options)
\begin{itemize}
\item[(a)] 
either to accept the derivation of political and economical 
decisions from an anarchic standpoint 
that eventually claims that there is 
a connection to experiment and observation,
and, hence, the real world, when there is 
no such connection;
\item[(b)] 
or to call in the derivation of political and economical
decisions from verifiable research results  within the frame of physics,
where there is a connection to experiment and
observation, and hence, the real world.
\end{itemize}
Evidently, the option (b) defines
a pragmatic approach to science,
defining a minimum of common features, 
such that engineers, managers and 
policymakers have something to rely on: Within the frame of exact sciences
a theory should
\begin{itemize}
\item[(a)] be logically consistent;
\item[(b)] be consistent with observations;
\item[(c)] have a grounding in empirical evidence;
\item[(d)] be economical in the number of assumptions;
\item[(e)] explain the phenomena;
\item[(f)] be able to make predictions;
\item[(g)] be falsifiable and testable;
\item[(h)] be reproducible, at least for the colleagues;
\item[(i)] be correctable;
\item[(j)] be refinable;
\item[(k)] be tentative;
\item[(l)] be understandable by other scientists.
\end{itemize}
Can these criteria ever be met by a computer model approach of global climatology?
%
%
\subsubsection{Evaluation of Climatology and Climate Modelling}
In contrast to meteorology climatology studies the averaged 
  behavior of the local weather. There are several branches,
such as paleoclimatology, historical climatology, and climatology
involving statistical methods which more or less fit into
the realm of sciences. The problem is, what climate modelling
is about, especially if it does refer to chaotic dynamics on the
one hand, and the greenhouse hypothesis on the other.

The equations discussed in 
Section~\ref{Sec:TheEquations} 
may give an idea what the final defining 
equations of the atmospheric and/or oceanic system may look 
like. It has been emphasized that in a more realistic albeit 
phenomenological description of nature the system of the 
relevant equations may be huge. But even by simplifying 
the structure of equations one cannot determine solutions 
numerically, and this will not change, if one does not 
restrict oneself on small space-time domains.

There are serious solvability questions 
in the theory of non-linear partial 
differential equations and the shortage of numerical recipes 
leading to sufficient accurate results will remain 
in the nearer or farer future - for fundamental mathematical 
reasons. The Navier-Stokes equations are something like 
the holy grail of theoretical physics, and a brute force 
discretization with the aid of lattices with very wide meshes 
leads to models, which have nothing to do with the original puzzle 
and thus have no predictability value.

In problems involving partial differential equations the boundary 
condition determine the solutions much more than the differential 
equations themselves. The introduction of a discretization is 
equivalent to an introduction of artificial boundary conditions, 
a procedure, that is characterized in von\,Storch's statement 
\lq\lq The discretization \textit{is} the model\rq\rq\TschSpace%
%
\cite{Storch2005}. 
%
In this context a correct statement of a mathematical 
or theoretical physicist would be: 
\lq\lq A discretization is a model with unphysical 
boundary conditions.\rq\rq\ 
Discretizations of continua problems will be allowed 
if there is a strategy to compute stepwise refinements. 
Without such a renormalization group analysis 
a finite approximation does not lead to 
a physical conclusion. 
However, in
%
\ifthenelse{\equal{IJMPB}{\TschStyle}}
           {Ref.~\refcite{Storch2005}}
           {}%
\ifthenelse{\equal{arXiv}{\TschStyle}}
           {Ref.~\cite{Storch2005}}
           {}%
\ifthenelse{\equal{TeX4ht}{\TschStyle}}
           {Ref.~\cite{Storch2005}}
           {} 
%
von\,Storch emphasized that 
this is by no means the strategy he follows, 
rather he takes the finite 
difference equations are as they are. 
Evidently, this would be a grotesque standpoint, 
if one considered the heat conduction equation, 
being of utmost relevance to the problem and 
being a second order partial differential 
equation, that cannot be replaced by a finite difference model 
with a lattice constant in the range of kilometers.

Generally, it is \textit{impossible} to derive differential equations for averaged functions
and, hence, 
averaged non-linear dynamics
\cite{GerlichSolo1976,GerlichKagermann1977,Emmerich1978,Wulbrand1978}.

Thus there is simply no physical foundation of 
global climate computer models, for which still 
the chaos paradigma holds: Even in the case of 
a well-known deterministic dynamics nothing is 
predictable\TschSpace%
%
\cite{Lorenz1963}.
%
That discretization has neither 
a physical nor a mathematical basis in non-linear 
systems is a lesson that has been taught in the 
discussion of the logistic differential equation, 
whose continuum solutions differ fundamentally 
from the discrete ones
%
\cite{Draper1998,Sprott2003}.
%

Modern global climatology has confused and continues 
to confuse fact with fantasy by introducing the concept 
of a scenario replacing the concept of a model.
In
%
\ifthenelse{\equal{IJMPB}{\TschStyle}}
           {Ref.~\refcite{IPCC2000}}
           {}%
\ifthenelse{\equal{arXiv}{\TschStyle}}
           {Ref.~\cite{IPCC2000}}
           {}%
\ifthenelse{\equal{TeX4ht}{\TschStyle}}
           {Ref.~\cite{IPCC2000}}
           {} 
%
a clear definition of what scenarios are
is given:
\begin{quote}
Future greenhouse gas (GHG) emissions are the product of
very complex dynamics systems, determined by driving forces
such as demographic development, socio-economic
development, and technological change. Their future evolution
is highly uncertain, Scenarios are alternative images of how the
future might unfold and are an appropriate tool with which to
analyze how driving forces may influence future emission
outcomes and to access the associated uncertainties. They assist
in climate change analysis, including climate modeling and the
assessment of impacts, adaptation and mitigation. The 
possibility that any single emissions path will occur as
described in scenarios is highly uncertain. 
\end{quote}
Evidently, this is a description 
of a pseudo-scientific (i.e.\ non-scientific) method
by the experts at the IPCC. 
The next meta-plane 
beyond physics would be a questionnaire among 
scientists already performed by von\,Storch\TschSpace%
%
\cite{Storch2007}
%
or, finally, a democratic vote about 
the validity of a physical law.
Exact science is going to be replaced by 
a sociological methodology involving 
a statistical field analysis and 
by \lq\lq democratic\rq\rq\ rules 
of order. This is in harmony with 
the definition of science advocated 
by the \lq\lq scientific\rq\rq\ website
\textit{RealClimate.org}
that has integrated 
inflammatory statements,
personal attacks and offenses 
against authors as a part 
of their \lq\lq scientific\rq\rq\
workflow.  
%
%
\subsubsection{Conclusion}
A statistical analysis, no matter how sophisticated it is, 
heavily relies on underlying models and if the latter are 
plainly wrong then the analysis leads to nothing. 
One cannot detect and attribute something that 
does not exist for reason of principle like the
${\rm CO}_2$ greenhouse effect. 
There are so many unsolved and unsolvable problems 
in non-linearity and the climatologists believe to 
beat them all by working with crude approximations
leading to unphysical results that have been corrected 
afterwards by mystic methods, flux control in the past,
obscure ensemble averages 
over different climate institutes today, 
by excluding accidental global cooling results by hand\TschSpace%
\cite{Stainforth2005},
continuing the greenhouse 
inspired global climatologic tradition of
\textit{physically meaningless} averages
and 
\textit{physically meaningless} 
applications of mathematical statistics.

In conclusion, the derivation of statements on the 
${\rm CO}_2$ induced anthropogenic global warming 
out of the computer simulations lies 
\textbf{outside any science}.

\newpage%
\section{Physicist's Summary}
A thorough discussion of the 
planetary heat transfer problem  
in the framework of theoretical 
physics and engineering thermodynamics 
leads to the following results:
\begin{enumerate}
\item
There are no common physical laws between 
the warming phenomenon in glass houses and 
the fictitious atmospheric greenhouse effect, 
which explains the relevant physical phenomena.
The terms \lq\lq greenhouse effect\rq\rq\
and \lq\lq greenhouse gases\rq\rq\ are
deliberate misnomers.
\item
There are no calculations to determinate 
an average surface temperature of a planet 
\begin{itemize}
\item with or without an atmosphere, 
\item with or without rotation, 
\item with or without infrared light 
      absorbing gases. 
\end{itemize}
The frequently mentioned difference 
of 33~$^\circ{\rm C}$ for the fictitious 
greenhouse effect of the atmosphere 
is therefore a meaningless number.
\item 
Any radiation balance for the average 
radiant flux is completely irrelevant 
for the determination of the ground level 
air temperatures and thus for the average 
value as well.
\item
Average temperature values cannot 
be identified with
the fourth root of average values 
of the absolute temperature's fourth 
power.
\item 
Radiation and heat flows do not determine 
the temperature distributions and their 
average values.
\item
Re-emission is not reflection and can in no 
way heat up the ground-level air against 
the actual heat flow without mechanical work.
\item 
The temperature rises in the climate model 
computations are made plausible by a perpetuum 
mobile of the second kind. This is possible by setting 
the thermal conductivity in the atmospheric models 
to zero, an unphysical assumption. 
It would be no longer a perpetuum mobile of the second kind,
if the \lq\lq average\rq\rq\ fictitious radiation balance, 
which has no physical justification anyway, was given up.
\item
After Schack 1972 water vapor is responsible for most 
of the absorption of the infrared radiation 
in the Earth's atmosphere. 
The wavelength of the part of radiation, 
which is absorbed by carbon dioxide
is only a small part of the full 
infrared spectrum and does not change 
considerably by raising its partial pressure.  
\item
Infrared absorption does not imply \lq\lq backwarming\rq\rq.
Rather it may lead to a drop of the temperature 
of the illuminated surface.
\item
In radiation transport models with the assumption 
of local thermal equilibrium, it is assumed that 
the absorbed radiation is transformed into the 
thermal movement of all gas molecules. 
There is no increased selective re-emission 
of infrared radiation at the low temperatures 
of the Earth's atmosphere.
\item
In climate models, planetary or astrophysical
mechanisms are not accounted for properly.
The time dependency of the gravity acceleration 
by the Moon and the Sun (high tide and low tide) 
and the local geographic situation, which is important 
for the local climate, cannot be taken into account.
\item 
Detection and attribution studies, 
predictions from computer models 
in chaotic systems,
and the concept of scenario analysis
lie outside the framework 
of exact sciences,
in particular theoretical physics.
\item
The choice of an appropriate 
discretization method and 
the definition of appropriate
dynamical constraints (flux control)
having become a part of computer modelling 
is nothing but another form of 
data curve fitting. 
The mathematical physicist v.\,Neumann
once said to his young collaborators: 
\lq\lq If you allow me four free parameters 
       I can build a mathematical model that 
       describes exactly everything that 
       an elephant can do. If you allow me 
       a fifth free parameter, the model 
       I build will forecast that the elephant 
       will fly.{\rq\rq}\
%
\ifthenelse{\equal{IJMPB}{\TschStyle}}
           {(cf.\ Ref.~\refcite{Zichichi2007}.)}
           {}%
\ifthenelse{\equal{arXiv}{\TschStyle}}
           {(cf.\ Ref.~\cite{Zichichi2007}.)}
           {}%
\ifthenelse{\equal{TeX4ht}{\TschStyle}}
           {(cf.\ Ref.~\cite{Zichichi2007}.)}
           {} 
%
\item    
Higher derivative operators 
(e.g.\ the Laplacian) can never
be represented on grids with wide meshes. 
Therefore a description of heat conduction
in global computer models is impossible. 
The heat conduction equation is not and cannot properly
be represented on grids with wide meshes.

\item 
Computer models of higher dimensional 
chaotic systems, best described by 
non-linear partial differential equations
(i.e.\ Navier-Stokes equations),
fundamentally differ from calculations 
where perturbation theory is applicable
and successive improvements of the
predictions - by raising the computing power -
are possible. At best, these computer models
may be regarded as a heuristic game.
\item 
Climatology misinterprets unpredictability of
chaos known as butterfly phenomenon as another
threat to the health of the Earth.
\end{enumerate}
In other words: 
Already the natural greenhouse effect 
is a myth beyond physical reality.
The ${\rm CO}_2$-greenhouse effect, 
however is a \lq\lq mirage\rq\rq\TschSpace%
%
\cite{Thieme2007}.
%
The horror visions of a risen sea level, 
melting pole caps and developing deserts 
in North America and in Europe 
are \textit{fictitious consequences} 
of \textit{fictitious physical mechanisms} 
as they cannot be seen even 
in the climate model computations.
The emergence of hurricanes and tornados
cannot be predicted by climate models, 
because all of these deviations 
are ruled out. 
The main strategy of modern 
${\rm CO}_2$-greenhouse gas defenders 
seems to hide themselves behind more
and more pseudo-explanations, 
which are not part of the academic education 
or even of the physics training. 
A good example are the radiation transport 
calculations, which are probably not 
known by many. Another example are
the so-called feedback mechanisms, which are 
introduced to amplify an effect
which is not marginal but does not
exist at all.
Evidently, the defenders of the
${\rm CO}_2$-greenhouse thesis refuse to accept 
any reproducible calculation as an explanation 
and have resorted to unreproducible ones.
A theoretical physicist must complain 
about a lack of transparency here, 
and he also has to complain about 
the style of the scientific discussion, 
where advocators of the greenhouse thesis 
claim that the discussion is closed, 
and others are discrediting justified arguments
as a discussion of \lq\lq questions of yesterday
and the day before yesterday\rq\rq%
\footnote{%
a phrase used by von\,Storch in
%
\ifthenelse{\equal{IJMPB}{\TschStyle}}
           {Ref.~\refcite{Stockholm2006}}
           {}%
\ifthenelse{\equal{arXivOrTeX4ht}{\TschStyles}}
           {Ref.~\cite{Stockholm2006}}
           {}
%
}.
In exact sciences, in particular in theoretical 
physics, the discussion is never closed and
is to be continued \textit{ad infinitum}, 
even if there are proofs of theorems available. 
Regardless of the specific field of studies 
a minimal basic rule should be fulfilled 
in natural science, 
though, even if the scientific fields 
are methodically as far apart as physics 
and meteorology: At least among experts, 
the results and conclusions should 
be understandable or reproducible. 
And it should be strictly distinguished
between a theory and a model on the one hand, and between a model and a scenario
on the other hand, as clarified in the philosophy of science.

That means that 
if conclusions out of computer simulations 
are to be more than simple speculations, 
then in addition to the examination of the 
numerical stability and the estimation 
of the effects of the many vague 
input parameters, at least the 
simplifications of the physical original 
equations should be critically exposed. 
Not the critics have to estimate the effects 
of the approximation, but the scientists 
who do the computer simulations. 

\lq\lq Global warming is good $\dots$
       The net effect of a modest global warming
       is positive.\rq\rq\ (Singer).%
\footnote{cf.\ Singer's summary at the Stockholm 2006 conference\TschSpace%
%
\cite{Stockholm2006}.
%
}
In any case, it is extremely interesting to understand
the dynamics and causes of the long-term fluctuations 
of the climates. However, it was not the purpose of this 
paper to get into all aspects of the climate variability 
debate.
 
The point discussed here was to answer the question,
whether the supposed atmospheric effect
has a physical basis. This is not the case. 
In summary, there is no atmospheric greenhouse 
effect, in particular ${\rm CO}_2$-greenhouse effect, 
in theoretical physics and engineering thermodynamics.
Thus it is illegitimate to deduce predictions
which provide a consulting solution
for economics and intergovernmental policy.%

\newpage%
\addcontentsline{toc}{section}{Acknowledgement}
\section*{Acknowledgements}
This work is dedicated 
(a) to the late 
Professor S.\ Chandrasekhar,
whom 
R.D.T.\ 
met in Chicago in 1991, 
(b) to the late Professor 
C.\ F.\ v.\ Weizs\"acker, 
a respected discussion partner 
of both authors, and 
(c) the late investigative science journalist 
H.\ Heuseler, 
whom G.G.\ owes valuable information on the topic.

Both authors would like to thank many people 
for discussions, email exchanges, and support
at various stages of this work, in particular 
%
StD Dipl.-Biol.\ Ernst-Georg Beck,
H.\ J.\ Labohm,
Professor B.\ Peiser, 
H.\ Thieme, 
Dr.\,phil.\ Wolfgang Th\"une,
and 
Professor A.\ Zichichi
%
for sending them the manuscript of his
talk presented at the Vatican conference.
Mrs.\ S.\ Feldhusen's first translation of
%
\ifthenelse{\equal{IJMPB}{\TschStyle}}
           {Ref.~\refcite{Gerlich1995}}
           {}%
\ifthenelse{\equal{arXiv}{\TschStyle}}
           {Ref.~\cite{Gerlich1995}}
           {}%
\ifthenelse{\equal{TeX4ht}{\TschStyle}}
           {Ref.~\cite{Gerlich1995}}
           {}
%
is greatly appreciated. 

Gerhard Gerlich would like to express his gratitude
to all those who contributed to this study 
either directly or indirectly: 
Students, Staff Members, 
Research and Teaching Assistants,
even collegues,
who listened to his lectures and talks,
who read his texts critically,
who did some successful literature search. 
In particular, he is indebted to
the Diploma Physicists (Diplomphysiker) 
Dr.\ V.\ Blahnik, 
Dr.\ T.\ Dietert, 
Dr.\ M.\ Guthmann,
Dr.\ F.\ Hoffmann,
Dr.\ G.\ Linke,  
Dr.\ K.\ Pahlke, 
Dr.\ U.\ Schom\"acker, 
H.\ Bade,
M.\ Behrens,
C.\ Bollmann,
R.\ Fl\"ogel, 
StR D.\ Harms,
J.\ Hauschildt,
C.\ Mangelsdorf, 
D.\ Osten,
M.\ Schmelzer,
A.\ S\"ohn,
and
G.\ T\"or\"o,
the architects 
P.\ Bossart 
and 
Dipl.-Ing.\ K.\ Fischer.
%
Gerhard Gerlich extends his special gratitude to  
Dr.\ G.-R.\ Weber 
for very early bringing 
his attention to the outstanding DOE 1985 report\TschSpace%
%
\cite{DOE1985}
%
to which almost no German author contributed.
Finally, he is pleased about the interest 
of the many scientific laymen who enjoyed 
his talks, his letters, and his comments.

Ralf D.\ Tscheuschner thanks all his students 
who formulated and collected a bunch of questions 
about climate physics, in particular Elvir Donl\'ic. 
He also thanks 
Professor A.\ Bunde 
for email correspondence.
Finally he is indebted to 
Dr.\ M.\ Dinter,  
C.\ Kloe\ss,
M.\ K\"ock,
R.\ Schulz 
for interesting discussions,
and 
Professor H.\ Gra\ss l 
for an enlightening discussion 
after his talk on Feb.\,2, 2007 at 
Planetarium Hamburg. A critical reading by 
M.\ Mross and Dr.\ M.\ Dinter 
and a translation
of Fourier's 1824 paper 
in part by Melanie Willer's team 
and by Dr.\ M.\ Dinter are especially 
acknowledged.

The authors express their hope that in the schools around the world
the fundamentals of physics will be taught correctly and not by using 
award-winning  
\lq\lq Al Gore\rq\rq\ movies shocking every straight physicist
by confusing 
absorption/emission with reflection,
by confusing the tropopause with the ionosphere,
and by confusing microwaves with shortwaves.

\newpage
\addcontentsline{toc}{section}{List of Figures}
\listoffigures
\newpage
\addcontentsline{toc}{section}{List of Tables}
\listoftables
\newpage
\ifthenelse{\equal{IJMPB}{\TschStyle}}{\section*{References}}{}


%
%
%
\end{document}